\documentclass[11pt]{article}
\usepackage{graphicx}
\usepackage{amssymb}
\usepackage{amsmath}
\usepackage{graphicx}
\usepackage{amstext}
\usepackage{leftidx}
\usepackage{esint}
\usepackage[utf8]{inputenc}
\usepackage{color}
\usepackage{cite}
\usepackage{hyperref}
\usepackage{multido}
\usepackage{comment}
\usepackage{appendix}
\usepackage{float}
\usepackage{placeins}
\makeatletter
\ams@newcommand{\vardot}[2]{%
  {\mathop{#2\kern0pt}\limits^{\vbox to-1.4\ex@{\kern-\tw@\ex@
   \hbox{\normalfont\multido{}{#1}{.}}\vss}}}}
\makeatother
\newcommand{\CC}{\Lambda}

\newcommand{\rv}{\rho_{\rm vac}}

\newcommand{\rvo}{\rho^0_{\rm vac}}

\newcommand{\nueff}{\nu_{\rm eff}}

\newcommand{\mpl}{m_{\rm Pl}}

\newcommand{\be}{\begin{equation}}
\newcommand{\ee}{\end{equation}}

\newcommand{\jtext}[1]{{\textcolor{black}{#1}}}

\newcommand{\wv}{w_{\rm vac}}

\newcommand{\zstar}{z_{*}}

\begin{document}

\hyphenation{theo-re-ti-cal gra-vi-ta-tio-nal theo-re-ti-cally mo-dels stu-dies e-vol-ving la-test li-te-ra-tu-re cos-mo-lo-gi-cal va-cu-um pa-ra-me-ter}

\begin{center}
%{\bf \LARGE Running Vacuum from QFT in curved spacetime}\\
\vskip 2mm
{\bf \LARGE   Running vacuum in the Universe:  phenomenological status in light of the latest observations, and its impact\\ on the  $\sigma_8$ and $H_0$ tensions} \vskip 2mm
%{\bf \Large The dynamics of $\rv(H)$ from the quantized matter fields}

 \vskip 8mm

\textbf{ \Large Joan Sol\`a Peracaula$^1$, Adri\`a G\'omez-Valent$^2$, \\  Javier de Cruz P\'erez$^3$  and Cristian Moreno-Pulido$^1$}

\vskip 5mm
$^1$ Departament de F\'isica Qu\`antica i Astrof\'isica, \\
and   Institute of Cosmos Sciences,\\ Universitat de Barcelona,
Av. Diagonal 647, 08028 Barcelona, Catalonia, Spain

\vskip 5mm

$^2$ INFN, Sezione di Roma 2, and Dipartimento di Fisica, Universit\`a degli Studi di Roma Tor Vergata, via della Ricerca Scientifica 1, 00133 Roma, Italy

\vskip 5mm

$^3$ Departamento de F\' \i sica Te\'orica,
Universidad Complutense de Madrid, 28040 Madrid, Spain

\vskip 5mm

E-mails: sola@fqa.ub.edu, agvalent@roma2.infn.it, jadecruz@ucm.es, cristian.moreno@fqa.ub.edu

 \vskip2mm

\end{center}
\vskip 15mm

\begin{quotation}
\noindent {\large\it \underline{Abstract}}.
A substantial body of phenomenological and theoretical work over the last few years strengthens  the possibility that the vacuum energy density (VED) of the universe is dynamical, and in particular that it adopts  the `running vacuum model' (RVM) form,   in which  the VED evolves  mildly as  $\delta \rho_{\rm vac}(H)\sim \nu_{\rm eff} m_{\rm Pl}^2{\cal O}\left(H^2\right)$, where $H$ is the Hubble rate and $\nu_{\rm eff}$ is a (small) free parameter. This dynamical scenario is grounded on recent studies of quantum field theory (QFT) in curved spacetime and also on string theory.  It turns out  that what we call the `cosmological constant', $\Lambda$,   is no longer a rigid parameter but the nearly sustained value of  $8\pi G(H)\rho_{\rm vac}(H)$  around (any)  given epoch $H(t)$, where $G(H)$ is the  gravitational coupling, which can  also be very mildly running (logarithmically). Of particular interest is the possibility suggested in past works that such a running may help to cure the cosmological tensions afflicting the $\Lambda$CDM.  In the current study, we reanalyze it in full and we find it becomes further buttressed. Using the modern cosmological data, namely a compilation of the latest SNIa+BAO+$H(z)$+LSS+CMB observations, we probe to which extent the RVM provides a quality fit better than the concordance $\Lambda$CDM model, paying particular emphasis on its  impact on the $\sigma_8$ and $H_0$ tensions.  We utilize the Einstein-Boltzmann system solver  \texttt{CLASS} and the Monte Carlo sampler \texttt{MontePython} for the statistical analysis, as well as the statistical DIC criterion to compare the running vacuum against the rigid vacuum ($\nu_{\rm eff} = 0$). On fundamental grounds,  $\nu_{\rm eff}$ receives contributions from all the quantized matter fields in FLRW spacetime. We show that with a tiny amount of vacuum dynamics ($|\nu_{\rm eff}|\ll 1$)  the global fit can improve significantly with respect to  the $\CC$CDM and the mentioned tensions may  subside to inconspicuous levels.

\end{quotation}

\newpage

\tableofcontents

%\newpage
\section{Introduction}
The vanilla concordance model of cosmology, or standard $\CC$CDM model  (the current standard model of cosmology with flat three-dimensional geometry), is based on the Friedmann-Lemaître-Robertson-Walker (FLRW) metric and has been a rather successful  paradigm for the phenomenological description of the universe  for more than three decades\,\cite{peebles:1993,Peebles:1984ge}.  Its consolidation after a solid observational underpinning, however,  was only possible in the late nineties\,\cite{Turner:2022gvw}. The vanilla model has remained robust and unbeaten for a long time, as  it is essentially consistent with a large body of observations.  These have indeed provided strong support for a spatially flat and accelerating universe in the present time. The ultimate cause of such an acceleration is currently unknown, but it is attributed to an energy component in the universe  popularly  called  ``dark energy" (DE), which may adopt a large number of picturesque forms depending on the favorite theoretical preference of different cosmologists\, see e.g. \cite{Amendola:2015ksp} for a large variety of options. The DE constitutes $\sim70\%$ of the total energy density of the universe and presumably possesses enough negative pressure as to produce the observed cosmic acceleration.   Nevertheless, the nature of the DE  remains still a complete mystery.  The simplest candidate is the cosmological term in Einstein's equations, $\CC$, usually assumed to be constant, which is  why it is usually called the cosmological constant (CC) \cite{Peebles:2002gy,Padmanabhan:2002ji}.
Consistent observational measurements of $\CC$  (treating it as a mere fit parameter)
%\cristiantext{[CMP] Tècnicament no prenem $\CC$ directament, sinó un paràmetre relacionat.})
made independently
in the last quarter of a century using distant type Ia supernovae (SNIa), the baryonic acoustic oscillations (BAO) and the anisotropies of the cosmic microwave background (CMB), have put the very foundations of the concordance  $\CC$CDM model of cosmology\,\cite{Riess:1998cb,Perlmutter:1998np,Boomerang:2000efg,Eisenstein:2005su,SDSS:2006lmn,WMAP:2008lyn,Riess:2011yx,Planck:2015fie,Planck:2018vyg}.

Despite the vanilla $(\Lambda$CDM) model  fares relatively well with the current observational data, it  traditionally suffers from a variety of problems of different kinds which seriously challenge its credibility. For a long time people somehow decided to turn a blind eye on the deepest questions and also on different spots and wrinkles which  perturb that flawless and immaculate condition.   The profound theoretical problems (and the practical wrinkles as well) are nonetheless still there alive and kicking, whether we wish to look at them or not. First and foremost, the hypothetical existence  of dark matter (DM)  still lacks of direct observational evidence. On a deeper level of mystery, the nature and origin of the DE (the dominant component of the cosmic energy budget)  still lies in the limbo of the most unfathomable cosmological riddles.  Basically because if we admit the simplest proposal for the DE, that is to say,  the cosmological constant $\CC$,  one has to cope with the `cosmological constant problem'\cite{Weinberg:1988cp},  perhaps the most inscrutable of all problems in theoretical physics and cosmology ever\,\cite{Sola:2013gha}. It manifests itself in a dual manner,  to wit:  the fine-tuning problem associated with the large value of $\CC$ predicted by most theoretical approaches  (``the old CC problem''\,\cite{Weinberg:1988cp});  and also what has become customary to call the   `cosmic coincidence problem'\,\cite{Steinhardt:2003st}, see also \cite{Sola:2013gha,Sola:2015rra,SolaPeracaula:2022hpd} for a discussion of these enigmas, which lie in the interface between cosmology and quantum field theory (QFT).

The toughest conundrum of all is probably that of explaining the relation between $\CC$ and the vacuum energy density (VED): $\rv=\CC/(8\pi G_N)$, where $G_N$ is Newton's constant.  \jtext{Traditionally one assumes that the corresponding pressure is $p_{\rm vac}=-\rv$.  In this respect, we note that in 1934 G. Lema\^\i tre pointed out the following\,\cite{lemaitre1934evolution}: ``Everything happens as though the energy in vacuum would be different from
zero. In order that it shall not be possible to measure motion relative to the
vacuum, we have to associate a pressure to the energy density of the vacuum.
This is essentially the meaning of the cosmological constant.'', see also \cite{Gron:2018mpg}.}  The interrelationship between VED and $\CC$ in the general quantum theory context has been assessed by theoretical physicists since more than a century ago, as of  the days of  W. Nernst  and  W. Pauli. At that time the issue  was already troublesome\,\cite{Sola:2013gha}.  But the most severe implications in the cosmological arena took shape only with the development of the formal aspects of QFT. It is in this modern theoretical context where the notion of VED seems to cause a serious conflict with the cosmological measurements, the reason being that the typical contribution from the vacuum fluctuations of any quantum field of mass $m$ is expected (on mere dimensional grounds) to be proportional to the quartic power of its mass: $\rv\propto m^4$, \jtext{as noted by Zel'dovich\,\cite{zeldovich:1967,Zeldovich:1968}.}  Such a prediction must be compared with the measured value of $\CC$ expressed in terms of the corresponding VED, which is  $\rv\sim 10^{-47}$ GeV$^4$ in natural units. This is extremely small in comparison with the energy density that one may estimate using any particle physics' mass $m$   from, say,  the electronvolt scale to the mass scale of the weak gauge bosons  in electroweak theory,  $W^\pm$ and  $Z$  ($\sim 80, 90$ GeV), the Higgs mass ($\sim 125$ GeV) and  the top quark mass ($\sim 170$ GeV). The exception would be, of course, a millielectronvolt neutrino\,\cite{Sola:2013gha}, but for any typical standard model particle the value of $\rv\propto m^4$ is mind-bogglingly too large, being indeed dozens of orders  of magnitude astrayed as compared to the measured value of $\CC$, not to speak of the situation in the Grand Unified Theories (GUT's), where the characteristic energy scale can reach $\sim 10^{16}$ GeV.  It is because of the cosmological constant problem phrased on these grounds that the VED option became outcast as if were to be blamed of all evils. The aforesaid notwithstanding, the criticisms usually have nothing
better to offer, except to defend tooth and nail  particular forms of the DE without providing any explanation about the genuine subject involved in the original discussion of this problem, which is, of course, to understand the  role played by the VED in QFT and its fundamental relation with $\CC$.  Most, if not all, proposed forms of DE, are actually  plagued with the same  (purported) fine tuning illness that is  attributed (in a way by fiat) to the vacuum option exclusively.  This is certainly the case e.g. with the popular family of quintessence models, phantom fields and generalizations thereof, see e.g. \cite{Peebles:2002gy,Padmanabhan:2002ji,Copeland:2006wr,Tsujikawa:2013fta} and references therein.  \jtext{Attempts to understand the vacuum energy as a form of repulsive gravitation capable of driving the slow accelerating expansion of the universe, notwithstanding its exceedingly large value,  have been made in the literature, see e.g.\cite{Gron:1986fv,Wang:2017oiy} and \cite{Gron:2018mpg} for a review.}

In recent years, new approaches to the notion of vacuum energy in QFT and its relation with the $\CC$-term suggest  that these problems can be smoothed out to a large extent. In fact, the VED can be properly renormalized in QFT in curved spacetime, thereby offering  a tamer theoretical context for the traditional vacuum energy approach to fit in with the observations.  In the light of these developments, the quantum vacuum energy could well be after all the most fundamental explanation for the DE in our universe. See e.g. \cite{Sola:2013gha,Sola:2015rra} as well as the latest  formal developments in\,\cite{Moreno-Pulido:2020anb,Moreno-Pulido:2022phq,Moreno-Pulido:2022upl,Moreno-Pulido:2023ryo}, summarized in  \cite{SolaPeracaula:2022hpd}.

The vanilla $\CC$CDM model, to which modern cosmological observations have converged in the last decades, is certainly an important triumph in our description of the main background features of the cosmic expansion   and the large-scale structure formation processes in the universe.  However, it is only a partial success. Its exceeding simplicity eventually turned into a perilous double-edge sword; in fact,  the absence of any  connection with fundamental physics is the literal expression of such a simplicity and is most likely at the root of its many shortages. In truth, the $\CC$CDM does not possess enough theoretical structure to explain the successfulness of the  observations (e.g. the measured value of $\CC$)  on a fundamental context, and at the same time it cannot even provide an explanation for other measurements that are threatening its viability.  If we pay attention to the existing conflicts on several active fronts,  the observational situation of the $\CC$CDM  in the last decade or so does not seem to paint a fully rosy picture  anymore.  Beyond formal theoretical issues, a series of practical problems of more mundane nature than those mentioned above are piling up as well\,\cite{Perivolaropoulos:2021jda}.  On a mere phenomenological perspective, it is particularly worrisome the situation with some ``tensions'' existing with the data.  For example, it has long been known that  there appear to exist potentially  serious discrepancies  between the CMB observations (based on the vanilla $\Lambda$CDM),  and the local direct (distance ladder) measurements of the Hubble parameter today\cite{DiValentino:2020zio}. The persisting mismatch between these measurements is what has been called the ``$H_0$-tension''. It is arguably the most puzzling open question within the current cosmological paradigm and it leads, if taken at face value, to a severe discrepancy of $\sim 5 \sigma$ c.l. or more, between the mentioned observables.  Many proposals have been put forward to shed some spark of light into that puzzling cosmological imbalance. Among the possibilities debated in the literature, it has been conjectured e.g. that it could stem from a possible intrinsic ``running  of $H_0(z)$ with the redshift'' presumably  connected  with the differences that may appear  in the (total) effective equation of state  (EoS) of the universe between the vanilla cosmology and the actual FLRW model underlying the observations\,\cite{Krishnan:2020vaf,Dainotti:2021pqg}.
While these are an interesting  possibilities, we are probably still far away from understanding the resolution of this conundrum on fundamental grounds. At the same time, there exists a smaller but appreciable  ($\sim 2-3\, \sigma$) tension in the realm of the large-scale structure (LSS) growth data,  called the  ``$\sigma_8$-tension''\,\cite{DiValentino:2020vvd}.  It is concerned with the measurements of weak gravitational lensing at low redshifts ($z<1$). Such a tension is usually evaluated with the help of the parameter $S_8$ or, alternatively, by means of $\sigma_8$; recall that $S_8\equiv\sigma_8\sqrt{\Omega^0_m/0.3}$. It turns out that these measurements favor matter clustering weaker than that expected from the vanilla model  using parameters determined by CMB measurements, see e.g. \cite{Macaulay:2013swa,Nesseris:2017vor,Lin:2017bhs,Gomez-Valent:2018nib,Garcia-Quintero:2019cgt,KiDS:2020suj,DES:2021wwk,Nguyen:2023fip}. Recently, it has been claimed that $S_8$ values determined from $f\sigma_8$ increase  with redshift in the $\CC$CDM\cite{Adil:2023jtu}, which, according to these authors, provides additional support to the fact that such a  discrepancy  may be physical in origin and with a value in the enhanced $2-4\,\sigma$
range. In the constant pursue for a possible late-time solution to these tensions, it has been argued that within the large class of models where the DE is treated as a fluid with EoS $w(z)$, solving the $H_0$ tension demands the phantom condition $w(z)<-1$
at some $z$, while solving both the $H_0$ and $\sigma_8$ tensions requires $w(z)$ to cross the phantom divide and/or other sorts of transitions, see e.g. \cite{Heisenberg:2022gqk,Marra:2021fvf,Alestas:2020zol,Perivolaropoulos:2021bds,Alestas:2021luu,Perivolaropoulos:2022khd}. Specific realizations of the noticed  double condition for the DE fluid can be found in the literature, e.g. in the context of the $\CC$XCDM model\,\cite{Grande:2006nn,Grande:2008re}, closely related to the idea of running vacuum to be discussed in the present work, see below.
For detailed reviews on these tensions and other challenges afflicting the concordance $\CC$CDM model, see e.g. ~\cite{Perivolaropoulos:2021jda,Abdalla:2022yfr,Dainotti:2023yrk} and the long list of references quoted there bearing relation to these matters.

The severity of some of these tensions and the huge number of proposals existing in the literature trying to explain them through a large disparity of ideas suggest that it is perhaps time to come to grips anew with the  fundamentals of the theoretical formulations, such as quantum field theory and string theory.  We have already pointed out to recent calculations claiming a more adequate renormalization prescription for the VED in quantum field theory in FLRW spacetime,  and leading to the  ``running vacuum model'' (RVM). It turns out that this QFT approach may have real impact not only on the more formal theoretical problems described in the beginning, but also on the practical issues concerning the aforementioned tensions.  In fact, the resulting VED from the RVM   leads to a time-varying vacuum energy density,  and hence a time-varying (physical) $\CC$ as well, in which $\CC$  acquires a dynamical component through the quantum vacuum effects: $\CC\to \CC+\delta\CC$. The shift $\delta\CC$ is calculable in QFT since it depends on the contributions from the quantized matter fields (bosons and fermions).  Upon appropriate renormalization one finds that  $\delta\CC$  depends on a term of order of the Hubble rate $H(t)$ squared\,\cite{Moreno-Pulido:2020anb}: $\delta\Lambda \sim \nu\, {\cal O}\left(H^2\right)$  ($|\nu|\ll1$).  This is the typical form of the  RVM. The connection of the latter with QFT can be motivated  from semi-qualitative renormalization group arguments on scale-dependence, see  the reviews \cite{Sola:2013gha,Sola:2015rra}. In particular, we mention the old works \cite{Shapiro:2000dz,Babic:2004ev} and also recent approaches along these lines, such as \cite{Alvarez:2020xmk}. However, an explicit QFT calculation leading to that form of `running $\CC$' (associated to  the `running VED')  appeared only very recently\,\cite{Moreno-Pulido:2020anb,Moreno-Pulido:2022phq,Moreno-Pulido:2022upl,Moreno-Pulido:2023ryo}.

A note of caution is in order here. Over time a large variety of cosmological models have been proposed to describe the DE and its possible dynamics.  Apart from the aforesaid quintessence models and the like\cite{Peebles:2002gy,Padmanabhan:2002ji,Copeland:2006wr,Tsujikawa:2013fta}, there is a very populated habitat of models with generic time-dependent cosmological constant, the so-called ``$\CC(t)$-cosmologies''.  Many of these models, however, are of pure phenomenological nature since the time dependence of $\Lambda(t)$ is parameterized in an \textit{ad hoc} manner. They might have a connection with fundamental theory, but it is not implemented in an explicit way in the corresponding papers. The list of models of this type is large and we will cite here only a few of them \cite{Ozer:1985ws,Bertolami:1986bg,Freese:1986dd,Peebles:1987ek,Chen:1990jw,Abdel-Rahman:1992msa,Carvalho:1991ut,Arcuri:1993pb,Waga:1992hj,Lima:1994gi,Lima:1995ea,Arbab:1997ph,Espana-Bonet:2003qjh,Wang:2004cp,Borges:2005qs,Alcaniz:2005dg,Barrow:2006hia,Costa:2009wv,Bessada:2013maa,Gomez-Valent:2014fda}; see also the old review\,\cite{Overduin:1998zv}.  In some cases, the parameterization is performed through a direct function of the cosmic time or of the scale factor, and sometimes  as a function of the Hubble parameter, or even  an hybrid combination of these various possibilities.  Be as it may, the general and rather nonspecific class of the ``$\CC(t)$-cosmologies''  should not be confused with the  ``running vacuum models'' (RVMs) discussed above, in which the  running of $\CC$ stems from the  quantum effects on the effective action of QFT in curved (FLRW) spacetime.  In other words, the RVMs are to be understood in a much more restricted sense; in fact, one that is   closer to fundamental aspects of QFT, and only this precise type of time-evolving VED cosmologies will be dealt with here.
 Let us finally note that, apart from the QFT formulation, a `stringy' version of the RVMs is also available which can be very promising too\,\cite{Basilakos:2019acj,Basilakos:2020qmu,Mavromatos:2020kzj,Mavromatos:2021urx}. The potential dynamics of the cosmic vacuum is, therefore, well motivated from different theoretical perspectives, and this fact further enhances the interest for the current study, whose main purpose is to focus exclusively on the phenomenological implications of the class of RVM models.

We should mention that the running vacuum framework  has already been tested with considerable success in previous works over the years. It has been known for quite some time that the RVM-type of cosmological models can help in improving the overall fit to the cosmological observations and also in smoothing out the mentioned  tensions as compared to the $\CC$CDM;  see for instance  \cite{SolaPeracaula:2021gxi,Sola:2017znb,Gomez-Valent:2017idt,Gomez-Valent:2018nib,Sola:2016ecz,Sola:2016zeg,sola2017first,Sola:2017jbl,Sola:2015wwa,Gomez-Valent:2014rxa}, and \cite{SolaPeracaula:2022mlg,SolaPeracaula:2018xsi} for a short summary.  For this reason  we believe it is worthwhile to  keep on exploring the phenomenological consequences of the running vacuum  in the light of the latest observations on all the main data sources: SNIa+BAO+$H(z)$+LSS+CMB. The state-of-the-art-phenomenological performance of the RVM has been reported not too long ago, in \cite{SolaPeracaula:2021gxi}. In the current work, however,  we definitely enhance the scope of the results presented in that paper by considering an updated cosmological data set  in combination with an extended analysis of the  CMB part.  In point of fact, the  main focus in this paper is to delve into the practical ability of the RVM to tackle the $\sigma_8$ and $H_0$ tensions versus the vanilla $\CC$CDM model. It is reassuring to find that the global fit to the cosmological observations can be improved within  the running vacuum framework with respect to the $\CC$CDM. The optimal situation is when the VED presents a threshold in the recent past, where its dynamics becomes activated,  and/or when the gravitational coupling is mildly running.

All in all, the dynamical DE models  may offer a clue not only to relieve some high-brow aspects of the cosmological constant and coincidence problems, but also to straighten out some very practical ones, such as helping to  modulate the processes of structure formation which may impinge positively on the $\sigma_8$-tension. Last but not least, they can help to explain the existing mismatch   between the distinct values of $H_0$ derived from measurements of the local and the early universe.

\vspace{0.25cm}

\indent The paper is organized as follows. In Sec. \ref{eq:IRRVM}, we present the running vacuum model (RVM) from a phenomenological point of view and emphasize its connection with QFT in curved spacetime. For convenience, we introduce the model variant of the RVM which we call RRVM, as we did in \cite{SolaPeracaula:2021gxi}. In it, the VED can be expressed entirely in terms of the curvature scalar ${\cal R}$ (which is of order $\sim H^2$) at the background level and study two types of RRVMs: type I  and type II,  depending on whether the gravitational coupling $G$ is fixed at its current local gravity value, $G=G_N$,  or evolving mildly with the expansion, $G=G(H)$, a feature which in our case is respectively linked  to the interaction or not of the evolving vacuum energy with cold dark matter (CDM). Type I is studied at length in Sec. \ref{eq:TypeIRRVM}, where we describe the background cosmological equations and its solution under appropriate conditions.   At the same time we discuss the corresponding perturbations equations.  Type II, on the other hand,  is studied in detail in  Section \ref{sec:RRVMTypeII}, where again we provide the background solution and analyze the perturbations.  In Sec. \ref{sec:Data_and_methodology}  we enumerate and briefly describe the different sources of observational data employed in this paper and the methodology used to constrain the free parameters of the models under discussion.  We also define the four characteristic datasets (Baseline and Baseline+SH0ES with and without CMB polarization data) that will be used to test the running vacuum models and their comparison with the vanilla $\CC$CDM model. The outcome of our analyses under the different datasets is presented and discussed in detail in Sec. \ref{sec:Discussion_of_the_results}.  Finally, in Section \ref{sec:Conclusions}  we summarize our findings and present the main conclusions of this study.  In  the Appendix A at the end of our work we include additional tables with a detailed breakdown of the different $\chi^2$ contributions from each observable.

\section{Running Vacuum in the Universe}\label{eq:IRRVM}

As indicated, throughout our study we will assume that the background spacetime is FLRW with flat three-dimensional hypersurfaces. The general low-energy form of the vacuum energy density (VED) within the running vacuum model (RVM) has been explored phenomenologically on several previous occasions and with a remarkable degree of success, in the sense that in all cases it has proven to be rather competitive with the $\CC$CDM and even able to surpass the fitting performance of the latter,  see e.g.\cite{SolaPeracaula:2021gxi,Sola:2017znb,Gomez-Valent:2017idt,Gomez-Valent:2018nib,Sola:2016ecz,Sola:2016zeg,Sola:2015wwa,sola2017first,Sola:2017jbl,Gomez-Valent:2014rxa}.  Herein we shall test if it is still the case with the current wealth of observations and  using the state-of-the-art methods of analysis of the cosmological data.  The dynamical structure of the running VED adopts the perspicuous form\cite{Moreno-Pulido:2020anb,Moreno-Pulido:2022phq}
\begin{equation}\label{eq:RVMvacuumdadensity}
\rv(H) = \frac{3}{8\pi G_N}\left(c_{0} + \nu{H^2+\tilde{\nu}\dot{H}}\right)+{\cal O}(H^4)\,.
\end{equation}
\jtext{For $\nu=\tilde{\nu}=0$, this expression reduces to  $\rv= \CC_{\rm phys}/(8\pi G_N)$, where  $\CC_{\rm phys}=3 c_0$ retakes the traditional role of the cosmological constant term.  However, for nonvanishing values of the coefficients $\nu$ and $\tilde{\nu}$, the vacuum acquires a certain amount of dynamics which is provided by the $H^2$ and $\dot{H}$ contributions}.  Here,  the dot indicates derivative with respect to the cosmic time and  $H=\dot{a}/a$ is the Hubble function.  As we can see, the two leading dynamical terms of $\rv$ in Eq.\,\eqref{eq:RVMvacuumdadensity} are both of ${\cal O}(H^2)$ since $\dot{H}\sim H^2$, this being true both in the matter-  and radiation- dominated epochs.  Despite the fact that the  higher order powers ${\cal O}(H^4)$ in the above expression are also predicted in the QFT context along with the lower order ones ${\cal O}(H^2)$\cite{Moreno-Pulido:2022phq}, the former are unimportant for the current universe and will be hereafter ignored in front of the latter. The additive parameter  $c_0$ is  constrained to satisfy  $\rho_{\rm vac}(H_0)=\rvo$, where $\rvo$ is the value of the VED today, and hence is connected with the physical value of the measured cosmological constant through $\rvo= \CC_{\rm phys}/(8\pi G_N)$. \jtext{In this case, however, $\CC_{\rm phys}\neq 3 c_0$  is a quantity nontrivially connected to the dynamical terms in Eq.\,\eqref{eq:RVMvacuumdadensity} since a formal renormalization of the theory becomes necessary within the QFT context\,\cite{Moreno-Pulido:2020anb,Moreno-Pulido:2022phq}.} Upon renormalization,  the bulk of the physical value is still provided by $c_0$ since the two (dimensionless) coefficients $\nu$ and $\tilde{\nu}$  adjoined to the two dynamical terms in \eqref{eq:RVMvacuumdadensity} are expected to be small  ($|\nu,\tilde{\nu}|\ll1$) \,\cite{Sola:2007sv}. They encode the running character of the vacuum at low energy and can be computed in QFT in curved spacetime,  receiving contributions from the quantized bosons and fermion fields.  The explicit calculation was first presented in \cite{Moreno-Pulido:2020anb,Moreno-Pulido:2022phq} and was recently completed in \cite{Moreno-Pulido:2023ryo}. In these references, it is shown that the above VED structure can be formally derived from quantum effects on the effective action of  QFT in FLRW spacetime.

\jtext{Indeed, as shown in the works\cite{Moreno-Pulido:2020anb,Moreno-Pulido:2022phq,Moreno-Pulido:2022upl,Moreno-Pulido:2023ryo}, fully within the spirit of the renormalization group (RG) analysis inherent to the RVM structure\,\cite{Sola:2013gha}, the Hubble rate  $H$ (with natural dimension of energy) can be viewed as a RG scaling parameter. For the sake of simplicity, let us illustrate the type of effects that contribute to the coefficient $\nu$ of the $H^2$ terms in the VED \eqref{eq:RVMvacuumdadensity}, albeit focusing only on one quantized scalar field of mass $m$ and non-minimal coupling $\zeta$.  The independent contributions to $\tilde{\nu}$  will not be shown\cite{Moreno-Pulido:2022phq} for this summarized discussion of the QFT aspects, and will be assumed to vanish.  Then one may write for the RVM energy density $\rv(H)$ which connects  two given values of the Hubble parameter; say, one at the $H$ era and the other at another epoch (the current one,  $H_0$, for example}):
\begin{align}\label{QFTresult}
\rv(H) = \rvo + \frac{3\, \nu}{8\pi G_N} \, \Big(H^2 - H_0^2 \Big)\,,
\end{align}
where
\begin{align}\label{QFTnu}
\nu= \frac{1}{2\pi} \Big(\zeta - \frac{1}{6}\Big) \, \frac{m^2}{\mpl^2}\,  \, {\rm ln}\left(\frac{m^2}{H_0^2}\right)  \,,
\end{align}
\jtext{in which $\mpl$ is the Planck mass.  According to the QFT calculation,  $\nu$ actually appears as a very mildly (logarithmically) dependent function of $H=H(t)$ -- see the exact expression in \cite{Moreno-Pulido:2022phq}. Since it remains essentially invariable with $H$ for fitting purposes, we can fix $H=H_0$ in it and take $\nu$ as a constant fitting parameter. Notice also that in arriving at Eq.\,\eqref{QFTresult}, ${\rm ln}(m^2/H_0^2) \gg 1$ has been used (an inequality which is valid for virtually any massive particle).  The previous considerations justify in part the structure of the VED in Eq.\,\eqref{eq:RVMvacuumdadensity}. Regarding the $\dot{H}$ term, it has a similar structure~\cite{Moreno-Pulido:2022phq}, but the above considerations should suffice to grasp the kind of QFT contributions that are found.
As previously noted,
$\zeta$ is the non-minimal coupling of the (quantized) scalar matter fields  with gravity (in general, one expects a different coupling for each scalar field). In the conformal limit (certainly not our case) one would have  $\zeta=1/6$ and the running of the VED from the scalar sector would disappear, as could be expected. The quantity $\rvo$ in \eqref{QFTresult} denotes the vacuum energy density at the current era and hence it is connected with the measured value of the cosmological `constant' through the aforementioned relation $\rvo=\Lambda_{\rm phys}/(8\pi G_N)$.}

\jtext{We should also mention, for the sake of a more complete summary of this QFT part, that the exact formula for scalars (including an arbitrary number of them) can be found in  \cite{Moreno-Pulido:2022phq}, and that the calculation of the scalar contribution recently became complemented with the QFT calculation of the (quantized) fermionic contributions, also in an arbitrary number, see \cite{Moreno-Pulido:2023ryo}.  It is therefore clear that at present a theoretical quantitative prediction for the coefficients $\nu$ and $\tilde{\nu}$ cannot be performed in practice since they depend on the contributions from the masses and non-minimal couplings of the various scalar fields, as well as from the masses of the different species of fermions. Furthermore, from the above formulas it should be clear that the relevant fields contributing significantly (at one-loop and higher order) to the VED running are not the low-energy fields of the standard model of particle physics but the very heavy  fields belonging to the given Grand Unified Theory (usually accompanied with large multiplicity factors). All that said, and despite the fact that a quantitative QFT prediction of the cosmic running of the VED  cannot be presently furnished, the great virtue of these formal calculations is,  at least from our point of view,  that they provide a theoretical link between cosmologically relevant quantities with the QFT framework, and hence contribute to establish a deeper connection of cosmology with the fundamental principles of theoretical physics.}

In practice,  therefore, the values of $\nu,\tilde{\nu}$ must be fitted  to the cosmological observations.  Thus, in what follows we will focus exclusively on the phenomenological consequences of the RVM.  What is important is that these coefficients  are expected to be small and of order $\sim M_X^2/\mpl^2\ll 1$, where  $\mpl\simeq 1.22\times 10^{19}$ GeV is the Planck mass and $M_X\sim 10^{16-17}$ GeV  is of order of a typical GUT scale (or even a string scale slightly above it) times a multiplicity factor accounting for the number of heavy particles in the GUT\,\cite{Sola:2007sv}.
For  $\nu=\tilde{\nu}=0$  we recover the $\CC$CDM smoothly.  This is a very welcome property of the RVM since  DE models having no smooth  $\CC$CDM limit, e.g. predicting  a VED of the form $\rv(H)\propto H^2$ or a combination of $H^2$ and $\dot{H}$  (without any additive term),   would  be excluded owing to their absence of  an inflexion point from deceleration into acceleration in the cosmic evolution, see \cite{Basilakos:2012ra,Basilakos:2014tha,Gomez-Valent:2014rxa}.  The presence of the nonvanishing additive term $c_0$ is therefore crucial for the RVM to avoid this unwanted situation, something that other models (e.g. entropic and  ghost models of the DE) cannot avoid and thereby get into trouble\,\cite{Gomez-Valent:2015pia,Rezaei:2019xwo}. Holographic models with dynamical cutoff $L=H^{-1}$  also lack of an additive term in the DE and are also unfavored already at a pure cosmographic level \cite{Rezaei:2022bkb}.  In stark contrast,  the condition $c_0\neq 0$ is always warranted within the class of the RVMs.

It is important to realize that the dynamics of the VED must preserve, of course,  the Bianchi identity satisfied by the Einstein tensor. In practice this means that the total energy-momentum tensor (EMT), which receives the contributions from nonrelativistic matter, radiation and vacuum (assumed here to be ideal fluids), must be covariantly conserved, namely $\nabla^\mu T_{\mu\nu}^{\rm tot} =0$. The total EMT reads,
\begin{equation}\label{totalEMT}
 T_{\mu\nu}^{\rm tot} =  (p_t-\rv)\,g_{\mu\nu}+\big(\rho_t+p_t \,\big)\,U_{\mu}\,U_{\nu}\,,
\end{equation}
where $U^\mu$ is the $4$-velocity vector of the cosmic fluid. We have defined $\rho_t=\rho_m+\rho_{\rm ncdm}+\rho_\gamma$, where $\rho_m=\rho_{\rm cdm}+\rho_{b}$ denotes the contribution to the proper density of nonrelativistic matter from cold dark matter and baryons, $\rho_{\rm ncdm}$ (non-cold dark matter) corresponds to the energy density of  neutrinos and $\rho_\gamma$ designates the energy density associated with photons. \jtext{In other words, $\rho_t$ refers to the sum of all the species in the universe excluding the vacuum. Analogous notations apply to the pressures.} We shall, however,  be more specific in our treatment of the various contributions to the EMT in the next section.  Notice that in the above expression we have used $p_{\rm vac}=-\rv$ for the EoS of the  vacuum fluid, as indicated in the introduction.  Even though  this condition may be violated slightly by quantum effects within a formal treatment of the subject in QFT\cite{Moreno-Pulido:2022upl}, we shall nonetheless stick for now to the traditional EoS of the vacuum.  We shall come back to this point later on. Upon expanding $\nabla^\mu T_{\mu\nu}^{\rm tot} =0$, it amounts to the local covariant conservation law in a FLRW universe
\begin{equation}\label{BianchiGeneral}
\frac{d}{dt}\,\left[G(\rho_t+\rv)\right]+3\,G\,H\,(\rho_t+p_t)=0\,,
\end{equation}
where, in general, not only $\rv$ but also  $G$  may be functions of the cosmic time.  This will depend on the particular implementation assumed for  the matter sector.  If we assume that there is an interaction of the VED with matter, then $G$ can stay fixed at the usual value $G_N$ (the local gravity value), whereas if matter is locally conserved, then $G$ must vary accordingly in order to preserve the covariant conservation law  \eqref{BianchiGeneral}.

In order to ease the comparison with previous results, we shall adhere to the approach of \cite{SolaPeracaula:2021gxi} and assume  $\tilde{\nu}=\nu/2$.  In this way  the RVM model is left with one single parameter and at the same time  adopts the suggestive  form
\begin{equation}\label{eq:RRVM}
\rv(H) =\frac{3}{8\pi{G_N}}\left(c_0 + \frac{\nu}{12} {\cal R}\right)\equiv \rv({\cal R})\,,
\end{equation}
in which  ${\cal R} = 12H^2 + 6\dot{H}$ is the curvature scalar. That particular implementation  is called, for obvious reasons, the RRVM, since it is a version of the RVM which involves the scalar of curvature \cite{SolaPeracaula:2021gxi}.  One additional advantage is that it is automatically well-behaved in  the radiation dominated epoch since in it   $\mathcal{R}/H^2\ll 1$, and the standard  BBN is not perturbed at all by the presence of vacuum energy.  In the general case  \eqref{eq:RVMvacuumdadensity} such condition can also be fulfilled  on assuming sufficiently small  (absolute) values of $\nu,\tilde{\nu}$\,\cite{sola2017first}.

Finally, despite the general structure  of the running VED  is of the form \eqref{eq:RVMvacuumdadensity}, for convenience we define two types of RRVM scenarios.  In type-I scenario the vacuum is in interaction with matter, whereas in type-II  matter is conserved at the expense of an exchange between the vacuum and a slowly evolving gravitational coupling $G (H)$.   The combined cosmological `running' of these quantities  insures the accomplishment of the Bianchi identity (and the associated local conservation law).  In the following sections we study these two cases separately.

\section{Type I: running vacuum interacting with dark matter}\label{eq:TypeIRRVM}

In this section,  we consider the type-I  RRVM scenario, in which the vacuum can be running at the expense of exchanging energy with matter.  We will assume that only cold dark matter (CDM) is involved in such an exchange (therefore no baryons, neutrinos or photons are transferred to or from the vacuum).  Whether it is the vacuum that generates new CDM  or  the CDM that disappears into the vacuum depends on the sign of the parameter $\nu$ in Eq.\,\eqref{eq:RRVM}. For $\nu>0$, the vacuum  decays into tiny amounts of  CDM, whilst for $\nu<0$ some dark matter disappears into the vacuum.  We do not presume which of these situations hold,  we will  fit  the value  (and sign) of $\nu$ to the cosmological data.  This requires to solve the background and linear perturbation equations of the type-I running vacuum model, which we do in Secs. \ref{sec:backRRVMI} and \ref{sec:pertRRVMI}.

\subsection{Background equations}\label{sec:backRRVMI}
The VED expression \eqref{eq:RRVM}  can be cast more explicitly as follows,
\begin{equation}\label{eq:rhov}
\rv = \frac{3}{8\pi{G_N}}\left[c_0 + \nu\left(H^2 + \frac{1}{2}\dot{H}\right)\right].
\end{equation}
The above energy component becomes now part of the Friedmann and the pressure equations  written  in terms of the energy densities and the pressures for the different species in game, which read
\begin{align}
3H^2 &= 8\pi{G_N}\left(\rho_t + \rv\right)=8\pi{G_N} \left(\rho_m+\rho_{\rm ncdm}+\rho_\gamma+\rv \right)\,,\label{FriedmannEquation} \\
3H^2 + 2\dot{H} &= -8\pi{G_N}\left(p_t+ p_{\textrm{vac}}\right)=-8\pi{G_N}\left( p_{\rm ncdm}+p_\gamma+p_{\rm vac}\right) \,.\label{AccelerationEquation}
\end{align}
The following comment is in order. As it is well known there is a transfer of energy from the relativistic neutrinos to the nonrelativistic ones throughout the whole cosmic history. It is difficult to make a perfect separation of the relativistic and nonrelativistic phases and, strictly speaking, this splitting can be a little bit inaccurate at those epochs of the expansion history at which a neutrino species is in an intermediate step, between the ultra-relativistic and nonrelativistic regimes, since in this case one cannot classify such neutrino species in any of these two categories. Nevertheless, it is useful to obtain approximate formulas for the two components, as we shall see in a moment. We proceed as in the Einstein-Boltzmann solver \texttt{CLASS}\footnote{\url{https://lesgourg.github.io/class_public/class.html}} \cite{Lesgourgues:2011re,Blas:2011rf}, where we have implemented our model. \texttt{CLASS} solves the Einstein and Boltzmann differential equations at any value of the scale factor and, in particular, provides the functions $\rho_{\rm ncdm}(a)$ and $p_{\rm ncdm}(a)$. \texttt{CLASS} performs then a rather artificial splitting of these quantities, as if they came from the sum of an ultra-relativistic fluid (denoted with a subscript $\nu$) and a nonrelativistic one (denoted with a subscript $h$),
\begin{align}
& \rho_h = \rho_{\rm ncdm} - 3p_{\rm ncdm}\qquad ;\qquad p_h=0 \,;\\
& \rho_\nu = 3p_{\rm ncdm}\qquad ;\qquad p_\nu=p_{\rm ncdm}\,.
\end{align}
At this point we can rewrite the combination $H^2 + (1/2)\dot{H}$ appearing in \eqref{eq:rhov} in terms of the energy densities and pressures using \eqref{FriedmannEquation} and \eqref{AccelerationEquation},
\begin{equation}\label{combination}
H^2 + \frac{1}{2}\dot{H} = \frac{2\pi{G_N}}{3}\left(\rho_m + 4\rv + \rho_{\rm ncdm} - 3p_{\rm ncdm}\right)\,.
\end{equation}
In this expression, we can appreciate that the nonrelativistic contribution from massive neutrinos, namely $\rho_h = \rho_{\rm ncdm} - 3p_{\rm ncdm}$ is present. This is a problem if we want to solve the background equations, since it carries a complicated (non-analytic) dependence on the scale factor. In order to solve this problem we can consider a reasonable approximation, which is the following:
\begin{equation}
r \equiv\frac{\rho_h}{\rho_m} = \frac{\rho_h}{\rho_{\textrm{cdm}} + \rho_{b}} \simeq 0\,.
\end{equation}
We have checked explicitly the validity of this approximation with \texttt{CLASS}, computing the ratio $r =\rho_h/\rho_m$ for the whole cosmic history. We have found that $r$  varies smoothly from $10^{-7}$ at redshift $z=10^{14}$ to $10^{-3}$ at $z=0$ considering a massive neutrino with mass $\sim \mathcal{O}(0.1)$ eV. In addition, $r$ is multiplied by $\nu$ in Eq. \eqref{eq:rhov}, so the resulting quantity is of order $\mathcal{O}\left(10^{-5}\right)$ at most. Therefore, we deem it natural and licit to drop this term to make things easier without any significant loss of accuracy in our calculation.

Under this very good approximation, we can express the vacuum energy density \eqref{eq:rhov} as follows,
\begin{equation}\label{eq:rhov1}
\rv(a) = \rvo + \frac{\nu}{4(1-\nu)}(\rho_m(a) - \rho^0_m)\,,
\end{equation}
with $\rv(a=1) = \rho^{0}_{\textrm{vac}}$ and $\rho_m(a=1) = \rho^{0}_m$.  We still need to find $\rho_m(a)$, though.
The starting point is the equation \eqref{BianchiGeneral} which yields the interaction law between vacuum and matter in the general case.
Now since  we assume that $G$ is strictly constant for the type-I models, Eq.\,\eqref{BianchiGeneral}  boils down to
\begin{equation}
\dot{\rho}_m + 3H\rho_m = -\dot{\rho}_{\rm vac}\,,
\end{equation}
where we are neglecting the pressure of the matter components.
Notice that the previous equation is entirely equivalent to the interaction law between CDM and vacuum,
\begin{equation}
\dot{\rho}_{\textrm{cdm}} + 3H\rho_{\textrm{cdm}} = -\dot{\rho}_{\rm vac}\,, \label{ConservationEquation}
\end{equation}
owing to the fact that we are assuming that baryons do not interact at all with the vacuum, which entails the relation $\dot{\rho}_b+3H\rho_b=0$. In this way we have obtained the conservation equation of matter (baryons+CDM) for type-I models.

Combining the above equations, we arrive at the final result
\begin{equation}
\dot{\rho}_{m} + 3H\xi\rho_m = 0\,,
\end{equation}
where we have defined the dimensionless parameter
\begin{equation}
\xi \equiv \frac{1 -\nu}{1 - \frac{3}{4}\nu}\,.
\end{equation}
It is then straightforward to find out the expressions for the various energy densities:
\begin{align}\label{eq:definition}
\rho_m(a) &= \rho^0_m{a^{-3\xi}}\,,\\
\rho_{\textrm{cdm}}(a) &= \rho^{0}_m{a^{-3\xi}}  - \rho^0_{\rm b}{a^{-3}}\,,  \\
\rv(a) &= \rho^0_{\textrm{vac}} + \left(\frac{1}{\xi} -1\right)\rho^0_m\left(a^{-3\xi} -1\right)\label{eq:rhovac}\,.
\end{align}
In the limit  $\xi\to 1$ ($\nu\to 0$) we recover the expected forms of these equations in the $\Lambda$CDM. It is also possible to encode the deviations with respect to the standard cosmological model in terms of an effective parameter $\nu_{\rm eff}$, defined as
\begin{equation}\label{eq:nueff_definition_typeI}
\xi = \frac{1-\nu}{1-\frac{3}{4}\nu} \simeq 1 - \frac{\nu}{4} + \mathcal{O}\left(\nu^2\right) \equiv 1-\nu_{\rm eff} + \mathcal{O}\left(\nu_{\rm eff}^2\right)\,.
\end{equation}
We will report all our fitting results in terms of parameter $\nueff$. As with $\nu$, positive values of $\nu_{\rm eff}$ lead to a decay of the vacuum into CDM, whereas negative values source an energy transfer from CDM to the vacuum.

\subsection{Perturbation equations}\label{sec:pertRRVMI}

We have implemented the perturbation equations in \texttt{CLASS}, using the synchronous gauge. Denoting by $\tau$ the conformal time, the perturbed (flat three-dimensional) FLRW metric in the conformal frame reads \cite{Ma:1995ey},
\begin{equation}\label{eq:LineElementSyn}
ds^2=a^2(\tau)[-d\tau^2+(\delta_{ij}+h_{ij})dx^idx^j]\,,
\end{equation}
with
\begin{equation}\label{eq:hFourier}
h_{ij}(\tau,\vec{x})=\int d^3k\, e^{-i\vec{k}\cdot\vec{x}}\left[\hat{k}_i\hat{k}_j h(\tau,\vec{k})+\left(\hat{k}_i\hat{k}_j-\frac{\delta_{ij}}{3} \right)6\eta(\tau,\vec{k})\right]\,,
\end{equation}
and $\hat{k}_i=k_i/k$. The above formula represents   the perturbation expressed as a Fourier integral  on the two fields in $k$-space, $h(\tau,\vec{k})$ and $\eta(\tau,\vec{k})$, which parameterize the non-traceless and traceless parts, respectively. The nonvanishing trace is the $h$ function.
The perturbed Einstein equations in Fourier space adopt the same form as in the $\Lambda$CDM. They read as follows:

\begin{equation}\label{eq:PerturbTypeIEq1}
\mathcal{H}h^\prime-2\eta k^2=8\pi {G_N}a^2\sum_l\delta\rho_l\,,
\end{equation}
\begin{equation}\label{eq:PerturbTypeIEq2}
\eta^\prime k^2=4\pi {G_N} a^2\sum_l (\bar{\rho}_l+\bar{p}_l)\theta_l\,,
\end{equation}
\begin{equation}\label{eq:PerturbTypeIEq3}
h^{\prime\prime}+2\mathcal{H}h^\prime-2\eta k^2=-24\pi {G_N}a^2\sum_l\delta p_l\,.
\end{equation}
\begin{equation}\label{eq:PerturbTypeIEq4}
h^{\prime\prime}+6\eta^{\prime\prime}+2\mathcal{H}(h^\prime+6\eta^\prime)-2k^2\eta=-24\pi G_N a^2(\bar{\rho}+\bar{p})\sigma\,,
\end{equation}
where $\mathcal{H}\equiv aH$, the sums run over the different matter components,  the primes denote derivatives with respect to the conformal time, and

\begin{equation}
(\bar{\rho}+\bar{p})\sigma\equiv -\left(\hat{k}_i\hat{k}_j-\frac{\delta_{ij}}{3}\right)\left(T^i_{j}-\frac{\delta^i_{j}}{3}T^{k}_{k}\right)
\end{equation}
carries the information of the anisotropic stress, with $T_{\mu\nu}$ the total energy-momentum tensor. The bars in these equations indicate background quantities and $\theta_l$ is the divergence of the perturbed velocity of the fluid $l$. Eqs. \eqref{eq:PerturbTypeIEq1} and \eqref{eq:PerturbTypeIEq2} are obtained from the $00$ and $0i$ components of Einstein's equations, respectively, whereas Eqs. \eqref{eq:PerturbTypeIEq3} and \eqref{eq:PerturbTypeIEq4} are the trace and the part proportional to $\hat{k}_i\hat{k}_j$ of the $ij$ component.

All the perturbed conservation equations are also the same as in the standard model, except those that relate CDM and the vacuum, which take the following form,

\begin{equation}
\theta_{\textrm{cdm}}^\prime+\mathcal{H}\theta_{\textrm{cdm}}=\frac{\bar{\rho}_{\rm vac}^\prime}{\bar{\rho}_{\textrm{cdm}}}\theta_{\textrm{cdm}}-k^2\frac{\delta\rho_{\rm vac}}{\bar{\rho}_{\textrm{cdm}}}\,,
\end{equation}
\begin{equation}\label{eq:pertcons2}
\delta_{\textrm{cdm}}^\prime-\frac{\bar{\rho}_{\rm vac}^\prime}{\bar{\rho}_{\textrm{cdm}}}\delta_{\textrm{cdm}}+\frac{\delta\rho_{\rm vac}^\prime}{\bar{\rho}_{\textrm{cdm}}}+\theta_{\textrm{cdm}}+\frac{h^\prime}{2}=0\,,
\end{equation}
with $\delta_{\textrm{cdm}}=\delta\rho_{\textrm{cdm}}/\bar{\rho}_{\textrm{cdm}}$ the CDM density contrast and $\theta_{\textrm{cdm}}$ the divergence of the perturbed CDM 3-velocity. We consider a vacuum–geodesic CDM interaction such that there is no net momentum transfer between the vacuum and cold dark matter \cite{Wang:2013qy,Wang:2014xca,Gomez-Valent:2018nib}. Thus, we can fix the gauge by setting $\theta_{\textrm{cdm}}=0$, as in the $\Lambda$CDM. This automatically sets $\delta\rho_{\rm vac}=0$. In this setup, Eq. \eqref{eq:pertcons2} simplifies,

\begin{equation}
\delta_{\textrm{cdm}}^\prime+\frac{h^\prime}{2}-\frac{\bar{\rho}_{\rm vac}^\prime}{\bar{\rho}_{\textrm{cdm}}}\delta_{\textrm{cdm}}=0\,.
\end{equation}
This is actually the only perturbation equation that must be modified in \texttt{CLASS} in order to accommodate the dynamical character of the VED, i.e. the fact that $\bar{\rho}_{\rm vac}^\prime\neq 0$. In this work, we consider adiabatic perturbations for the various matter and radiation species.
\subsection{Type I with threshold}\label{subsec:TypeIthreshold}
Once we have obtained the background and the perturbation equations we are in position to study the cosmological evolution of the RVM in different situations. The conventional option would be to assume that the above equations are valid throughout the entire cosmic history (subsequent, of course,  to the inflationary period, which will not be dealt with here at all). The phenomenological analyses of Refs.\cite{Sola:2016ecz,Sola:2016zeg,sola2017first,Sola:2017jbl,Sola:2015wwa,Gomez-Valent:2014rxa}, for example, were based on that standard assumption.  However, we may also entertain the intriguing possibility that the interaction between the vacuum and dark matter is only relatively recent. In that case we could have a scenario where the dynamics of the vacuum starts approximately at the time when it becomes dominant over matter (i.e. at about the outset of what is usually referred to as the DE epoch). Such  a situation should be characterized by the presence of a `threshold point' for the vacuum dynamics at some redshift value $z_*$, where the transition occurs.  The idea is to study the response of our fit to the overall cosmological data when we switch off the interaction between the VED and the CDM for most of the cosmic history, except when we approach the usual epoch of vacuum dominance.  Since the conventional DE epoch  in the late universe is usually assumed to commence at around a redshift value $z_{*}\simeq 1$, we will assume that its evolution also gets kicked off at around that point (see below).

We will implement the simplest version of such a threshold scenario through a Heaviside $\Theta$-function, and for definiteness it will be restricted to type-I models only. Thus,  let  $a_{*}$  be  the value of the scale factor where the  activation of the vacuum dynamics occurs ($z_*=a_*^{-1}-1$ being the corresponding redshift value). Before reaching that point (that is, at earlier epochs $a<a_*$)  the vacuum is rigid, whereas after that point (hence nearer to our present time) the vacuum evolves with the expansion following the type-I running vacuum behavior, see Eq.\,\eqref{eq:rhovac}. We have checked that the optimal value for this parameter is $a_{*}\simeq 0.5$, which indeed corresponds to $z_*\simeq 1$ \footnote{In practice this means that we have first fitted the value of $z_*$ as one more free parameter in our analysis. Subsequently we have assumed that the threshold point remains fixed at that point. See also \cite{Salvatelli:2014zta,Martinelli:2019dau,Hogg:2020rdp,Goh:2022gxo}  for a binned/tomographic approach to the DE. In our case we have just one threshold whose existence might be motivated by QFT calculations \cite{Moreno-Pulido:2022phq,Moreno-Pulido:2022upl}. }
It should be noted that while the VED function $\rv(a)$ will remain continuous in our implementation of the step function procedure, its time derivative $\dot{\rho}_{\rm vac} $ does not, and in fact it gets modified through a Heaviside function factor $\Theta(a-a_{*})$.
Accordingly, the derivative of the CDM energy density must also change in a discontinuous way. In contradistinction, all the energy densities are continuous at $a=a_{*}$. Consequently, we find that in order to fulfill these requirements we must implement the analytical expressions for the various density functions in the following way:
\newline
\newline
{\bf \underline {$a < a_{*}$}\ ($z>z_*$)}
\begin{align}
&\rho_{\textrm{cdm}}(a) = \rho_{\textrm{cdm}}(a_{*})\left(\frac{a}{a_{*}}\right)^{-3}\\
&\rho^{*}_{\textrm{vac} } = \rho^0_{\textrm{vac}} + \left(\frac{1}{\xi} -1\right)\rho^0_m\left(a_{*}^{-3\xi} -1\right) = {\rm const.}
\end{align}
{\bf \underline {$a > a_{*}$}}\ ($z<z_*$)
\begin{align}
&\rho_{\textrm{cdm}}(a) = \rho^{0}_m{a^{-3\xi}}  - \rho^0_{\rm b}{a^{-3}}  \\
&\rho_{\textrm{vac}}(a) = \rho^0_{\textrm{vac}} + \left(\frac{1}{\xi} -1\right)\rho^0_m\left(a^{-3\xi} -1\right).
\end{align}
In the above expression, we have defined
\begin{equation}
\rho_{\textrm{cdm}}(a_{*}) = \rho^{0}_m{a_{*}^{-3\xi}}  - \rho^0_{\rm b}{a_{*}^{-3}}= {\rm const.}
\end{equation}
It goes without saying that the same modifications have to be applied in the perturbation sector.We denote this version of the type-I RRVM with threshold as type-I RRVM$_{\rm thr.}$.

We do not wish to speculate here on the possible origin of the threshold postulated above, it  could be a manifestation of a late-time interaction in the dark sector.  However, we mention that a  fundamental microscopical explanation might  come from the  RVM framework emerging from QFT in curved spacetime\cite{Moreno-Pulido:2022upl}, which  indicates that the EoS of the quantum vacuum stays in the characteristic DE range ($0<\wv<-1/3$) only below  a redshift value (threshold) in our recent past $z_*\simeq 1$. For $z>z_*$, instead, one has $\wv>-1/3$ and the vacuum  does no longer behave as DE. This is of course impossible for the classical vacuum, for which $\wv=-1$ all the time.  Additional studies will obviously  be necessary to gauge the impact of such EoS behavior on the global fits to the cosmological data and its potential relation with the type-I scenario with threshold that we have defined above. Finally, we mention that the behavior of the type-I RRVM with threshold should be essentially the same as that of the `canonical RVM' with threshold, namely the original RVM form with the dynamical component $\sim H^2$, see\cite{Sola:2013gha,Sola:2015rra} and references therein. The latter corresponds to \eqref{eq:RVMvacuumdadensity} with only the single parameter $\nu$ (with $\tilde{\nu}=0$). At  low $z$, the two models are expected to be indistinguishable from the phenomenological  point of view.

\section{Type II: running vacuum with running $G$}\label{sec:RRVMTypeII}

For type-II RRVM we have an entirely different sort of scenario, in which matter is strictly conserved, in particular dark matter, and hence no interaction of any sort is permitted between matter and vacuum. However, to switch off the energy exchange between matter and vacuum in a fully consistent way with the Bianchi identity, we must allow for the running of the gravitational coupling with the expansion,  $G=G(H)$.  Therefore, for type-II models we have both the running of the vacuum energy density $\rv$ and the running of $G$. Let us briefly see how it comes about.
If matter is conserved we have $\dot{\rho}_t+3H(\rho_t+p_t)=0$, where as in the previous sections the subscript $t$ refers to the sum of all the species in the universe excluding the vacuum. Whereupon the general Bianchi identity \eqref{BianchiGeneral} boils down to\footnote{If (dark) matter is not conserved, but $G$ remains constant, we retrieve of course our previous scenario \eqref{ConservationEquation}. In general, we may expect a mixture of both situations, but we shall refrain from dealing with the general case since it would introduce extra parameters. See, however, \cite{Fritzsch:2012qc,Fritzsch:2015lua} for additional discussions that can be relevant for studies on the possible variation of the fundamental constants of Nature.}
\begin{equation}\label{eq:Bianchidiv}
\dot{G}\left(\rho_t+\rv\right)+G\,\dot{\rho}_{\rm vac }=0\,.
\end{equation}
Since the running of $\rv$ is still fixed by \eqref{eq:rhov}, the previous equation is essential to determine the running of $G$. Being $\dot{\rho}_{\rm vac}\propto\nu$, the sign of $\nu$ determines the sign of $\dot{G}$, i.e. if $\nu>0$ ($\nu<0$) $G$ increases (decreases) with the cosmic expansion. Of course, we have to make sure that $G$ evolves in a very mild way, which in fact turns out to be the case as we shall verify explicitly.

In what follows we bring forth the relevant background and linear perturbation equations for the  type-II RRVM  in Secs. \ref{sec:backRRVMII} and \ref{sec:pertRRVMII}, respectively. As it will become clear, the solution of this model type is more complicated.

\subsection{Background equations}\label{sec:backRRVMII}

We come back to the following form of the vacuum energy density,
\begin{equation}\label{eq:vacuum}
\rv(\mathcal{R})=C_0+\frac{\nu}{32\pi G_N}\mathcal{R}\,.
\end{equation}
which is of course equivalent to our original RRVM expression \eqref{eq:RRVM}, with  $C_0\equiv 3c_0/(8\pi G_N)$ and $\mathcal{R}=12 H^2+6\dot{H}$. The Friedmann and pressure equations read, respectively,
\begin{equation}
3H^2=8\pi G\left[\rho_t+C_0+\frac{3\nu}{16\pi G_N}(2H^2+\dot{H})\right]\,,
\end{equation}
\begin{equation}
-(3H^2+2\dot{H})=8\pi G\left[p_t-C_0-\frac{3\nu}{16\pi G_N}(2H^2+\dot{H})\right]\,,
\end{equation}
\jtext{where $G_N$ is  Newton’s gravitational constant}, whereas $G$ stands for the running gravitational coupling. For type-II models $G$ evolves with the expansion, and hence it is generally different from $G_N$.
At the same time, we remind the reader that for type-II models the background energy densities and pressures of the matter species evolve as a function  of the scale factor exactly  as in the $\Lambda$CDM, since now there is no energy exchange between them and the vacuum.

A practical consideration is now in order which will make clear immediately why solving the type-II models is more difficult. Recall that for the numerical analysis we are using the \texttt{CLASS} system solver \cite{Lesgourgues:2011re,Blas:2011rf}. Now the point is that for the standard cosmological model, \texttt{CLASS}  computes $H$ and $\dot{H}$ after computing the energy densities of the various components that fill the universe. In the model under consideration, though, we cannot proceed in the same way, just because we first need to compute $G$. Before explaining how it is still possible to solve the system in the \texttt{CLASS} platform,  it is useful to rewrite the above equations in terms of the auxiliary variable\footnote{It should be clear that $\varphi$ is not a dynamical degree of freedom, in contradistinction to Brans-Dicke type theories of gravitation\,\cite{BransDicke1961}, and therefore $\varphi$ does not mediate any sort of long-range interaction that should be subdued by screening mechanisms.}  $\varphi \equiv G_N/G$.  One expects $\varphi\simeq 1$ at present and one may even impose this condition (see, however, below). In terms of $\varphi$,  the set of relevant equations read
\begin{equation}\label{eq:fried}
3H^2=\frac{8\pi G_N}{\varphi}\left[\rho_t+C_0+\frac{3\nu}{16\pi G_N}(2H^2+\dot{H})\right]\,,
\end{equation}
\begin{equation}\label{eq:pres}
-(3H^2+2\dot{H})=\frac{8\pi G_N}{\varphi}\left[p_t-C_0-\frac{3\nu}{16\pi G_N}(2H^2+\dot{H})\right]\,,
\end{equation}
\begin{equation}\label{eq:KG}
\frac{\dot{\varphi}}{\varphi}=\frac{\dot{\rho}_{\rm vac} }{\rho_t+\rv}\,,
\end{equation}
where the last one, Eq. \eqref{eq:KG}, is of course nothing but a rephrasing of Eq.\,\eqref{eq:Bianchidiv}.
This equation is rather complicated since it is obvious from \eqref{eq:rhov} that it involves not only $H$ and $\dot{H}$, but also $\ddot{H}$.  In order to compute the latter it is possible, of course, to differentiate the pressure equation and use it together with \eqref{eq:fried} and \eqref{eq:pres}. Unfortunately, in doing so one obtains $\ddot{H}$ as a function of the derivative of the neutrinos energy density and pressure, which should then be computed numerically. This approach looks too complicated, and therefore we opt for the following alternative and simpler method.  We first obtain a differential equation for $H$. Dividing out equations \eqref{eq:fried} and \eqref{eq:pres} we can get rid of $\varphi$, and after some rearrangement we are led to the following differential equation:
\begin{equation}
0=\frac{3\nu}{8\pi G_N}\dot{H}^2+\dot{H}\left(2C_0+2\rho_t+\frac{3\nu H^2}{4\pi G_N}\right)+3H^2(\rho_t+p_t)\,.
\end{equation}
The previous equation can be restated in the more convenient form
\begin{equation}\label{eq:Hdot}
\dot{H}=\frac{4\pi G_N}{3\nu}\left(-B+\sqrt{B^2-\nu\frac{9H^2}{2\pi G_N}(p_t+\rho_t)}\right)\,,
\end{equation}
where we have defined the function
\begin{equation}\label{eq:B_definition}
B\equiv 2C_0 + 2\rho_t + \frac{3\nu{H^2}}{32{G_N}} \,.
\end{equation}
Equation \eqref{eq:Hdot} can be solved much more easily than \eqref{eq:KG}, although we still need to employ a numerical procedure. We have all the necessary ingredients. As initial condition (at $z_{\rm ini}\sim 10^{14}$) we can use
\begin{equation}
H^2_{\rm ini}=\frac{8\pi G_N}{3\varphi_{\rm ini}}\rho_r(z_{\rm ini})\,,
\end{equation}
because the radiation energy density clearly dominates over the nonrelativistic matter and the vacuum. We can tell \texttt{CLASS} to apply the finite difference method to solve step by step the system. In each of these steps \texttt{CLASS} computes $H_{n+1}$ from $H_n$ and $\dot{H}_n$ \eqref{eq:Hdot}. For the latter it takes the various energy densities and pressures. Then we can employ the Friedmann equation to compute $\varphi_{n+1}$,

\begin{equation}
\varphi=\frac{8\pi G_N}{3H^2}\left[\rho_t+C_0+\frac{3\nu}{16\pi G_N}(2H^2+\dot{H})\right],
\end{equation}
and iterate the process till we have the complete expansion history to the necessary degree of accuracy.

Next we show what is the evolution of $\varphi$ during the radiation-dominated (RDE) epoch. First, let us write $\rv$ in terms of the energy densities, pressures, and $\varphi$. Notice that using \eqref{eq:fried} and \eqref{eq:pres} we obtain:

\begin{equation}
H^2=\frac{8\pi G_N}{3(\varphi-\nu)}\left[\rho_t+C_0-\frac{3}{4}\frac{\nu}{\varphi}(\rho_t+p_t)\right]\,,
\end{equation}

\begin{equation}
\dot{H}=-\frac{4\pi G_N}{\varphi}(\rho_t+p_t)\,.
\end{equation}
Introducing these expressions into \eqref{eq:vacuum} we get:

\begin{equation}\label{eq:RDEapprox}
\rv(a)=\frac{\varphi(a)C_0+\frac{\nu}{4}[\rho_t(a)-3p_t(a)]}{\varphi(a)-\nu}\,.
\end{equation}
Even though we have solved the type-II model in an exact way using the aforementioned numerical strategy, the following approximate analytical considerations may help to better understand the behaviour of the solution.
In the RDE, we have
\begin{equation}\label{eq:VDEm}
\rv(a)=\frac{\nu\rho_m^{0}}{4\varphi(a)}a^{-3}+\mathcal{O}(\nu^2)
\end{equation}
since only the nonrelativistic component $\rho_m=\rho_m^0 a^{-3}$ contributes in the term proportional to $\nu$ in the numerator  of \eqref{eq:RDEapprox} after the pressure and the radiation densities cancel in the difference $\rho_r(a)-3p_r(a)$.
Finally, using \eqref{eq:VDEm} in  \eqref{eq:KG} we can easily integrate the resulting equation since in the denominator we have $\rv(a)\ll\rho_t(a)\simeq\rho_r^0 a^{-4}$ in the radiation epoch. Integrating from  the initial scale factor value  $a_{\rm ini}=10^{-14}$ (see above) up to an arbitrary value,  we find the evolution of the gravitational coupling within this analytical  approximation:
\begin{equation}\label{eq:varphiRDE}
\varphi(a)=\varphi_{\rm ini}-\frac{3\nueff}{a_{\textrm{eq}}}(a-a_{\rm ini})\,,
\end{equation}
with $a_{\textrm{eq}}=\Omega^{0}_{r,*}/\Omega_m^{0}$ and $\Omega^{0}_{r,*}$   being the radiation density parameter computed assuming that the neutrinos are all massless.  Similarly to what we have done for the type-I RRVM (cf. Sec.\,\ref{sec:backRRVMI}), we have defined
\begin{equation}\label{eq:nueff_definitoin_typeII}
\nu_{\rm eff}\equiv\frac{\nu}{4}\,,
\end{equation}
as in \cite{SolaPeracaula:2021gxi}.  We actually report the fitting value of this parameter in our tables and contour plots, see the discussion in Sec. \ref{sec:Discussion_of_the_results}. The term $(a-a_{\rm ini})/a_{\rm eq}$ in \eqref{eq:varphiRDE} is much smaller than 1, since \eqref{eq:varphiRDE} is valid for $a\ll a_{\rm eq}$. Hence, the total variation of $\varphi$ during the RDE is small and of order $\nu_{\rm eff}$, $\Delta\varphi\approx -3\nu_{\rm eff}\sim \mathcal{O}(\nu)$, despite being linear with the scale factor. It is easy to see that the relation \eqref{eq:VDEm} still holds good in the MDE,  but the denominator of \eqref{eq:KG} is now dominated by the term $\rho_t(a)\simeq \rho_m^0 a^{-3}$. Integration now gives
\begin{equation}\label{eq:varphia}
\varphi(a)=C-3\nueff\ln\,a\,,
\end{equation}
where $C$ is a constant to be fixed by some initial condition in the MDE. It is easy to check that the total variation of $\varphi$ in the MDE will be of the order of $\sim \mathcal{O}(10)\nu$, i.e. ten times larger than in the RDE. In the vacuum-dominated epoch, the \textit{r.h.s.} of Eq.\,\eqref{eq:KG} goes to zero and $\varphi\to \textrm{const}$, so $G$ stays constant as well.

Traditional limits on the relative time variation of $G$
can be found e.g. in the review \cite{Uzan:2010pm}.  More recent determinations, e.g. those based on measurements on the double pulsar
PSR J0737–3039A/B, yield rather tight bounds\,\cite{Kramer:2021jcw}:
\begin{equation}\label{eq:lc1}
\frac{\dot{G}}{G}=\left(-0.8\pm 1.4\right)\times 10^{-13} \frac{1}{{\cal F}_{AB}} \,{\rm yr}^{-1}\,,
\end{equation}
where ${\cal F}_{AB}\simeq 1$ for weakly self-gravitating bodies. A previous limit from a binary pulsar  (PSR J1713+0747) provided
\cite{Zhu:2018etc}
\begin{equation}\label{eq:lc2}
\frac{\dot{G}}{G}=\left(-0.1\pm 0.9\right)\times 10^{-12}\,{\rm yr}^{-1}\,,
\end{equation}
which is a bit weaker but essentially in the same ballpark.
On the other hand, the best current limit on the relative variation of $G$ obtained in the solar system is \cite{Genova2018}
\begin{equation}\label{eq:lc3}
\frac{\dot{G}}{G}<4\times 10^{-14}\,{\rm yr}^{-1}\,.
\end{equation}
Assuming that \eqref{eq:lc1}-\eqref{eq:lc3} can be used to constrain the cosmological evolution of $G$, let us tentatively use these limits (in order of magnitude) in combination with our equation \eqref{eq:varphia} to constrain the parameter $\nueff$. The last equation implies $\dot{\varphi}\simeq -3\nueff H_0$ for $H\simeq H_0\simeq 1.023 h\times 10^{-10}$ yr$^{-1}$ around our time ($h\simeq 0.7$)  The previous relation is equivalent to $\dot{G}/G\simeq 3\nueff H_0$ to order  $\nueff$. It is then easy to check that in order to fulfill the above limits we must require $|\nueff|\lesssim (2-5)\times 10^{-4}$, a condition which is  satisfied by our fitting values of $\nueff$ for the type-II models, even in the most restrictive case (see the fitting tables of  Sec. \ref{sec:Discussion_of_the_results}) \footnote{Let us emphasize that Eq.\,\eqref{eq:varphia} is valid only in the MDE, and we have also pointed out that $\varphi\to$ const. in the DE epoch. This means that $G$ gets more and more rigid when it transits from the MDE to the DE epoch, and therefore the actual limits on $\nueff$ are weaker than those that we have roughly estimated. This works on our benefit, of course. In fact, a detailed calculation would require computing $\varphi$ in the DE epoch, but it proves unnecessary once we have shown that even in the most unfavorable case (i.e. when $\varphi$ evolves more rapidly than it actually does in the DE epoch) the obtained limits on $\nueff$ are nonetheless preserved by our fits. Notice that type-I models are totally unaffected by these limits since $G$ is in this case constant, so $\nueff$ can be, in principle, larger for them. }.

From the above considerations it is clear that $\varphi=G_N/G$  changes very slowly (logarithmically) with the expansion and proportionally to the small parameter $\nu$. The predicted variations of $G$ within the type-II running vacuum models lie in fact within the most restrictive experimental limits existing in the literature.  This demonstrates our contention, mentioned previously,  that the running of $G$ within these models is consistent with the observations.

Finally we mention that with the purpose of giving more freedom to the model, in this work we will not impose  the condition $G(a=1)=G_N$ or, equivalently, $\varphi(a=1)=1$, as we did in the previous studies, such as \cite{SolaPeracaula:2021gxi} and also in  \cite{SolaPeracaula:2019zsl,SolaPeracaula:2020vpg,deCruzPerez:2023wzd} within the context of the Brans-Dicke model with a cosmological constant.  We naturally expect $\varphi(a=1)\simeq 1$, of course. The obtained fitting values for $\varphi$ at present are indicated in our tables as $\varphi(0)\equiv\varphi(z=0)$. Let us finally note that cosmologies with variable $G$ may have to rely on an efficient screening mechanism that allows to recover $G_N$ at the Solar System. We will not focus on this issue, which has been addressed in several places in the literature\,(see e.g. \cite{Amendola:2015ksp,Clifton:2011jh,Avilez:2013dxa} and references therein), and more recently in \cite{Gomez-Valent:2021joz}, as it deserves a more devoted study which is certainly beyond the scope of this work. Let us however note that this situation affects mainly the Brans-Dicke type models\,\cite{BransDicke1961}, where the effective gravitational coupling is tied to a dynamical scalar field that could mediate long range interactions as a true degree of freedom of the underlying  gravitational framework. As indicated before, this is not our case.  These situations may have an impact only in the non-linear scales, which are anyway not strongly affected by the cosmological observables employed in this study.  We close this section by noticing that while in this paper we have assumed that  $G$ may change with the cosmic time in type-II models,  it could also evolve with a distance scale in a galactic domain, $G=G(r)$ ($0<r<L$). Such an extension of the running of the gravitational coupling was considered in \cite{Shapiro:2004ch} and  could be helpful to connect the running of $G$ between the galactic/astrophysical and cosmological domains.

\subsection{Perturbation equations}\label{sec:pertRRVMII}

The actual implementation of the linear perturbation equations for type-II models in the \texttt{CLASS} computing platform is more difficult than the background part (which was already nontrivial), and certainly much more involved than the one carried out for the type-I variant (viz. the one with vacuum-CDM interaction). We use again the synchronous gauge, but in this case the gauge unfortunately does not fix $\delta\rv=0$, in contrast to the situation in Sec. \ref{sec:pertRRVMI}, so we have to keep the contribution of the vacuum perturbation in our equations.

The $00$, $0i$ and $ii$ components of the Einstein equations in Fourier space read, respectively,
\begin{equation}\label{eq:PerturbTypeIIEq1}
\mathcal{H}h^\prime-2\eta k^2=8\pi G_Na^2\sum_l\left(\frac{\delta\rho_l}{\bar{\varphi}}-\bar{\rho}_l\frac{\delta\varphi}{\bar{\varphi}^2}\right)\,,
\end{equation}
\begin{equation}\label{eq:PerturbTypeIIEq2}
\eta^\prime k^2=\frac{4\pi G_N}{\bar{\varphi}} a^2\sum_l (\bar{\rho}_l+\bar{p}_l)\theta_l\,,
\end{equation}
\begin{equation}\label{eq:PerturbTypeIIEq3}
h^{\prime\prime}+2\mathcal{H}h^\prime-2\eta k^2=\frac{24\pi G_N a^2}{\bar{\varphi}}\sum_l\left(\frac{\bar{p}_l}{\bar{\varphi}}\delta\varphi-\delta p_l\right)\,,
\end{equation}
where we have of course split also $\varphi$ into a background contribution and a perturbation, $\varphi=\bar{\varphi}+\delta\varphi$. The equation coming from the part that is proportional to $\hat{k}_i\hat{k}_j$ of the Einstein $ij$ component reads,

\begin{equation}
h^{\prime\prime}+6\eta^{\prime\prime}+2\mathcal{H}(h^\prime+6\eta^\prime)-2\eta k^2=-\frac{24\pi G_N}{\bar{\varphi}} a^2(\bar{\rho}+\bar{p})\sigma\,.
\end{equation}
The zero and spatial component of the covariant conservation equation of the vacuum are, respectively,

\begin{equation}\label{eq:deltavarphi}
0=\frac{\bar{\varphi}^\prime}{\varphi}\left(\frac{\delta\rho}{\bar{\rho}}-\frac{\delta\varphi}{\bar{\varphi}}\right)+\frac{\delta\varphi^\prime}{\bar{\varphi}}-\frac{\delta\rho^\prime_{\textrm{vac}}}{\bar{\rho}}\,,
\end{equation}
\begin{equation}\label{eq:extra}
0=\bar{\varphi}^\prime(\bar{p}+\bar{\rho})\theta-k^2(\bar{p}\delta\varphi+\bar{\varphi}\delta\rv)\,,
\end{equation}
where in the last two equations
\begin{equation}
\delta\rho=\sum_l\delta\rho_l\ ;\qquad  \bar{p}=\sum_l\bar{p}_l\ ;\qquad \bar{\rho}=\sum_l\bar{\rho}_l\ ;\qquad (\bar{p}+\bar{\rho})\theta =\sum_l(\bar{p}_l+\bar{\rho}_l)\theta_l\equiv g\,.
\end{equation}
It is possible to isolate $\delta\rv$ from \eqref{eq:extra} and substitute it in the other equations. The only difficulty is found when the substitution is performed in \eqref{eq:deltavarphi}, since in order to do this we need to evaluate the quantity $g^\prime \equiv [(\bar{p}+\bar{\rho})\theta]^\prime$. \texttt{CLASS} computes $g=[(\bar{p}+\bar{\rho})\theta]$ but not its derivative. If we knew how to write $[(\bar{p}+\bar{\rho})\theta]^\prime$ in terms of $\delta\varphi$, $\delta\varphi^\prime$, $g$ and other accessible quantities, then we could obtain the differential equation for $\delta\varphi$ from \eqref{eq:deltavarphi} and implement it in \texttt{CLASS} without defining explicitly $\delta\rv$, just incorporating the effect of the vacuum perturbation directly into the equations. This is actually possible. Let us start differentiating \eqref{eq:extra}. We obtain:

\begin{equation}\label{eq:interm}
\delta\rv^\prime=-\frac{\bar{p}}{\bar{\varphi}}\delta\varphi^\prime-\frac{\bar{p}^{\prime}}{\bar{\varphi}}\delta\varphi+\frac{\bar{\varphi}^{\prime\prime}g+\bar{\varphi}^\prime g^\prime-(\bar{\varphi}^\prime)^2g/\bar{\varphi}}{k^2\bar{\varphi}}+\frac{\bar{\varphi}^\prime\bar{p}\delta\varphi}{\bar{\varphi}^2}\,.
\end{equation}
Notice that the quantity $g^\prime$, which we have defined previously,  appears in this expression. As \texttt{CLASS} does not provide it to us we need to evaluate it by adding some supplementary piece in the code. It is possible to obtain $g^\prime$ upon differentiating \eqref{eq:PerturbTypeIIEq2} and combining it with \eqref{eq:PerturbTypeIIEq3}. The result reads

\begin{equation}\label{eq:gprime}
g^\prime = \frac{k^2\bar{\varphi}\mathcal{H}}{12\pi G_N a^2}(h^\prime-k^2f^\prime)+k^2\left(\delta p-\frac{\bar{p}\delta\varphi}{\bar{\varphi}}\right)+\left(\frac{\bar{\varphi}^\prime}{\bar{\varphi}}-2\mathcal{H}\right)g\,,
\end{equation}
where $f\equiv (h+6\xi)/k^2$ and its derivative, $f^\prime$, are quantities that we can get from \texttt{CLASS}. Also $h^\prime$ is provided by \texttt{CLASS}. Now we only need to substitute $g^\prime$ from \eqref{eq:gprime} in \eqref{eq:interm}, and substitute the resulting expression for $\delta\rv^\prime$ in \eqref{eq:deltavarphi}. In doing so we finally get the equation for the $\delta\varphi$ perturbation:

\begin{equation}
\delta\varphi^\prime=\frac{\bar{\rho}}{\bar{\rho}+\bar{p}}\left[\delta\varphi\left(\frac{\bar{\varphi}^\prime}{\bar{\varphi}}-\frac{\bar{p}^\prime}{\bar{\rho}}+\frac{\bar{\varphi}^\prime}{\bar{\varphi}}\frac{\bar{p}}{\bar{\rho}}\right)-\bar{\varphi}^\prime\frac{\delta\rho}{\bar{\rho}}+\frac{\bar{\varphi}^{\prime\prime}g+\bar{\varphi}^\prime g^\prime-g(\bar{\varphi}^\prime)^2/\bar{\varphi}}{k^2\bar{\rho}}\right]\,.
\end{equation}
The initial condition for $\delta\varphi$ is easy to find. Deep in the RDE we can neglect all the terms proportional to $\bar{\varphi}^\prime$ and $\bar{\varphi}^{\prime\prime}$, so we are left with the following simple equation:
\begin{equation}
\delta\varphi^\prime=-\delta\varphi\frac{\bar{p}^\prime}{\bar{\rho}+\bar{p}}\longrightarrow  a\frac{d\delta\varphi}{da}=\delta\varphi\longrightarrow \delta\varphi=\tilde{C}a\,,
\end{equation}
where in the RDE we have used $\bar{p}=\frac13\bar{\rho}=\frac13\rho_r^0 a^{-4}$, of course. The background value of $\varphi$, i.e. $\bar{\varphi}$, remains almost constant during the RDE. It only grows very mildly with $a$, with a constant of proportionality which is small. Hence we would expect the constant $\tilde{C}$ to be small as well. In any case, we expect $\delta\varphi\approx 0$ at $z_{\rm ini}=10^{14}$, which is the initial condition that we use for $\delta\varphi$ in our modified version of  \texttt{CLASS}.

 %%%%%%%%%%%%%%%%%%%%%%%%%%%%%%%%%%%%%%%%%%%%%%%%%%%%%%%%%%%%%%%%%%%%%%%%%%%%%%%%%%%%%%%%%
%%%%%%%%%%%%%%%%%%%%%%%%%%%%%%%%%%%%%%%%%%%%%%%%%%%%%%%%%%%%%%%%%%%%%%%%%%%%%%%%%%%%%%%%%%%%%%%%%%%%%%%%%%

%
\begin{table}[t!]
\begin{center}
\resizebox{14.5cm}{!}{
\begin{tabular}{| c | c |c | c |c|}
\multicolumn{1}{c}{Survey} &  \multicolumn{1}{c}{$z$} &  \multicolumn{1}{c}{Observable} &\multicolumn{1}{c}{Measurement} & \multicolumn{1}{c}{{\small References}}
\\\hline
6dFGS+SDSS MGS & $0.122$ & $D_V(r_d/r_{d,\rm fid})$[Mpc] & $539\pm17$[Mpc] &\cite{Carter:2018vce}
\\\hline
DR12 BOSS & $0.32$ & $Hr_d/(10^{3}\rm km/s)$ & $11.549\pm0.385$   &\cite{Gil-Marin:2016wya}\\ \cline{3-4}
 &  & $D_A/r_d$ & $6.5986\pm0.1337$ &\tabularnewline \cline{3-4}
 \cline{2-2}& $0.57$ & $Hr_d/(10^{3} \rm km/s)$  & $14.021\pm0.225$ &\\ \cline{3-4}
 &  & $D_A/r_d$ & $9.389\pm0.1030$ &
\\\hline
WiggleZ & $0.44$ & $D_V(r_d/r_{d,\rm fid})$[Mpc] & $1716.4\pm 83.1$[Mpc] &\cite{Kazin:2014qga} \tabularnewline
\cline{2-4} & $0.60$ & $D_V(r_d/r_{d,\rm fid})$[Mpc] & $2220.8\pm 100.6$[Mpc]&\tabularnewline
\cline{2-4} & $0.73$ & $D_V(r_d/r_{d,\rm fid})$[Mpc] &$2516.1\pm 86.1$[Mpc] &
\\\hline
DESY3 & $0.835$ & $D_M/r_d$ & $18.92\pm 0.51$ &\cite{DES:2021esc}
\\\hline
eBOSS Quasar & $1.48$ & $D_M/r_d$ & $30.21\pm 0.79$   &\cite{Hou:2020rse}
\\ \cline{3-4} &  & $D_H/r_d$ & $13.23\pm 0.47$
&\\\hline
Ly$\alpha$-Forests & $2.334$ & $D_M/r_d$ & $37.5^{+1.2}_{-1.1}$   &\cite{duMasdesBourboux:2020pck}
\\ \cline{3-4} &  & $D_H/r_d$ & $8.99^{+0.20}_{-0.19}$
&\\\hline
\end{tabular}}
\caption{\scriptsize Published values of BAO data, see the quoted references for details and for the corresponding covariance matrices. The fiducial values of the comoving sound horizon appearing in the table are $r_{d,{\rm fid}} = 147.5$ Mpc for \cite{Carter:2018vce} and $r_{d,{\rm fid}} = 148.6$ Mpc for \cite{Kazin:2014qga}.}\label{tab:BAO_table}
\end{center}
\end{table}
\begin{table}[t!]
\begin{center}
\begin{tabular}{| c | c | c |}
\multicolumn{1}{c}{$z$} &  \multicolumn{1}{c}{$H(z)$[km/s/Mpc]} & \multicolumn{1}{c}{{\small References}}
\\\hline
$0.07$ & $69.0\pm 19.6$ & \cite{Zhang:2012mp}
\\\hline
$0.09$ & $69.0\pm 12.0$ & \cite{Jimenez:2003iv}
\\\hline
$0.12$ & $68.6\pm 26.2$ & \cite{Zhang:2012mp}
\\\hline
$0.17$ & $83.0\pm 8.0$ & \cite{Simon:2004tf}
\\\hline
$0.1791^{*}$ & $77.72\pm 6.01$ & \cite{Moresco:2012jh}
\\\hline
$0.1993^{*}$ & $77.79\pm 6.83$ & \cite{Moresco:2012jh}
\\\hline
$0.2$ & $72.9\pm 29.6$ & \cite{Zhang:2012mp}
\\\hline
$0.27$ & $77.0\pm 14.0$ & \cite{Simon:2004tf}
\\\hline
$0.28$ & $88.8\pm 36.6$ & \cite{Zhang:2012mp}
\\\hline
$0.3519^{*}$ & $85.45\pm 15.75$ & \cite{Moresco:2012jh}
\\\hline
$0.3802^{*}$ & $86.17\pm 14.61$ & \cite{Moresco:2016mzx}
\\\hline
$0.4$ & $95.0\pm 17.0$ & \cite{Simon:2004tf}
\\\hline
$0.4004^{*}$ & $79.90\pm 11.38$ & \cite{Moresco:2016mzx}
\\\hline
$0.4247^{*}$ & $90.39\pm 12.76$ & \cite{Moresco:2016mzx}
\\\hline
$0.4497^{*}$ & $96.24\pm 14.38$ & \cite{Moresco:2016mzx}
\\\hline
$0.47$ & $89.0\pm 49.6$ & \cite{Ratsimbazafy:2017vga}
\\\hline
$0.4783^{*}$ & $83.74\pm 10.18$ & \cite{Moresco:2016mzx}
\\\hline
$0.48$ & $97.0\pm 62.0$ & \cite{Stern:2009ep}
\\\hline
$0.5929^{*}$ & $106.80\pm 15.06$ & \cite{Moresco:2012jh}
\\\hline
$0.6797^{*}$ & $94.875\pm 10.600$ & \cite{Moresco:2012jh}
\\\hline
$0.75$ & $89.0\pm 49.6$ & \cite{Borghi:2021rft}
\\\hline
$0.7812^{*}$ & $96.27\pm 12.72$ & \cite{Moresco:2012jh}
\\\hline
$0.8754^{*}$ & $124.70\pm 17.13$ & \cite{Moresco:2012jh}
\\\hline
$0.88$ & $90.0\pm 40.0$ & \cite{Stern:2009ep}
\\\hline
$0.9$ & $117.0\pm 23.0$ & \cite{Simon:2004tf}
\\\hline
$1.037^{*}$ & $133.35\pm 18.12$ & \cite{Moresco:2012jh}
\\\hline
$1.3$ & $168.0\pm 17.0$ & \cite{Simon:2004tf}
\\\hline
$1.363^{*}$ & $163.95\pm 34.61$ & \cite{Moresco:2015cya}
\\\hline
$1.43$ & $177.0\pm 18.0$ & \cite{Simon:2004tf}
\\\hline
$1.53$ & $140.0\pm 14.0$ & \cite{Simon:2004tf}
\\\hline
$1.75$ & $202.0\pm 40.0$ & \cite{Simon:2004tf}
\\\hline
$1.965^{*}$ & $191.10\pm 51.91$ & \cite{Moresco:2015cya}
\\\hline
\end{tabular}
\end{center}
\caption{\scriptsize Values of $H(z)$ from cosmic chronometers and their $1 \sigma$ uncertainties, which include the contribution of statistical and systematic effects \cite{Moresco:2020fbm}. They are expressed in km/s/Mpc. We have considered the correlations between the data points marked with a $^{*}$, as discussed in \cite{Moresco:2020fbm}. In some of the quoted references, the authors provide measurements obtained with two different stellar population synthesis (SPS) models. In these cases, we have employed the mean of the two central values and statistical errors. The systematic uncertainty already accounts for the choice of SPS model.}
\label{tab:CC_table}
\end{table}
\begin{table}[t!]
\begin{center}
\resizebox{10cm}{!}{
\begin{tabular}{| c | c |c | c |}
\multicolumn{1}{c}{Survey} &  \multicolumn{1}{c}{$z$} &  \multicolumn{1}{c}{$f(z)\sigma_8(z)$} & \multicolumn{1}{c}{{\small References}}
\\\hline
ALFALFA & $0.013$ & $0.46\pm 0.06$ & \cite{Avila:2021dqv}
\\\hline
6dFGS+SDSS & $0.035$ & $0.338\pm 0.027$ & \cite{Said:2020epb}
\\\hline
GAMA & $0.18$ & $0.29\pm 0.10$ & \cite{Simpson:2015yfa}
\\ \cline{2-4}& $0.38$ & $0.44\pm0.06$ & \cite{Blake:2013nif}
\\\hline
 WiggleZ & $0.22$ & $0.42\pm 0.07$ & \cite{Blake:2011rj} \tabularnewline
\cline{2-3} & $0.41$ & $0.45\pm0.04$ & \tabularnewline
\cline{2-3} & $0.60$ & $0.43\pm0.04$ & \tabularnewline
\cline{2-3} & $0.78$ & $0.38\pm0.04$ &
\\\hline
DR12 BOSS & $0.32$ & $0.427\pm 0.056$  & \cite{Gil-Marin:2016wya}\\ \cline{2-3}
 & $0.57$ & $0.426\pm 0.029$ &
\\\hline
VIPERS & $0.60$ & $0.49\pm 0.12$ & \cite{Mohammad:2018mdy}
\\ \cline{2-3}& $0.86$ & $0.46\pm0.09$ &
\\\hline
VVDS & $0.77$ & $0.49\pm0.18$ & \cite{Guzzo:2008ac},\cite{Song:2008qt}
\\\hline
FastSound & $1.36$ & $0.482\pm0.116$ & \cite{Okumura:2015lvp}
\\\hline
eBOSS Quasar & $1.48$ & $0.462\pm 0.045$ & \cite{Hou:2020rse}
\\\hline
 \end{tabular}}
\end{center}
\caption{\scriptsize Published values of $f(z)\sigma_8(z)$, see the quoted references and text in Sec. \ref{sec:Data_and_methodology}.}
\label{tab:fs8_table}
\end{table}
\section{Data and methodology}\label{sec:Data_and_methodology}
We fit the $\Lambda$CDM model, the running vacuum models under consideration (the type-I RRVM, the type-I RRVM$_{\textrm{thr.}}$ and the type-II RRVM) and finally the  XCDM \cite{Turner:1998ex} (also called $w$CDM), a generic parameterization of the dynamical DE whose dark energy EoS, $w_0$, is constant and is one parameter of the fit (expected to lie near $-1$). To test the response of the XCDM along with the relevant models under consideration can be useful, as it serves as a benchmark scenario for generic models of dynamical dark energy. We fit all these models to a large, robust and updated set  of cosmological observations from all the main sources.  Our data set involves observations from: i) distant type Ia Supernovae (SNIa);  ii) baryonic acoustic oscillations (BAO);  iii) a compilation of (differential age) measurements of the Hubble parameter at different redshifts ($H(z_i)$);  iv)  large-scale structure (LSS) formation data  (specifically, an updated list of data points on the observable  $f(z_i)\sigma_8(z_i)$), and, finally,  v) CMB Planck 2018 data of different sorts.  A brief description now follows of each of these datasets along with the corresponding references:
\newline
\newline
{\bf SNIa}: We consider the data from the so-called  `Pantheon+' compilation \cite{Scolnic:2021amr}, which contains the apparent magnitudes and redshifts associated to 1701 light curves obtained from 1550 SNIa in the redshift range $0.001 \leq z \leq 2.26$. See Sec. 2.2 of \cite{Favale:2023lnp} for details of the theoretical formulae employed to take into account these data points. %In order to avoid possible biases due to the model-dependent estimation of peculiar velocities we perform a cut in the sample, removing those data points with $z<0.01$, as it is done in \cite{Brout:2022vxf}.
Interestingly, the new Pantheon+ compilation also includes the 77 light curves from the 42 SNIa in the host galaxies employed by the SH0ES team in their analysis \cite{Riess:2021jrx,Brout:2022vxf}. The distance to the host galaxies has been measured with calibrated Cepheids. The inclusion of these luminosity distances in our dataset will be made clear by adding the label ``+SH0ES''. They break the existing full degeneracy between $H_0$ and the absolute magnitude of SNIa, $M$, when only SNIa are considered in the analysis. The SH0ES calibration of the supernovae in conjunction with the cosmic distance ladder leads to larger preferred values of the Hubble parameter  of $73.04\pm 1.04$ km/s/Mpc\,\cite{Riess:2021jrx}. This large value as compared to Planck's measurement ($67.36\pm 0.54$ km/s/Mpc,  obtained from the TT,TE,EE+lowE+lensing data\cite{Aghanim:2018oex}), is at the root of the $\sim 5\sigma$ $H_0$-tension.
\newline
\newline
{\bf BAO}: We employ 13 data points on isotropic and anisotropic BAO estimators. See Table \ref{tab:BAO_table} to know the exact values and the corresponding references.
\newline
\newline
\textbf{Cosmic chronometers}: In our analyses we use 32 data points on the Hubble parameter $H(z_i)$ measured with the differential age technique \cite{Jimenez:2001gg}. They span the redshift range $0.07 \leq z \leq 1.965$. We provide the complete list of data points and the corresponding references in Table \ref{tab:CC_table}. We have considered the effect of the known correlations between the various data points, as explained in \cite{Moresco:2020fbm}. See also Table \ref{tab:CC_table} and its caption. The covariance matrix has been computed using the script provided in the following link \footnote{ \url{https://gitlab.com/mmoresco/CCcovariance/-/blob/master/examples/CC_covariance.ipynb}}.
\newline
\newline
{\bf LSS}: 15 large-scale structure (LSS) data points between $0.01\lesssim z\lesssim 1.5$ embodied in the observable $f(z_i)\sigma_8(z_i)$, which is known as the weighted linear growth rate, being $f(z)$ the so-called growth factor and $\sigma_8(z)$ the root mean square mass fluctuations on the $R_8=8h^{-1}$ Mpc scale. See Table \ref{tab:fs8_table} for the complete list of data points and the corresponding references. We can take advantage of the relation $f(z)\sigma_8(z) = -(1+z)\frac{d\sigma_8(z)}{dz}$ to compute this quantity. The function $\sigma_8(z)$ involves the matter power spectrum, which is computed numerically by our modified version of the Einstein-Boltzmann code \texttt{CLASS}. It is important to note that this way of computing $f(z)\sigma_8(z)$ can only be used provided that we are in the linear regime, since in this case and in our models the matter density contrast can be written as $\delta_m(a,k) = D(a)F(k)$, where the dependence on the scale factor and the comoving wave number $k$ is factored out. The term $D(a)$ is known as the growth function and $F(k)$ encodes the initial conditions \footnote{\jtext{While it is common to rescale the measured values of $f\sigma_8$ by a factor $\frac{H(z)D_A(z)}{\tilde{H}(z)\tilde{D}_A(z)}$ to account for the Alcock-Paczynski (AP) effect \cite{Alcock:1979mp} (in which the tildes denote the quantities computed in the fiducial cosmology employed by the galaxy surveys),
%It is not clear, though, to what extent is this expression correct, since there is no formal derivation of it in the literature and there is actually more than one correction factor available,
there does not not seem to exist a general consensus on the exact correction to apply, see e.g. \cite{Kazantzidis:2018rnb} and references therein. In this sense the above formula should be considered as just a rough estimate. We have actually checked that the AP-rescaling introduces negligible shifts in our fitting results, a conclusion that is well in accordance with previous analyses in the literature \cite{Macaulay:2013swa,Nesseris:2017vor,Kazantzidis:2018rnb}. For this reason we have opted to not include this correction in our work.}}.
\newline
\newline
{\bf CMB}: For the cosmic microwave background data, we utilize the full Planck 2018  TT,TE,EE+lowE likelihood \cite{Planck:2018vyg}. It incorporates the information of the CMB temperature and polarization power spectra, and their cross-correlation. We refer to this dataset simply as ``CMB''. We also test separately the Planck 2018  TT+lowE likelihood,  which does not include the effect of the high-$\ell$ multipoles of the CMB polarization spectrum. This is done to check the impact of this particular dataset on our fitting results. It is also useful to compare with our previous analyses\,\cite{SolaPeracaula:2021gxi}, in which only this type of CMB data were used. In our fitting scenarios, we indicate the removal of the high-$\ell$ CMB polarization data from the complete CMB likelihood with the label ``CMB (No pol.)''.
\newline
\newline
As described in the preceding lines, for the SNIa we may or may not include the information provided by the SH0ES team whereas for the CMB we can consider the effect of the high-$\ell$ polarization data or not. An alternative calibration method of the absolute magnitude of SNIa based on the tip of the red giant branch \cite{Freedman:2019jwv,Freedman:2021ahq} instead of Cepheids yields a measurement of $H_0$ somewhat in the middle of those provided by Planck \cite{Planck:2018vyg} and SH0ES \cite{Riess:2021jrx}, $H_0=69.8\pm 1.7$ km/s/Mpc \footnote{This region is also preferred by late-time dynamical DE models when fitted to a very wide variety of background data that are independent from the direct cosmic distance ladder and CMB, $H_0=69.8\pm 1.3$ km/s/Mpc \cite{Cao:2023eja}. See, however, \cite{Yuan:2019npk,Soltis:2020gpl,Anand:2021sum,Scolnic:2023mrv} for measurements of $H_0$ more in accordance with SH0ES obtained also with the tip of the red giant branch method.}. In addition, it is also convenient to test the impact of the CMB polarization data from Planck, since previous works in the literature have found a moderate inconsistency between them and the Planck CMB temperature data, both in the 2015 \cite{Lin:2017bhs} and 2018 \cite{Garcia-Quintero:2019cgt} releases. This inconsistency could be due to a deficiency of the $\Lambda$CDM or the presence of unaccounted systematics in the data. Hence, these arguments motivate us to explore these four different datasets:

\begin{itemize}
\item {\bf Baseline}: In our Baseline dataset we consider the string SNIa+BAO+$H(z)$+LSS+CMB. Notice that here we do not include the SH0ES data.
\item {\bf Baseline+SH0ES}: The Baseline dataset is in this case complemented with the apparent magnitudes of the SNIa in the host galaxies and their distance moduli employed by SH0ES.
\item {\bf Baseline (No pol.)}: The same as in the Baseline case, but now removing the high-$\ell$ polarization data from the CMB likelihood.  That is to say, we have replaced the ``CMB'' dataset with ``CMB (No pol.)''.
\item {\bf Baseline (No pol.)+SH0ES}: The same as in ``Baseline (No pol.)'', but including also the data from SH0ES.
\end{itemize}
These are the four datasets that we employ to constrain our models. We present our fitting results in Tables \ref{tab:Baseline_results}-\ref{tab:BaselineNoPolSH0ES_results}, Figs. \ref{fig:base}-\ref{fig:base_nopol_SH0ES}, and also in the Tables \ref{tab:Baseline_chisquare}-\ref{tab:BaselineNoPolSH0ES_chisquare} of Appendix \ref{sec:appendix}.

\begin{table*}[t!]
\renewcommand{\arraystretch}{0.85}
\begin{center}
\resizebox{1\textwidth}{!}{
\begin{tabular}{|c  |c | c |  c | c | c  |c  |}
 \multicolumn{1}{c}{} & \multicolumn{4}{c}{Baseline}
\\\hline
{\scriptsize Parameter} & {\scriptsize $\Lambda$CDM}  & {\scriptsize type-I RRVM} & {\scriptsize type-I RRVM$_{\rm thr.}$}  &  {\scriptsize type-II RRVM} &  {\scriptsize XCDM}
\\\hline \hline
{\scriptsize $H_0$(km/s/Mpc)}  & {\scriptsize $68.27\pm 0.35$} & {\scriptsize  $68.22\pm 0.47$} & {\scriptsize  $67.65\pm 0.38$}  & {\scriptsize  $68.12\pm 0.97$} & {\scriptsize  $67.49\pm 0.56$}
\\\hline
{\scriptsize$\omega_{\rm b}$} & {\scriptsize $0.02251\pm 0.00013$}  & {\scriptsize  $0.02253\pm 0.00015$} & {\scriptsize  $0.02252\pm 0.00013$}  &  {\scriptsize  $0.02247\pm 0.00020$} & {\scriptsize  $0.02258\pm 0.00013$}
\\\hline
{\scriptsize$\omega_{\textrm{cdm}}$} & {\scriptsize $0.11803\pm 0.00078$}  & {\scriptsize $0.11807\pm 0.00078$} & {\scriptsize $0.1248\pm 0.0019$}  &  {\scriptsize  $0.1181\pm 0.0011$} &  {\scriptsize $0.11712\pm 0.00094$}
\\\hline
{\scriptsize$\Omega_{\rm m}^0$} & {\scriptsize $0.3029\pm 0.0045$}  & {\scriptsize  $0.3036\pm 0.0056$} & {\scriptsize $0.3235\pm 0.0071$ }  &  {\scriptsize $0.3032\pm 0.0089$} &  {\scriptsize $0.3082\pm 0.0055$}
\\\hline
{\scriptsize$w_0$} & {\scriptsize $-1$}  & {\scriptsize $-1$} & {\scriptsize $-1$}  &  {\scriptsize  $-1$} &  {\scriptsize $-0.962 \pm 0.022$}
\\\hline
{\scriptsize$\nu_{\rm eff}$} & {-}  & {\scriptsize  $0.00006\pm 0.00030$} & {\scriptsize $0.0227\pm 0.0055$ }  &  {\scriptsize  $-0.00008\pm 0.00035$} & {-}
\\\hline
{\scriptsize$\varphi_{\rm ini}$} & {-}  & {-} & {-}  &  {\scriptsize  $1.006\pm 0.024$} & {-}
\\\hline
{\scriptsize$\varphi (0)$} & {-}  & {-} & {-}  &  {\scriptsize  $1.008\pm 0.028$} & {-}
\\\hline
{\scriptsize$\tau_{\rm reio}$} & {{\scriptsize$0.0512\pm 0.0073$}} & {{\scriptsize $0.0511\pm 0.0080$}} & {{\scriptsize $0.0601\pm 0.0082$}}  &   {{\scriptsize $0.0505\pm 0.0078$}} & {\scriptsize $0.0546\pm 0.0077$}
\\\hline
{\scriptsize$\ln \left(10^{10} {\rm A}_{\rm s}\right)$} & {{\scriptsize $3.033\pm 0.015$}}  & {{\scriptsize $3.032\pm 0.016$}} & {{\scriptsize $3.053\pm 0.017$}} &   {{\scriptsize $3.031\pm 0.016$}} & {{\scriptsize $3.038\pm 0.016$}}
\\\hline
{\scriptsize$n_{\rm s}$} & {{\scriptsize $0.9698\pm 0.0035$}}  & {{\scriptsize $0.9701\pm 0.0038$}} & {{\scriptsize $0.9707\pm 0.0035$}} &   {{\scriptsize $0.9681\pm0.0069$}} & {\scriptsize $0.9722\pm 0.0038$}
\\\hline
{\scriptsize$M$} & {{\scriptsize $-19.415\pm 0.010$}}  & {{\scriptsize $-19.416\pm 0.014$}} & {{\scriptsize $-19.429\pm 0.011$}} &   {{\scriptsize $-19.420\pm 0.030$}} & {\scriptsize $-19.432\pm 0.014$}
\\\hline
{\scriptsize$\sigma_8$}  & {{\scriptsize$0.8003\pm 0.0064$}}  & {{\scriptsize $0.799\pm 0.011$}} & {{\scriptsize  $0.7733\pm 0.0092$}}  &   {{\scriptsize $0.801\pm 0.010$}} & {\scriptsize $0.7885\pm 0.0093$}
\\\hline
{\scriptsize$S_8$}  & {{\scriptsize$0.804\pm 0.010$}}  & {{\scriptsize $0.803\pm 0.011$}} & {{\scriptsize $0.803\pm 0.010$}}  &   {{\scriptsize $0.805\pm 0.015$}} & {\scriptsize $0.802\pm 0.011$}
\\\hline
{\scriptsize$r_{\rm d}$ (Mpc)}  & {{\scriptsize$147.46\pm 0.21$}}  & {{\scriptsize $147.47\pm 0.25$}} & {{\scriptsize $147.44\pm 0.21$}}  &   {{\scriptsize $147.9\pm 1.9$}} & {\scriptsize $147.62\pm 0.23$}
\\\hline
%{\scriptsize$\chi^2_{\rm min}$}  & {{\scriptsize $4241.01$}}  & {{\scriptsize $4239.26$}} & {{\scriptsize  $4224.12$}}  &   {{\scriptsize  $4240.27$}} & {-}
%\\\hline
{\scriptsize ${\rm \Delta DIC}$}  & {-}  & {{\scriptsize  -2.04}} & {{\scriptsize  +15.34}}  &   {{\scriptsize  -4.18}} & {\scriptsize $+1.74$}
\\\hline
\end{tabular}}
\end{center}
\caption{\scriptsize Mean values with $68\%$ confidence intervals obtained from our fitting analysis of our Baseline dataset, composed by the string SNIa+BAO+$H(z)$+LSS+CMB. We display the values of the different cosmological parameters: the Hubble parameter ($H_0$), the reduced baryon and CDM density parameters ($\omega_{\rm b} \equiv \Omega_{\rm b}^0 h^2$ and $\omega_{\textrm{cdm}}\equiv \Omega_{\textrm{cdm}}^0 h^2$, respectively, with $\Omega_i^0\equiv 8\pi G_N\rho_i^0/3H_0^2$), the current nonrelativistic matter density parameter  ($\Omega_{\rm m}^0$), the equation of state of the vacuum/DE fluid ($w_0$), the effective parameter of the running vacuum ($\nueff$) (see \eqref{eq:nueff_definition_typeI} and \eqref{eq:nueff_definitoin_typeII}), the initial and current values of the variable $\varphi\equiv G_N /G$, the optical depth to reionization ($\tau_{\rm reio}$), the amplitude and spectral index of primordial power spectrum ($A_s$ and $n_s$, respectively), the absolute magnitude of SNIa ($M$), the rms mass fluctuations at $8h^{-1}$ Mpc scale at present time ($\sigma_8$), the derived parameter $S_8 \equiv \sigma_8\sqrt{\Omega_{\rm m}^0/0.3}$ and the comoving sound horizon at the drag epoch ($r_d$). We also show the incremental value of DIC with respect the $\Lambda$CDM, denoted $\Delta$DIC.}
\label{tab:Baseline_results}
\end{table*}

In order to study the performance of the various models when they are confronted with the wealth of cosmological data, we define the joint $\chi^2$-function as follows,
\begin{equation}\label{eq:chi2_total}
\chi^2_{\textrm{tot}}= \chi^2_{\textrm{SNIa}} +\chi^2_{\textrm{BAO}}+\chi^2_{H}+\chi^2_{\textrm{LSS}}+\chi^2_{\textrm{CMB}}\,,
\end{equation}
where $\chi^2_{\rm CMB}$ and $\chi^2_{\rm SNIa}$ may include or not the contribution of the high-$\ell$ CMB polarization and SH0ES data, respectively, depending on the dataset that we consider.

%The various contributions to $\chi^2_{\textrm{tot}}$ are defined in the standard way, and this includes of course a nontrivial  covariance matrix if it is known.  Notice also that the term $\chi^2_H$ may contain or not the $H_0$-prior from \cite{Riess:2021jrx} integrated in the likelihood of the $H(z)$ data, depending on the dataset used for our analysis.

To solve the background and perturbation equations of the type-I RRVM, type-I RRVM$_{\textrm{thr.}}$ and type-II RRVM we make use of our own modified versions of the Einstein-Boltzmann system solver \texttt{CLASS} \cite{Lesgourgues:2011re,Blas:2011rf}, which is now equipped with the additional features that we have briefly described in the previous sections.  We explore and put constraints on the parameter spaces of our models with Markov chain Monte Carlo (MCMC) analyses. More specifically, we make use of the Metropolis-Hastings algorithm \cite{1953JChPh..21.1087M,Hastings:1970aa}, which is already implemented in the Monte Carlo sampler \texttt{MontePython} \footnote{\url{https://baudren.github.io/montepython.html}}\cite{Audren:2012wb,Brinckmann:2018cvx}. We stop the MCMC when the Gelman-Rubin convergence statistic is $R-1<0.02$ \cite{R1:1997,R2:1992}, and analyze the converged chains with the Python code \texttt{GetDist}\footnote{\url{https://getdist.readthedocs.io/en/latest/}}\cite{Lewis:2019xzd} to compute the mean values of the cosmological parameters, their confidence intervals and the posterior distributions.

We have set conservative flat priors for the input parameters in the MCMC, much wider than their marginalized posterior distributions. For the six primary cosmological parameters that are common in all the models, we use: $0.005 < \omega_{\rm b} < 0.1$, $0.001 < \omega_{\rm cdm} < 0.99$, $ 20 < H_0\, [{\rm km/s/Mpc}] < 100$, $1.61 < \ln(10^{10}A_s) < 3.91$, $0.8 < n_s < 1.2$, and $0.01 < \tau_{\textrm{reio}} < 0.8$. The type-I RRVM and type-I RRVM$_{\textrm{thr.}}$ have one additional degree of freedom ({\it d.o.f.}) compared to the $\Lambda$CDM, which is encoded in the parameter $\nu$. We use the flat prior $-0.5 < \nu_{\textrm{eff}} < 0.5$. On the other hand, the type-II RRVM is characterized by two extra parameters, $\nu_{\textrm{eff}}$ and the initial value of $\varphi$, for which we use the priors $-\frac{1}{6} < \nu_{\textrm{eff}} < \frac{1}{6}$ and $0.7 < \varphi_{\textrm{ini}} < 1.3$. Finally, for the constant dark energy EoS parameter of the XCDM model we employ the prior $-3 < w_0 < 0.2$.  In all our analyses we set the current temperature of the CMB to $T_0=2.7255$ K \cite{Fixsen:2009ug}, and consider three neutrinos species, approximated as two massless states and a massive neutrino of mass $m_\nu=0.06$ eV.

%The aforementioned datasets are to be used for constraining, or fitting up to a certain confidence level,  the value of the new free parameter $\nueff$ (along with the six standard free parameters of the vanilla model) for the type-I RRVM model, as well as the initial value parameter $\varphi_{\rm in}$ (defined in the previous section) for the type-II model, see Table XX.

To compare the fitting performance of the various models under study from a Bayesian perspective, we utilize the deviance information criterion (DIC) \cite{DIC}, which takes into account the presence of extra {\it d.o.f\,} by duly penalizing the inclusion of additional parameters in the model. See e.g. the review \cite{Liddle:2007fy} for a summarized discussion on how to use and interpret the information criteria in the cosmological context.
The DIC value can be computed through the following recipe:
\begin{equation}\label{eq:DIC}
\textrm{DIC} = \chi^2(\bar{\theta}) + 2p_D\,.
\end{equation}
In it,   $p_D = \overline{\chi^2} - \chi^2(\bar{\theta})$ represents the effective number of parameters and   $2p_D$ is the so-called `model complexity'. The latter is the quantity employed in the DIC criterion to  penalize the presence of extra {\it d.o.f}. The term $\overline{\chi^2}$ is the mean value of the $\chi^2$-function, which is  obtained from the Markov chains. In this sense the computation of the  DIC is a more sophisticated procedure of model comparison than other information criteria as the Akaike information criterion \cite{Akaike}. Finally, $\bar{\theta}$ in \eqref{eq:DIC} represents the mean value of the fitting parameters.

\begin{table*}[t!]
\renewcommand{\arraystretch}{0.85}
\begin{center}
\resizebox{1\textwidth}{!}{
\begin{tabular}{|c  |c | c |  c | c | c  |c  |}
 \multicolumn{1}{c}{} & \multicolumn{4}{c}{Baseline +SH0ES}
\\\hline
{\scriptsize Parameter} & {\scriptsize $\Lambda$CDM}  & {\scriptsize type-I RRVM} & {\scriptsize type-I RRVM$_{\rm thr.}$}  &  {\scriptsize type-II RRVM} &  {\scriptsize XCDM}
\\\hline \hline
{\scriptsize $H_0$(km/s/Mpc)}  & {\scriptsize $ 68.82\pm 0.33 $} & {\scriptsize  $69.17\pm 0.43 $} & {\scriptsize  $68.33\pm 0.35 $}  & {\scriptsize  $70.79\pm 0.69 $} & {\scriptsize  $68.67\pm 0.50 $}
\\\hline
{\scriptsize$\omega_{\rm b}$} & {\scriptsize $ 0.02264\pm 0.00013$}  & {\scriptsize  $0.02253\pm 0.00015 $} & {\scriptsize  $0.02266\pm 0.00013 $}  &  {\scriptsize  $ 0.02281\pm 0.00017 $} & {\scriptsize  $0.02265\pm 0.00013 $}
\\\hline
{\scriptsize$\omega_{\textrm{cdm}}$} & {\scriptsize $ 0.11697\pm 0.00073$}  & {\scriptsize $ 0.11685\pm 0.00075$} & {\scriptsize $ 0.01227\pm 0.0018 $}  &  {\scriptsize  $ 0.1178\pm 0.0011$} &  {\scriptsize  $ 0.11679\pm 0.00089$}
\\\hline
{\scriptsize$\Omega_{\rm m}^0$} & {\scriptsize $0.2961\pm 0.0041 $}  & {\scriptsize  $ 0.2928\pm 0.0049$} & {\scriptsize $0.3128\pm 0.0064 $ }  &  {\scriptsize $0.2808\pm 0.0058 $} &  {\scriptsize $0.2971\pm 0.0047 $}
\\\hline
{\scriptsize$w_0$} & {\scriptsize $-1$}  & {\scriptsize $-1$} & {\scriptsize $-1$}  &  {\scriptsize  $-1$} &  {\scriptsize $-0.993 \pm 0.020$}
\\\hline
{\scriptsize$\nu_{\rm eff}$} & {-}  & {\scriptsize  $-0.00037\pm 0.00029 $} & {\scriptsize $0.0197\pm 0.0055 $ }  &  {\scriptsize  $-0.00003\pm 0.00033$} & {-}
\\\hline
{\scriptsize$\varphi_{\rm ini}$} & {-}  & {-} & {-}  &  {\scriptsize  $ 0.949\pm 0.016 $} & {-}
\\\hline
{\scriptsize$\varphi (0)$} & {-}  & {-} & {-}  &  {\scriptsize  $ 0.950\pm 0.021$} & {-}
\\\hline
{\scriptsize$\tau_{\rm reio}$} & {{\scriptsize$0.0533\pm 0.0074$}} & {{\scriptsize $0.0501\pm 0.0079 $}} & {{\scriptsize $ 0.0617^{+0.0081}_{-0.0095}$}}  &   {{\scriptsize $0.0523\pm 0.0077 $}} & {\scriptsize $ 0.0539\pm 0.0078$}
\\\hline
{\scriptsize$\ln \left(10^{10} {\rm A}_{\rm s}\right)$} & {{\scriptsize $3.035\pm 0.015$}}  & {{\scriptsize $ 3.031\pm 0.016$}} & {{\scriptsize $ 3.053^{+0.017}_{-0.019}$}} &   {{\scriptsize $ 3.041\pm 0.016 $}} & {\scriptsize $ 3.036\pm 0.016 $}
\\\hline
{\scriptsize$n_{\rm s}$} & {{\scriptsize $0.9726\pm 0.0035$}}  & {{\scriptsize $ 0.9705\pm 0.0037$}} & {{\scriptsize $0.9736\pm 0.0034 $}} &   {{\scriptsize $0.9824 \pm 0.0058 $}} & {\scriptsize $0.9730 \pm 0.0037$}
\\\hline
{\scriptsize$M$} & {{\scriptsize $-19.3989\pm 0.0096$}}  & {{\scriptsize $-19.390\pm 0.012 $}} & {{\scriptsize $-19.410\pm 0.010 $}} &   {{\scriptsize $-19.339\pm 0.021 $}} & {\scriptsize $-19.402\pm 0.012 $}
\\\hline
{\scriptsize$\sigma_8$}  & {{\scriptsize$ 0.7978\pm 0.0064$}}  & {{\scriptsize $0.808\pm 0.011 $}} & {{\scriptsize  $0.7747\pm 0.0093 $}}  &   {{\scriptsize $ 0.807\pm 0.010$}} & {\scriptsize $ 0.7955\pm 0.0089$}
\\\hline
{\scriptsize$S_8$} & {{\scriptsize$0.7927\pm 0.0094$}}  & {{\scriptsize $0.799\pm 0.011 $}} & {{\scriptsize $ 0.7910\pm 0.0098$}}  &   {{\scriptsize $ 0.781\pm 0.013$}} & {\scriptsize $0.801\pm 0.010$}
\\\hline
{\scriptsize$r_{\rm d}$ (Mpc)}  & {{\scriptsize$147.59\pm 0.21$}}  & {{\scriptsize $147.44\pm 0.25 $}} & {{\scriptsize $147.60\pm 0.21 $}}  &   {{\scriptsize $ 143.3\pm 1.4$}} & {\scriptsize $ 147.63\pm 0.23$}
\\\hline
%{\scriptsize$\chi^2_{\rm min}$}  & {{\scriptsize $4325.20  $}}  & {{\scriptsize $ 4324.06$}} & {{\scriptsize  $ 4312.48 $}}  &   {{\scriptsize  $ 4315.36 $}} & {-}
%\\\hline
{\scriptsize$\Delta{\rm DIC}$}  & {-}  & {{\scriptsize $-0.64$ }} & {{\scriptsize $+10.94$ }}  &   {{\scriptsize  $+6.58$}} & {\scriptsize$-1.92$}
\\\hline
\end{tabular}}
\end{center}
\caption{\scriptsize Same as in Table\,\ref{tab:Baseline_results}, but adding the information from SH0ES to our Baseline dataset.}
\label{tab:BaselineSH0ES_results}
\end{table*}

Given a model $X$, we define the DIC difference with respect to the vanilla model (or concordance $\Lambda$CDM) in a way such that a positive difference of DIC implies that the new model ($X$) fares better than the vanilla model (and hence that $X$ provides smaller values of DIC than the concordance model), whereas negative differences mean just the opposite, that is, that model $X$  fares worse than the vanilla model. Therefore, the appropriate definition is
\begin{equation}
\Delta\textrm{DIC} \equiv \textrm{DIC}_{\CC\textrm{CDM}} - \textrm{DIC}_{\textrm{X}}\label{eq:differences_DIC}\,.
\end{equation}
In our case, X represents either the type-I or type-II running vacuum models in their RRVM implementation; and also the XCDM, which, as indicated before, is used as a benchmark scenario for dynamical DE. In the usual argot of the information criteria,  values $0 \leq \Delta\textrm{DIC}<2$ are said to entail {\it weak} evidence  in favor of the considered option beyond the standard model.  However, if  $2 \leq \Delta\textrm{DIC} < 6$ one then speaks of {\it positive} evidence, whilst if $6 \leq \Delta\textrm{DIC} < 10$ it is considered that there is {\it strong} evidence in favor of the non-standard model $X$. Finally, if it turns out that $\Delta\textrm{DIC}>10$  one may licitly  claim (according to the rules of these information criteria) that there is {\it very strong} evidence supporting the model under study as compared to the vanilla cosmology. In contrast,  if the statistical parameter \eqref{eq:differences_DIC} proves negative, it is an unmistakable sign that the vanilla cosmology is favored over model $X$ by the observational data.

\section{Discussion of the results}\label{sec:Discussion_of_the_results}
The class of running vacuum models (RVMs) has proven to be theoretically sound and thus worth being studied phenomenologically.  It emerges as a generic framework out of renormalizable QFT in curved spacetime; in fact, one which is capable of describing the expansion history of the universe from the early times to our days from first principles\cite{Sola:2013gha,SolaPeracaula:2022hpd,Sola:2015rra}. If we take the quantum vacuum seriously, the RVM framework is a natural consequence of it.  The predicted changes are not dramatic, but can be crucial to fit the pieces together. Indeed, the  phenomenological expectations from the running vacuum approach on the cosmological observables remain always very close to the $\CC$CDM, as can be seen from the fitting results displayed in Tables \ref{tab:Baseline_results}-\ref{tab:BaselineNoPolSH0ES_results}.  Nevertheless,  small departures are definitely predicted owing to the presence of vacuum fluctuations from the quantized matter fields in the FLRW background.  These vacuum effects must be properly renormalized in the QFT context, and as a result they bring about small ``radiative corrections'' on top of  the standard $\CC$CDM predictions -- recall their generic form in Eq.\,\eqref{eq:RVMvacuumdadensity}. They have been computed in detail in Refs.\,\cite{Moreno-Pulido:2020anb,Moreno-Pulido:2022phq,Moreno-Pulido:2022upl,Moreno-Pulido:2023ryo} and can help to fix the phenomenological hitches currently besetting the standard model of cosmology, which is strictly based on  GR and no quantum effects at all. The generic RVM contains a few free parameters amenable to fitting from the cosmological data, but the formal structure of the quantum effects is unambiguous and well defined. In fact, the quantum corrections at low energy appear to be proportional to $\sim H^2$ and $\sim \dot{H}$, as shown in Eq.\,\eqref{eq:RVMvacuumdadensity}. These corrections induce a dynamics in the physical value of the VED and the corresponding physical value of the cosmological term, $\CC$. In other words, in the RVM these quantities acquire a cosmological evolution rather than remaining strictly constant as in the $\CC$CDM. This fact  may have phenomenological consequences worth studying. In the present work, we have dwelled upon particular realizations of the RVM exhibiting a rich phenomenology and we have studied the conditions by which  they may offer a helping hand to curb one or both tensions ($\sigma_8$ and $H_0$) under study.

\begin{figure}[t!]
\begin{center}
 \includegraphics[width=6.5in, height=3.5in]{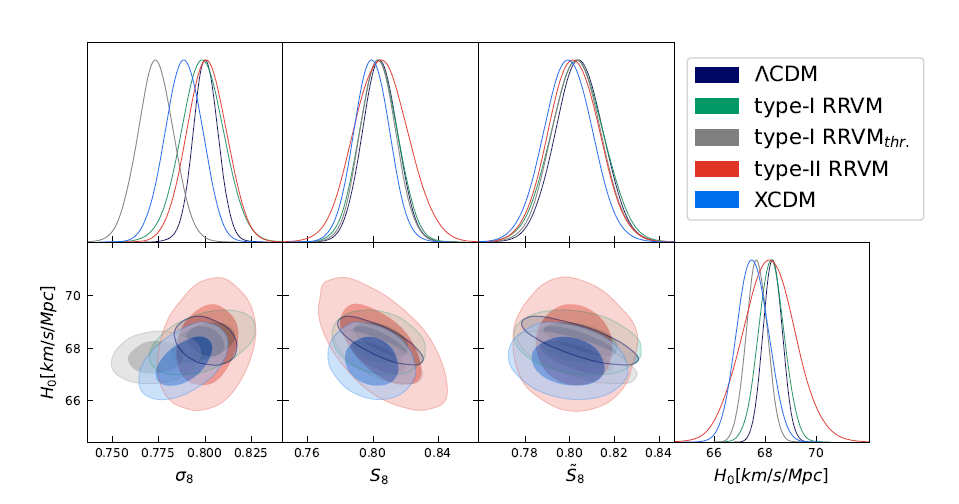}
 \caption{Contour plots at $1\sigma$ and $2\sigma$ c.l. in the $\sigma_8-H_0$, $S_8-H_0$
 and $\tilde{S}_8-H_0$ planes and their corresponding one-dimensional posteriors, obtained from the fit of the various models to the Baseline dataset (cf. Sec. \ref{sec:Data_and_methodology}). The parameter $\tilde{S}_8\equiv S_8/\sqrt{\varphi(0)}$ can only differ from the standard $S_8$ in the type-II RRVM, see the main text of Sec. \ref{sec:Discussion_of_the_results} and Refs. \cite{SolaPeracaula:2020vpg,SolaPeracaula:2021gxi}. The type-I RRVM$_{\rm thr.}$ can explain a value of $\sigma_8\sim 0.78$, much smaller than in the other models. This is accompanied by a $4.1\sigma$ evidence for a non-zero value of the RVM parameter $\nu_{\rm eff}$, see Table \ref{tab:Baseline_results}. We find in all cases similar values of $\tilde{S}_8$ and $H_0$ to those found in $\Lambda$CDM, but the type-II RRVM has a much wider posterior for this parameter, and hence this model can accommodate a larger Hubble constant. See also the comments in the main text.}\label{fig:base}
\end{center}
\end{figure}

\begin{figure}[t!]
\begin{center}
 \includegraphics[width=6.5in, height=3.5in]{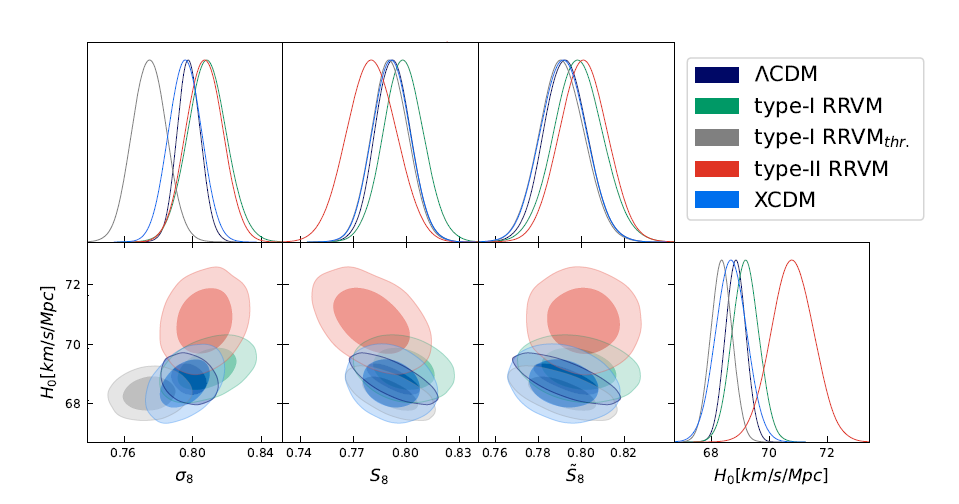}
 \caption{Same as in Fig. \ref{fig:base}, but using the Baseline+SH0ES dataset (cf. Sec. \ref{sec:Data_and_methodology}). The inclusion of the data from SH0ES shifts the one-dimensional posterior of $H_0$ towards $H_0\sim 71$ km/s/Mpc in the type-II RRVM, a region that is still allowed by the Baseline dataset, cf. Fig. \ref{fig:base}. Remarkably, the small values of $\sigma_8$ found in the type-I RRVM$_{\rm thr.}$ remain stable, and no important differences between the models are found regarding the value of $\tilde{S}_8$. The lower value of $S_8$ obtained in the type-II RRVM is due to the fact that this parameter does not account for the $2.4\sigma$ departure of $\varphi(0)$ from 1, $\varphi(0)=0.950\pm 0.021$, see the caption of Table \ref{tab:Baseline_results}.}\label{fig:base_SH0ES}
\end{center}
\end{figure}

\begin{figure}[t!]
\begin{center}
 \includegraphics[width=6.5in, height=3.5in]{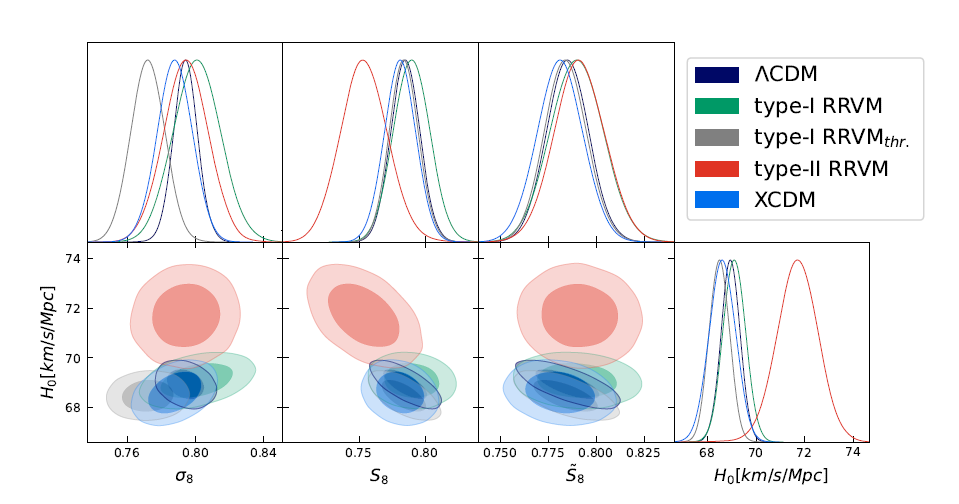}
 \caption{Same as in Fig. \ref{fig:base_SH0ES}, but removing the high-$\ell$ polarization data from Planck, i.e. considering the Baseline (No pol.)+SH0ES dataset (cf. Sec. \ref{sec:Data_and_methodology}). Again, as in the other fitting analyses, the value of $\sigma_8$ is kept small in the type-I RRVM$_{\rm thr.}$. The absence of CMB polarization data allows for even smaller values of $\varphi(0)$ in the type-II RRVM, $\varphi(0)=0.908^{+0.025}_{-0.028}$, which is now $3.5\sigma$ below the GR value $\varphi(0)=1$. This explains the large value of $H_0\sim 72$ km/s/Mpc, which basically renders the Hubble tension insignificant, below the $1\sigma$ c.l. }\label{fig:base_nopol_SH0ES}
\end{center}
\end{figure}

\begin{figure}[t!]
\begin{center}
 \includegraphics[width=7.0in, height=2.3in]{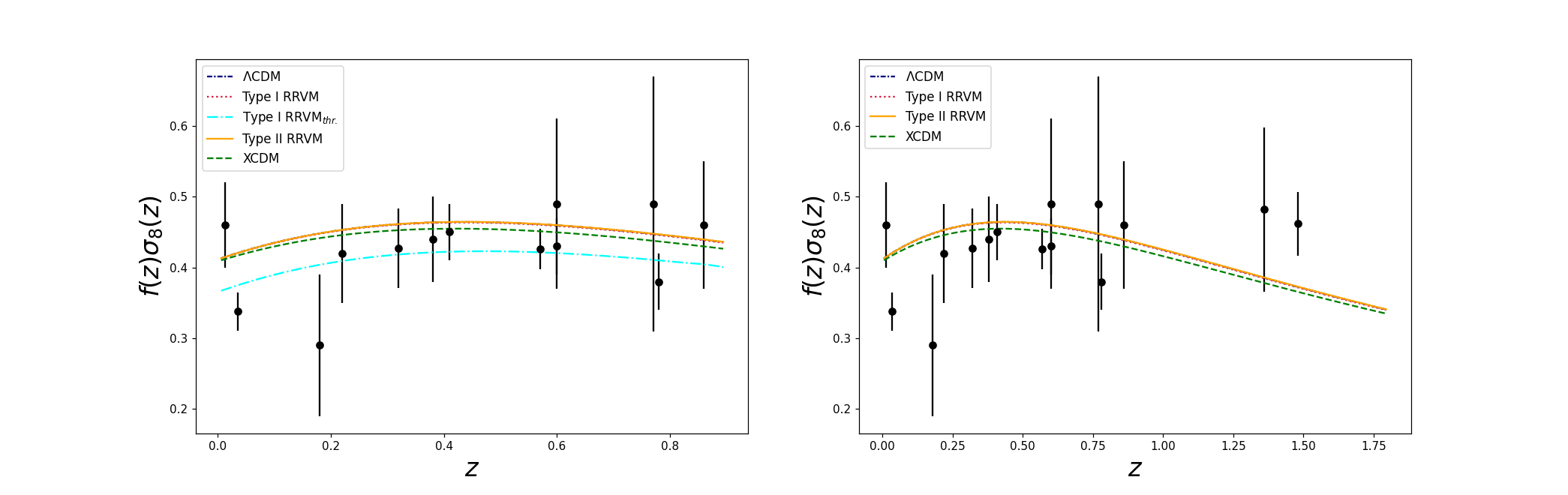}
 \caption{Theoretical curves of $f(z)\sigma_8(z)$ for the various models together with the observational data points listed in Table \ref{tab:fs8_table}. We have employed the central values of the Baseline fitting analysis (cf. Table \ref{tab:Baseline_results}). The type-I RRVM$_{\rm thr}$ has the ability to solve the $\sigma_8$ tension by suppressing the clustering at  $z<1$.}\label{fig:fsigma8_baseline}
\end{center}
\end{figure}

In this section, we discuss in detail the results we have obtained for particular RVM realizations, which in all cases are sourced by the same formal QFT structure indicated in Eq.\,\eqref{eq:RVMvacuumdadensity},    and compare them with those obtained with the $\Lambda$CDM and the popular XCDM parameterization of the dark energy EoS parameter\cite{Turner:1998ex}. Above all, we should remark at this point that the results obtained here are fully consistent with those reported in our last study confronting the RVMs against the overall cosmological observations\cite{SolaPeracaula:2021gxi}. In the present instance, however, we have  updated our datasets  and have extended significantly the reach of our considerations by displaying a much more comprehensive numerical study, see Tables \ref{tab:Baseline_results}-\ref{tab:BaselineNoPolSH0ES_results} and  Figs. \ref{fig:base}-\ref{fig:fsigma8_baseline}.  Most significantly, the current presentation includes for the first time the effect of the CMB polarization data from Planck.  In fact, we recall that the companion analysis of \cite{SolaPeracaula:2021gxi} focused exclusively on  the Planck 2018 TT+lowE data, and  hence without being sensitive to the influence from the high-$\ell$ polarizations.  In contrast, in the current study we use the two full likelihoods from Planck, namely  Planck 2018 TT+lowE and  Planck 2018  TT,TE,EE+lowE (cf. Sec.\ref{sec:Data_and_methodology}) and compare their distinct impact on the fitting results of each of the  RVM realizations under focus, viz. the type-I and type-II implementations.

\begin{table*}[t!]
\renewcommand{\arraystretch}{0.85}
\begin{center}
\resizebox{1\textwidth}{!}{
\begin{tabular}{|c  |c | c |  c | c | c  |c  |}
 \multicolumn{1}{c}{} & \multicolumn{4}{c}{Baseline (No pol.)}
\\\hline
{\scriptsize Parameter} & {\scriptsize $\Lambda$CDM}  & {\scriptsize type-I RRVM} & {\scriptsize type-I RRVM$_{\rm thr.}$}  &  {\scriptsize type-II RRVM} &  {\scriptsize XCDM}
\\\hline \hline
{\scriptsize $H_0$(km/s/Mpc)}  & {\scriptsize $68.29\pm 0.38 $} & {\scriptsize  $68.10\pm 0.48$} & {\scriptsize  $67.66\pm 0.41$}  & {\scriptsize  $68.8\pm 1.2$} & {\scriptsize  $67.31\pm 0.56$}
\\\hline
{\scriptsize$\omega_{\rm b}$} & {\scriptsize $0.02228\pm 0.00019 $}  & {\scriptsize  $ 0.02235\pm 0.00022$} & {\scriptsize  $0.02231\pm 0.00019$}  &  {\scriptsize  $0.02242\pm 0.00026$} & {\scriptsize  $0.02239\pm 0.00020$}
\\\hline
{\scriptsize$\omega_{\textrm{cdm}}$} & {\scriptsize $0.11746\pm 0.00085 $}  & {\scriptsize $0.11744\pm 0.00086$} & {\scriptsize $0.1242\pm 0.0019$}  &  {\scriptsize  $0.1166\pm 0.0016$} &  {\scriptsize  $0.1160\pm 0.0011$}
\\\hline
{\scriptsize$\Omega_{\rm m}^0$} & {\scriptsize $0.3011\pm 0.0048 $}  & {\scriptsize  $0.3029\pm 0.0056 $} & {\scriptsize $0.3215\pm 0.0072$ }  &  {\scriptsize $0.294\pm 0.011$} &  {\scriptsize $0.3068\pm 0.0055$}
\\\hline
{\scriptsize$w_0$} & {\scriptsize $-1$}  & {\scriptsize $-1$} & {\scriptsize $-1$}  &  {\scriptsize  $-1$} &  {\scriptsize $-0.948 \pm 0.022$}
\\\hline
{\scriptsize$\nu_{\rm eff}$} & {-}  & {\scriptsize  $0.00025\pm 0.00038$} & {\scriptsize $0.0223\pm 0.0056$ }  &  {\scriptsize  $0.00028\pm 0.00043$} & {-}
\\\hline
{\scriptsize$\varphi_{\rm ini}$} & {-}  & {-} & {-}  &  {\scriptsize  $0.982\pm 0.030$} & {-}
\\\hline
{\scriptsize$\varphi (0)$} & {-}  & {-} & {-}  &  {\scriptsize  $0.976\pm 0.035$} & {-}
\\\hline
{\scriptsize$\tau_{\rm reio}$} & {{\scriptsize$0.0489^{+0.0084}_{-0.0076}$}} & {{\scriptsize $0.0508\pm 0.0083$}} & {{\scriptsize $0.0581\pm 0.0082$}}  &   {{\scriptsize $0.0508\pm 0.0083$}} & {\scriptsize $0.0540\pm 0.0080$}
\\\hline
{\scriptsize$\ln \left(10^{10} {\rm A}_{\rm s}\right)$} & {{\scriptsize $3.026^{+0.018}_{-0.016} $}}  & {{\scriptsize $3.028\pm 0.016 $}} & {{\scriptsize $3.047\pm 0.017$}} &   {{\scriptsize $3.030\pm 0.017$}} & {\scriptsize $3.034\pm 0.016$}
\\\hline
{\scriptsize$n_{\rm s}$} & {{\scriptsize $0.9695\pm 0.0037 $}}  & {{\scriptsize $0.9712\pm 0.0045$}} & {{\scriptsize $0.9703\pm 0.0037$}} &   {{\scriptsize $0.9754\pm 0.0087$}} & {{\scriptsize $0.9736\pm 0.0041$}}
\\\hline
{\scriptsize$M$} & {{\scriptsize $ -19.415\pm 0.0011$}}  & {{\scriptsize $-19.420\pm 0.014 $}} & {{\scriptsize $-19.429\pm 0.012$}} &   {{\scriptsize $-19.397\pm 0.038$}} & {\scriptsize $ -19.436\pm 0.0014$}
\\\hline
{\scriptsize$\sigma_8$}  & {{\scriptsize$0.7965\pm 0.0069 $}}  & {{\scriptsize $0.790\pm 0.013$}} & {{\scriptsize  $0.7710\pm 0.0094$}}  &   {{\scriptsize $0.792\pm 0.012$}} & {\scriptsize $0.7799\pm 0.0098$}
\\\hline
{\scriptsize$S_8$}  & {{\scriptsize$0.798\pm 0.011 $}}  & {{\scriptsize $0.793\pm 0.013$}} & {{\scriptsize $0.798\pm 0.011 $}}  &   {{\scriptsize $0.783\pm 0.020$}} & {\scriptsize $0.793\pm 0.012$}
\\\hline
{\scriptsize$r_{\rm d}$ (Mpc)}  & {{\scriptsize$147.86\pm 0.30 $}}  & {{\scriptsize $148.00\pm 0.35$}} & {{\scriptsize $147.81\pm 0.30$}}  &   {{\scriptsize $146.5\pm 2.4$}} & {\scriptsize $148.15\pm 0.33$}
\\\hline
%{\scriptsize$\chi^2_{\rm min}$}  & {{\scriptsize $2650.36 $}}  & {{\scriptsize $2649.80 $}} & {{\scriptsize  $ 2634.45$}}  &   {{\scriptsize  $2650.26 $}} & {-}
%\\\hline
{\scriptsize$\Delta{\rm DIC}$}  & {-}  & {{\scriptsize $-1.84$ }} & {{\scriptsize $+14.54$ }}  &   {{\scriptsize  $-3.06$}} & {\scriptsize $+3.82$}
\\\hline
\end{tabular}}
\end{center}
\caption{\scriptsize Same as in Table\,\ref{tab:Baseline_results}, but without including the high-$\ell$ CMB polarization data from Planck in our combined dataset.}
\label{tab:BaselineNoPol_results}
\end{table*}

Let us start with the results obtained with the RRVM of type I, first under the assumption  that the vacuum interacts with dark matter during the entire cosmic history. The CMB data from Planck put very tight constraints on the amount of dark energy at the decoupling time (see e.g. \cite{Gomez-Valent:2021cbe}) and, therefore, on the RVM parameter that controls the exchange of energy in the dark sector, $\nueff$. We obtain central values of $\nu_{\rm eff}\sim \mathcal{O}(10^{-4})$ with all our datasets, with associated error bars that make our measurements compatible with 0 at $\lesssim 1\sigma$ c.l., indicating no statistical preference for a non-null vacuum dynamics in the universe in the context of this model. Although the type-I RRVM is fundamentally different from the $\Lambda$CDM, its phenomenology is in practice quite similar, due to the strong upper bounds on $|\nu_{\rm eff}|$. This explains why the constraints obtained on the other cosmological parameters are so similar in the two models, and also the small impact the type-I RRVM has on the cosmological tensions.  This is the conclusion that follows if we assume that the cosmological solution that we have found for the type-I models is valid all the way from the present time up to the point in the radiation-dominated epoch where we have placed our initial conditions following the standard setup of \texttt{CLASS} (see, however, below). We refer the reader to Tables \ref{tab:Baseline_results}-\ref{tab:BaselineNoPolSH0ES_results} for the detailed list of fitting results. In particular, we would like to mention that the results quoted in the last two tables (namely Tables \ref{tab:BaselineNoPol_results} and \ref{tab:BaselineNoPolSH0ES_results}, where the CMB data are used without polarizations) are perfectly compatible within error bars (both in order of magnitude and sign)  with the results obtained in our previous analysis\,\cite{SolaPeracaula:2021gxi}.

Nevertheless, we cannot exclude the possibility that the vacuum dynamics undergoes a transition between two (or more) epochs of the expansion history, e.g. through a change in the value of $\nu_{\rm eff}$ or of the effective EoS parameter $\wv$. As previously noted, the possibility of a tomographic behaviour of the DE throughout the cosmic expansion has been explored previously in the literature, see e.g. \cite{Salvatelli:2014zta,Martinelli:2019dau,Hogg:2020rdp,Goh:2022gxo}. In the RVM case, we are further motivated to think of  a scenario of this sort since it is actually suggested within the context of the QFT calculation supporting the RVM structure, see \cite{Moreno-Pulido:2022upl}.

\begin{table*}[t!]
\renewcommand{\arraystretch}{0.85}
\begin{center}
\resizebox{1\textwidth}{!}{
\begin{tabular}{|c  |c | c |  c | c | c  |c  |}
 \multicolumn{1}{c}{} & \multicolumn{4}{c}{Baseline (No pol.) +SH0ES}
\\\hline
{\scriptsize Parameter} & {\scriptsize $\Lambda$CDM}  & {\scriptsize type-I RRVM} & {\scriptsize type-I RRVM$_{\rm thr.}$}  &  {\scriptsize type-II RRVM} &  {\scriptsize XCDM}
\\\hline\hline
{\scriptsize $H_0$(km/s/Mpc)}  & {\scriptsize $68.94\pm 0.37$} & {\scriptsize  $69.10\pm 0.44$} & {\scriptsize  $68.48\pm 0.39$}  & {\scriptsize  $71.69\pm 0.80$} & {\scriptsize  $68.61\pm 0.51$}
\\\hline
{\scriptsize$\omega_{\rm b}$} & {\scriptsize $0.02247\pm 0.00018$}  & {\scriptsize  $0.02240\pm 0.00022$} & {\scriptsize  $0.02251\pm 0.00018$}  &  {\scriptsize  $0.02280\pm 0.00024$} & {\scriptsize  $0.02252\pm 0.00019$}
\\\hline
{\scriptsize$\omega_{\rm dm}$} & {\scriptsize $0.11630\pm 0.00083$}  & {\scriptsize $0.11632\pm 0.00083$} & {\scriptsize $0.1220\pm 0.0019$}  &  {\scriptsize  $0.1160\pm 0.0015$} &  {\scriptsize  $0.1157\pm 0.0010$}
\\\hline
{\scriptsize$\Omega_{\rm m}^0$} & {\scriptsize $0.2933\pm 0.0045$}  & {\scriptsize  $0.2919\pm 0.0062$} & {\scriptsize $0.3095\pm 0.0067$ }  &  {\scriptsize $0.2702\pm 0.0068$} &  {\scriptsize $0.2950\pm 0.0048$}
\\\hline
{\scriptsize$w_0$} & {\scriptsize $-1$}  & {\scriptsize $-1$} & {\scriptsize $-1$}  &  {\scriptsize  $-1$} &  {\scriptsize $-0.981 \pm 0.021$}
\\\hline
{\scriptsize$\nu_{\rm eff}$} & {-}  & {\scriptsize  $-0.00022\pm 0.00036$} & {\scriptsize $0.0193\pm 0.0055$ }  &  {\scriptsize  $0.00048\pm 0.00040$} & {-}
\\\hline
{\scriptsize$\varphi_{\rm ini}$} & {-}  & {-} & {-}  &  {\scriptsize  $0.919^{+0.019}_{-0.022}$} & {-}
\\\hline
{\scriptsize$\varphi (0)$} & {-}  & {-} & {-}  &  {\scriptsize  $0.908^{+0.025}_{-0.028}	$} & {-}
\\\hline
{\scriptsize$\tau_{\rm reio}$} & {{\scriptsize$ 0.0512\pm 0.0074$}} & {{\scriptsize $0.0494\pm 0.0084$}} & {{\scriptsize $0.0595^{+0.0082}_{-0.0092} $}}  &   {{\scriptsize $0.0528\pm 0.0085$}} & {\scriptsize$0.0533\pm 0.0079$}
\\\hline
{\scriptsize$\ln \left(10^{10} {\rm A}_{\rm s}\right)$} & {{\scriptsize $3.029\pm 0.016$}}  & {{\scriptsize $3.027\pm 0.017$}} & {{\scriptsize $3.047^{+0.017}_{-0.019}$}} &   {{\scriptsize $3.041\pm 0.017$}} & {\scriptsize $3.032\pm 0.016$}
\\\hline
{\scriptsize$n_{\rm s}$} & {{\scriptsize $0.9728\pm 0.0036$}}  & {{\scriptsize $0.9715\pm 0.0044$}} & {{\scriptsize $0.9739\pm 0.0037$}} &   {{\scriptsize $0.9915\pm 0.0070$}} & {\scriptsize $0.9744\pm 0.0041$}
\\\hline
{\scriptsize$M$} & {{\scriptsize $-19.396\pm 0.011$}}  & {{\scriptsize $-19.392\pm 0013$}} & {{\scriptsize $-19.406\pm 0.011$}} &   {{\scriptsize $-19.311\pm 0.024$}} & {\scriptsize $-19.403\pm 0.013$}
\\\hline
{\scriptsize$\sigma_8$}  & {{\scriptsize$0.7939\pm 0.0068$}}  & {{\scriptsize $0.801\pm 0.014$}} & {{\scriptsize  $0.7719\pm 0.0094$}}  &   {{\scriptsize $0.794\pm 0.012$}} & {\scriptsize $0.7876\pm 0.0096$}
\\\hline
{\scriptsize$S_8$} & {{\scriptsize$0.785\pm 0.010$}}  & {{\scriptsize $0.790\pm 0.014$}} & {{\scriptsize $0.784\pm 0.010$}}  &   {{\scriptsize $0.754\pm 0.017$}} & {\scriptsize $0.781\pm 0.011$}
\\\hline
{\scriptsize$r_{\rm d}$ (Mpc)}  & {{\scriptsize$147.97\pm 0.30$}}  & {{\scriptsize $147.85\pm 0.85$}} & {{\scriptsize $147.92\pm 0.30$}}  &   {{\scriptsize $141.3\pm 1.6$}} & {\scriptsize $148.08\pm 0.32$}
\\\hline
%{\scriptsize$\chi^2_{\rm min}$}  & {{\scriptsize $ 2735.80$}}  & {{\scriptsize $ 2736.46$}} & {{\scriptsize  $ 2726.17$}}  &   {{\scriptsize  $ 2719.88$}} & {-}
%\\\hline
{\scriptsize$\Delta{\rm DIC}$}  & {-}  & {{\scriptsize $-0.10$ }} & {{\scriptsize $+10.06$ }}  &   {{\scriptsize $+13.78$ }} & {{\scriptsize $-0.96$}}
\\\hline
\end{tabular}}
\end{center}
\caption{\scriptsize Same as in Table\,\ref{tab:Baseline_results}, but removing the high-$\ell$ polarization data from Planck and including the information provided by SH0ES.}
\label{tab:BaselineNoPolSH0ES_results}
\end{table*}

While we shall not go into theoretical details here, we have opted for mimicking such a (continuous, although quite abrupt) transition with a phenomenological $\Theta$-function approach. Thus, we have explored  the simplest scenario (with just one transition) in the context of what we have called the type-I RRVM$_{\rm thr.}$, i.e. the type-I model with a threshold. We thereby assume that the interaction between vacuum and dark matter is activated only at a threshold redshift lower than $z_*=1$\footnote{ See Sec. \ref{subsec:TypeIthreshold} for the practical implementation.}.  We have chosen this transition redshift after performing a fitting analysis allowing $z_*$ to vary freely in the Monte Carlo process.  The value $z_*\sim 1$ turns out to maximize the posterior. When the Baseline dataset is employed, we find $\nu_{\rm eff}=0.0227\pm 0.0055$ and, hence, a significant evidence for a late-time vacuum decay into dark matter at $4.1\sigma$ c.l. This allows to suppress the clustering in the universe at $z<z_*$, as it is clear from the fitting value $\sigma_8=0.773\pm 0.009$ reported in Table \ref{tab:Baseline_results} and the left-most plots in Fig. \ref{fig:base}. The small value of matter fluctuations at linear scales allows to essentially solve the tension with the $f\sigma_8$ data (see Fig. \ref{fig:fsigma8_baseline} and \cite{Nguyen:2023fip}), decreasing the value of $\chi^2_{f\sigma_8}$ by $\sim 9$ units with respect to the standard model, while keeping the good description of the other datasets (cf. Table \ref{tab:Baseline_chisquare}).

This is very remarkable and completely aligned with our previous results \cite{SolaPeracaula:2021gxi}, in which we already showed the outstanding capability of our model for producing lesser growth in the late universe\footnote{Notice that our work \cite{SolaPeracaula:2021gxi} is previous to \cite{Poulin:2022sgp}, in which the authors propose a friction mechanism\ between CDM and DE to suppress the clustering at $z\lesssim 1$.}. The vacuum decay leads to a decrease of the VED and an enhancement of $\rho_m$ at present. This produces larger values of the current $\Omega_m$, which somehow compensates the decrease of $\sigma_8$ and gives rise to values of $S_8(=\tilde{S}_8)$ of the same order to those obtained in the $\Lambda$CDM and the other models studied in this paper. The comparison of the results for the type-I RRVM$_{\rm thr.}$ reported in our Tables \ref{tab:Baseline_results}-\ref{tab:BaselineNoPolSH0ES_results} also demonstrates the robustness and stability of the fitting output under changes in the dataset. The values of DIC obtained with the Baseline configuration and also considering the SH0ES data with and without the use of the CMB polarization information from Planck are in the range $10\lesssim{\rm DIC}\lesssim 15$. Therefore we find in all cases {\it very strong} evidence for this model from a Bayesian perspective, i.e. after penalizing the use of the extra parameter $\nu_{\rm eff}$. The smallest values of DIC$\sim 10$ are obtained when the SH0ES data are also used in the analysis. This triggers a decrease of the evidence for non-zero vacuum dynamics, which still renders at the $\sim 3.5\sigma$ c.l. The model is able to solve the $\sigma_8$ tension, but does not alleviate the Hubble tension, since the values of $H_0$ stay close to those found in the $\Lambda$CDM.

Let us now move on to the type-II RRVM.  Here we switch off the exchange of energy between vacuum and the (dark) matter sector, but in compensation give allowance for a possible variation of $G$ at cosmological scales, which is induced by the running of the vacuum in accordance with the Bianchi identity.  In this type of model, therefore, the current value of the gravitational coupling may depart from $G_N$.   We have previously defined the auxiliary variable $\varphi(z)\equiv G_N/G(z)$ to parameterize such a departure. Let us also note that type-II models have two additional free parameters as compared to the $\CC$CDM (one more than type-I models): $\nueff$ and  $\varphi_{\textrm{ini}}$ (the latter being  the initial value of $\varphi$ at high redshift in the radiation dominated epoch, before it starts evolving very slowly with the cosmic evolution).  Since $G$ can vary for these models, stringent constraints on type-II models should apply from the existing limits on the relative variation of the gravitational coupling, if one assumes that the cosmological value of $G$ must satisfy them (cf. Sec. \ref{sec:backRRVMII}). These constraints are in fact satisfied by our fitting results for this type of running vacuum models. In fact, regardless of the dataset we use to fit the model, we obtain values of $\nu_{\rm eff}\sim \mathcal{O}(10^{-4})$ compatible with 0. There is no clear hint of vacuum dynamics in this case. However, in the limit $\nu_{\rm eff}\to 0$ we recover the $\Lambda$CDM only if $\varphi(0)\to 1$. This is actually the crucial ingredient that can make the type-II model a rather appealing framework for relieving the $H_0$-tension, but only in the presence of the SH0ES data as we now explain. In its absence, we obtain values of $\varphi(0)$ compatible with 1 (the strict GR value) at $<1\sigma$ c.l. Using the Baseline dataset we find $\varphi(0)=1.008\pm 0.028$, whereas we find $\varphi(0)=0.982\pm 0.030$ with the  Baseline (No pol.) alternative, which as we know is the same set but excluding the polarizations. It is obvious that the polarization data favor larger values of $\varphi$ (closer to 1) or, equivalently, smaller values of $G$ (closer to $G_N$) \footnote{This is something we already noted in previous studies within the context of the Brans-Dicke model with a cosmological constant \cite{SolaPeracaula:2019zsl,SolaPeracaula:2020vpg}.}. The improvement in the description of the data compared to the $\Lambda$CDM is in both cases only marginal within the Baseline scenario, with or without polarizations (cf. Tables \ref{tab:Baseline_chisquare} and \ref{tab:BaselineNoPol_chisquare}). This is indeed reflected in the negative  $\Delta$DIC values gathered in both cases, viz. $\Delta$DIC$\sim -(3-4)$, which point to a {\it positive} preference for the standard cosmological model (see Tables \ref{tab:Baseline_results}, \ref{tab:BaselineNoPol_results}). Now in stark contrast with the meager situation just described with the Baseline dataset, the inclusion of the SH0ES data produces a dramatic turnaround of the results in the desired direction.  It shifts the posterior of $\varphi(0)$ towards a region of lower values, which is more prominent in the absence of the CMB polarization likelihoods, to wit: $\varphi(0)=0.950\pm 0.021$ in the Baseline+SH0ES analysis and  $\varphi(0)=0.908^{+0.025}_{-0.028}$ in the Baseline (No pol.)+SH0ES one. This produces a significant decrease of the comoving sound horizon at the baryon-drag epoch $r_d$, which now lies in the ballpark $r_s\sim 141-143$ Mpc rather than in the usual higher range $147-148$ Mpc usually preferred by the $\CC$CDM model. This fact, together with a significant raise of the spectral index of the primordial power spectrum $n_s>0.98$ \cite{SolaPeracaula:2020vpg}, generates a noticeable increase in $H_0$, whose fitting constraints in the context of our analysis read now $H_0=70.8\pm 0.7$ km/s/Mpc and $H_0=71.7\pm 0.8$ km/s/Mpc, respectively\footnote{Noticeably, the central values of $r_d$, $H_0$ and the absolute magnitude of SNIa, $M$, obtained for the type-II RRVM when the CMB polarization data are excluded in the fitting analysis are in very good agreement with the model-independent measurements from low-$z$ data reported in  \cite{Favale:2023lnp}, which are also independent from the main drivers of the $H_0$ tension. For the Hubble constant these authors find $H_0=71.6 \pm3.1$ km/s/Mpc. However, these measurements have still large uncertainties and cannot arbitrate the Hubble tension yet. See also \cite{Gomez-Valent:2021hda}.}. The upshot is that the Hubble tension is basically washed out in this running vacuum model scenario\footnote{A similar phenomenology is found in the context of some modified gravity theories with a mild time evolution of $G$ and a non-negligible shift of its value with respect to $G_N$ \cite{SolaPeracaula:2019zsl,SolaPeracaula:2020vpg,Benevento:2022cql}. See also \cite{Ballesteros:2020sik,Braglia:2020iik}.}.  The incremental DIC values with respect to the vanilla model corroborate in fact a {\it strong}, or even {\it very strong}, evidence in favor of running vacuum depending on whether the CMB polarization data are  considered or not. As remarked, this happens only when we include the information from SH0ES and at the expense of worsening a bit the description of the CMB temperature spectrum -- cf. the supplementary Tables \ref{tab:BaselineSH0ES_chisquare} and \ref{tab:BaselineNoPolSH0ES_chisquare} in the appendix, in which we display the breakdown of the different $\chi^2$ contributions from each observable. Regarding the description of the LSS, the type-II RRVM is not able to improve the fit to the $f\sigma_8$ data with respect to the other models under study, as it is clear from Fig. \ref{fig:fsigma8_baseline} and the tables in the appendix. The model allows to shift the posterior values of $S_8$ towards the region preferred by the weak lensing measurements, more conspicuously in the analysis with the Baseline (No pol.)+SH0ES dataset. However, $S_8$ might not be the most correct quantity to make contact with observations in models with a renormalized gravitational coupling at cosmological scales. Alternative estimators, as $\tilde{S}_8$ (see the caption of Fig. \ref{fig:base} and also Figs. \ref{fig:base_SH0ES} and \ref{fig:base_nopol_SH0ES}), might be more appropriate. The values of $\tilde{S}_8$ are larger than those of $S_8$ and similar to those found in the other models explored in this work, including the $\Lambda$CDM.

Finally, we comment on the results obtained with the generic XCDM parameterization. It is well-known that a quintessence EoS parameter $w_0>-1$ allows to suppress the amount of structure in the universe due to the increase of dark energy in the past, which fights against the aggregation of matter. It is also known, though, that quintessence cannot alleviate the Hubble tension because the decaying nature of the DE makes in this case the critical density and, hence, also $H_0$ to be smaller at low redshifts, see e.g. \cite{Sola:2016hnq,Sola:2017znb,Sola:2018sjf}. If we do not employ the SH0ES data, i.e. if we use the Baseline and Baseline (No pol.) datasets, we find a $\sim 2\sigma$ deviation of the EoS parameter from $w_0=-1$ (a pure $\Lambda$), in the quintessence region. This is in accordance to what we have already mentioned. The LSS data, which points to a lower level of clustering than in the $\Lambda$CDM, prefers quintessence. Actually, we obtain small values of $\sigma_8\sim 0.78$. The XCDM is able to relieve the tension with the LSS data in the absence of SH0ES, but the decrease of $\chi^2_{f\sigma_8}$ is not as big as for the type-I RRVM$_{\rm thr.}$, see e.g. Table \ref{tab:Baseline_chisquare}. One can see in that table that the contribution to $\chi^2_{f\sigma_8}$ is significantly lesser (roughy a factor of two smaller) for the type-I model with threshold than in a generic XCDM parameterization. On the other hand, the improvement is less robust for the XCDM,  this being corroborated by the fact that the hints of DE dynamics disappear when we include the data from SH0ES, since the latter favors a phantom dark energy EoS parameter. This shifts $w_0$ towards smaller values. For instance, in the analysis with Baseline+SH0ES we obtain $w_0=-0.993\pm 0.020$. The phantom region, however, is not attained because we include LSS data in our analysis. In fact, the structure formation data do not favor the phantom region since in  that case the amount of DE is smaller in the past, and this does not help to prevent the excess of structure formation, which is tantamount to saying that it does not help to relieve the $\sigma_8$-tension. Thus, a compromise is needed and in the presence of SH0ES data the XCDM just provides a value of the EoS closer to $w=-1$ than in the absence of such data.  If LSS data were not used, the SH0ES data would succeed in pushing the EoS of the XCDM  to the phantom domain \cite{Gomez-Valent:2021cbe}.  In contrast to this voluble behavior of the EoS for a generic DE fluid, the type-I RRVM$_{\rm thr.}$ provides a substantially better overall fit and its effective DE behavior is quintessence-like in the structure formation region up to our days. Indeed, we find $\nueff>0$ (both with or without SH0ES data) at a large confidence level of $(3.5-4)\sigma$. Hence the vacuum energy density associated to that model is indeed decreasing with the expansion within the relevant region of structure formation for both data sets, Baseline or Baseline+SH0ES, with or without polarizations, cf. Tables \ref{tab:Baseline_results}-\ref{tab:BaselineNoPolSH0ES_results}.
%\newpage

\section{Conclusions}\label{sec:Conclusions}
In this work, we have put to the  test a class of dynamical dark energy (DDE) models that go under the name of running vacuum models (RVMs). These have been discussed for a long time in the literature, see e.g.\cite{Sola:2013gha,Sola:2015rra,SolaPeracaula:2022hpd} as well as \cite{Sola:2011qr,Sola:2014tta} and references therein. This kind of models have successfully withstood a number of litmus tests against all types of modern data, whence demonstrating its maturity and robustness as serious competitors to the concordance $\CC$CDM model, this being true not only in regard to their fitting power  but also, and indeed especially,  in regard to improving the status of the $\CC$CDM and generalizations thereof in the context of theoretical physics. The essential new feature of the RVM class is that it predicts the existence of DDE  associated with the vacuum, a fundamental concept in QFT.  Put another way: the running vacuum shows up here as if it were a form of DDE,  but in truth is (quantum) vacuum after all -- and not just another artifact extracted from  the blackbox of the DE aimed at mimicking or supplanting the fundamental notion of vacuum energy in QFT. In the RVM paradigm, there is no rigid cosmological term, $\CC$, owing to the fundamental need for renormalization of the VED in QFT. The scale of renormalization is dynamical and hence the computed quantum corrections produce a time-evolving VED with the expansion\cite{SolaPeracaula:2022hpd}. The general structure of the RVM has been recently buttressed  by explicit calculations in the context of QFT in curved spacetime. We should mention that the smooth VED dynamics in the RVM was long suspected from semi-qualitative renormalization group arguments, see the aforesaid references and corresponding bibliography, but it was  only  recently that it was substantiated in a full-fledged QFT context, see the detailed works \cite{Moreno-Pulido:2020anb,Moreno-Pulido:2022phq,Moreno-Pulido:2022upl}. Within the RVM, the gravitational coupling, $G$,  will also be running in general.  From its dynamical  interplay with the vacuum energy density (VED),  $\rv(H)$, we find that $G$ evolves very mildly as a logarithmic function of the Hubble rate, $G=G(\ln H)$. As it turns out, what we call $\Lambda$ (as a physical quantity, not just as a formal parameter) in the RVM formulation, is actually nothing but  the nearly sustained value of  $8\pi G(H)\rho_{\rm vac}(H)$  around (any)  given epoch.  There is no such thing as a true cosmological constant in the RVM framework, and as a matter of fact it is fair to say that a (physically measurable) rigid parameter of this sort is not to be expected in renormalizable QFT\,\cite{SolaPeracaula:2022hpd}.

As for the specific details of the phenomenological analysis put forward in this work and for the sake of a better comparison with previous studies -- particularly with the most recent one in Ref.\,\cite{SolaPeracaula:2021gxi} --,  in the current presentation we have focused on an implementation of the RVM which we have denoted RRVM. It has  one single (extra) parameter in the type-I formulation, $\nueff$, and two additional free parameters ($\nueff$ and $\varphi_{\rm ini}$) in the type-II RRVM with respect to the $\CC$CDM. The VED  has a dynamical component proportional to the scalar of curvature, ${\cal R}$,  being $\nueff$ its coefficient, see Eq.\,\eqref{eq:RRVM}. Such a coefficient can be accounted for analytically in QFT (see the above mentioned works) but it depends on the masses of all the quantized matter fields, so in practice it must be fitted to the overall cosmological data. This is actually the main task that we have undertaken in the present work.  In doing it,  we have found significant evidence that the VED is running with the cosmic expansion.
In fact, upon performing a global fit to the cosmological observations from a wealth of data sources of all the main sorts, thus involving the full string SNIa+BAO+$H(z)$+LSS+CMB of relevant cosmological observables, and  comparing the rigid option $\nueff=0$  (namely $\CC=$const. corresponding to the $\CC$CDM model),  with the running vacuum one ($\nueff\neq 0$) we find that a  mild dynamics  of the cosmic vacuum ($\nueff\sim 10^{-4}-10^{-2}$) is highly favored, depending on the model. For type-I RRVM with threshold we find {\it very strong direct} evidence of such vacuum dynamics through a nonvanshing value of $\nueff$ at more than $4\sigma$ c.l. and an overall statistical score of $\Delta$DIC$>+10$ with respect to the vanilla model, whereas for type-II RRVM the evidence is also {\it strong}, but indirect, through the change of $G$, which leads to a favorable scenario when we consider the SH0ES data at the level of $\Delta$DIC$>+6 (+10)$ depending on whether we use CMB polarization data or not.  We have also checked that the improvement of the fit is not just caused by a generic form of the DDE, meaning that when we test if a simple XCDM ($w$CDM) parameterization\cite{Turner:1998ex} would do a similar job we meet a negative result, i.e. in the latter case we do not observe any significant amelioration with respect to the $\CC$CDM fit.

This is in stark contrast to the fitting results from the running vacuum.  As previously indicated, for type-I models  the level of evidence turns out to be {\it very strongly} supported by the  DIC criterion (according to the conventional parlance used within the information criteria),  provided there exists a threshold  redshift for the DDE near our time ($\zstar\simeq 1$) where the vacuum evolution gets suddenly activated in the RRVM form. For the sake of simplicity, here we have mimicked it just through a $\Theta$-function. With a mild level of dynamics as that indicated above, the $\sigma_8$ tension is rendered essentially nonexistent. The relieve of the $H_0$ tension, on the other hand,  can be significantly accomplished only within the type-II model  with variable $G$ (the tension subsisting only at an inconspicuous level of $<2\sigma$). Finally, let us note that even though the type-I model cannot deal with the $H_0$ tension,  the overall fit quality that it offers in the presence of a  DDE threshold is really outstanding. Specifically, the DIC difference with respect to the vanilla $\CC$CDM is $\Delta$DIC$=+15.34$, cf. Table \ref{tab:Baseline_results}.  The type-I model with a threshold suppresses completely the $\sigma_8$ tension and provides a determination of $\nueff\neq0$ at a level of significance slightly more than $4\sigma$. We note that this intriguing result would stay even if the $H_0$ tension would fade away or would suddenly disappear. If the data on $f\sigma_8$ are free from unaccounted systematic errors, our results suggest, first of all, that it is very likely that the DE is dynamical and that it takes the running vacuum form; and, second, that such a vacuum dynamics started relatively recently ($z\sim 1$). This fact could be motivated by the same calculations underpinning the general RVM structure of the vacuum energy in QFT. The potential significance of \jtext{these considerations cannot be overemphasized} and we will certainly return to them in future studies.

%\newpage
\vspace{1cm}
\section*{Acknowledgements}
JSP and CMP are funded by projects PID2019-105614GB-C21 and FPA2016-76005-C2-1-P (MINE CO, Spain),  2021-SGR-249 (Generalitat de Catalunya) and CEX2019-000918-M (ICCUB). The work of
JSP is also partially supported by the COST Association Action (European Cooperation in Science and Technology) CA18108 “Quantum Gravity Phenomenology in the Multimessenger Approach (QG-MM)”. AGV is funded by the Istituto Nazionale di Fisica Nucleare (INFN) through the project of the InDark INFN Special Initiative: “Dark Energy and Modified Gravity Models
in the light of Low-Redshift Observations” (n. 22425/2020). JdCP is supported by the Margarita Salas fellowship funded by the European Union (NextGenerationEU). This article is also based upon work from COST Action CA21136 - Addressing observational tensions in cosmology with systematics and fundamental physics (CosmoVerse).

%%%%%%%%%%%%%%%%%%%%%%%%%%%%%%%%%%%%%%%%%%%%%%%%%%%%%%%%%%
%%%%%%%%%%%%%%%%%%%%%%%%%%%%%%%%%%%%%%%%%%%%%%%%%%%%%%%%%%
%%%%%%%%%%%%%%%%%%%%%%%%%%%%%%%%%%%%%%%%%%%%%%%%%%%%%%%%%%

\newpage

\begin{appendices}

\section{Additional tables}\label{sec:appendix}

In this appendix we present the Tables \ref{tab:Baseline_chisquare}-\ref{tab:BaselineNoPolSH0ES_chisquare}, with the individual contributions of each observable to the total $\chi^2$ for all the fitting analyses performed in this work, obtained from the mean values of the cosmological parameters. These results must be close to the true $\chi^2_{\rm min}$, since the underlying posteriors are Gaussian in very good approximation. We prefer not to use the latter, since the minimum $\chi^2$ found by \texttt{MontePython} is not always very precise, see the footnote 10 in \cite{Brinckmann:2018cvx}.

\begin{table*}[!ht]
\renewcommand{\arraystretch}{0.85}
\begin{center}
\resizebox{1\textwidth}{!}{
\begin{tabular}{|c  |c | c |  c | c | c  |c  |}
 \multicolumn{1}{c}{} & \multicolumn{4}{c}{Baseline}
\\\hline
{\scriptsize Experiment} & {\scriptsize $\Lambda$CDM}  & {\scriptsize type-I RRVM} & {\scriptsize type-I RRVM$_{\rm thr.}$}  &  {\scriptsize type-II RRVM} &  {\scriptsize XCDM}
\\\hline \hline
{\scriptsize CMB}  & {\scriptsize $2770.70$} & {\scriptsize  $2771.04$} & {\scriptsize  $2770.14$}  & {\scriptsize  $2770.48$} & {\scriptsize $2773.68$}
\\\hline
{\scriptsize SNIa} & {\scriptsize $1405.49$}  & {\scriptsize  $1405.39$} & {\scriptsize  $1403.24$}  &  {\scriptsize  $1405.64$} & {\scriptsize $1402.82$}
\\\hline
{\scriptsize $f\sigma_8$} & {\scriptsize $17.15$}  & {\scriptsize $16.92$} & {\scriptsize $8.29$}  &  {\scriptsize  $17.14$} &  {\scriptsize $15.08$}
\\\hline
{\scriptsize BAO-$f\sigma_8$ (correl.)} & {\scriptsize $19.96 $}  & {\scriptsize  $19.92$} & {\scriptsize $14.54$ }  &  {\scriptsize $19.92$} &  {\scriptsize $17.91$}
\\\hline
{\scriptsize $H(z)$ } & {\scriptsize $13.16$}  & {\scriptsize  $13.18$} & {\scriptsize $ 13.33$ }  &  {\scriptsize  $13.30$} & {\scriptsize $13.33$}
\\\hline
{\scriptsize BAO } & {\scriptsize $10.94$}  & {\scriptsize $10.97$} & {\scriptsize $10.70$}  &  {\scriptsize  $10.91$} & {\scriptsize $10.65$}
\\\hline
{\scriptsize $\chi^2_{\rm total}$} & {\scriptsize $4237.40$}  & {\scriptsize $4237.42$} & {\scriptsize $4220.24$}  &  {\scriptsize  $4237.39$} & {\scriptsize $4233.48$}
\\\hline
\end{tabular}}
\end{center}
\caption{\scriptsize Detailed breakdown of the different $\chi^2$ contributions from each observable in our Baseline dataset with the cosmological parameters reported in Table \ref{tab:Baseline_results}. We call BAO-$f\sigma_8$ (correl.) the contribution that contains the correlations between the BAO and LSS datasets (see the references of Table \ref{tab:BAO_table}), whereas the uncorrelated contributions are simply called BAO and $f\sigma_8$.}
\label{tab:Baseline_chisquare}
\end{table*}

%%%%%%%%%%%%%%%%%%%%%%%%%%%%%%%%%%%%%%%%%%%%%%%%%%%%%%%%%%%%%%%%%
%%%%%%%%%%%%%%%%%%%%%%%%%%%%%%%%%%%%%%%%%%%%%%%%%%%%%%%%%%%%%%%%%
\begin{table*}[!ht]
\renewcommand{\arraystretch}{0.85}
\begin{center}
\resizebox{1\textwidth}{!}{
\begin{tabular}{|c  |c | c |  c | c | c  |c  |}
 \multicolumn{1}{c}{} & \multicolumn{4}{c}{Baseline+SH0ES}
\\\hline
{\scriptsize Experiment} & {\scriptsize $\Lambda$CDM}  & {\scriptsize type-I RRVM} & {\scriptsize type-I RRVM$_{\rm thr.}$}  &  {\scriptsize type-II RRVM} &  {\scriptsize XCDM}
\\\hline \hline
{\scriptsize CMB}  & {\scriptsize $2774.02$} & {\scriptsize  $2771.46$} & {\scriptsize  $2774.02$}  & {\scriptsize  $2777.50$} & {\scriptsize  $2774.72$}
\\\hline
{\scriptsize SNIa} & {\scriptsize $1490.46$}  & {\scriptsize  $1488.22$} & {\scriptsize  $1491.62$}  &  {\scriptsize  $1474.38$} & {\scriptsize $1490.65$}
\\\hline
{\scriptsize $f\sigma_8$} & {\scriptsize $15.33$}  & {\scriptsize $16.92$} & {\scriptsize $8.21$}  &  {\scriptsize  $17.27$} &  {\scriptsize $14.92$}
\\\hline
{\scriptsize BAO-$f\sigma_8$ (correl.)} & {\scriptsize $19.82$}  & {\scriptsize  $21.68$} & {\scriptsize $13.35$ }  &  {\scriptsize $20.48$} &  {\scriptsize $19.21$}
\\\hline
{\scriptsize $H(z)$ } & {\scriptsize $12.97$}  & {\scriptsize  $12.85$} & {\scriptsize $13.07$ }  &  {\scriptsize  $12.47$} & {\scriptsize $13.00$}
\\\hline
{\scriptsize BAO } & {\scriptsize $10.77$}  & {\scriptsize $10.77$} & {\scriptsize  $10.43$}  &  {\scriptsize  $10.76$} & {\scriptsize $10.69$}
\\\hline
{\scriptsize $\chi^2_{\rm total}$} & {\scriptsize $4323.38$}  & {\scriptsize $4321.91$} & {\scriptsize $4310.71$}  &  {\scriptsize  $4312.85$} & {\scriptsize $4323.19$}
\\\hline
\end{tabular}}
\end{center}
\caption{\scriptsize Same as Table\,\ref{tab:Baseline_chisquare} but for the Baseline+SH0ES dataset. We have employed the parameters from Table\,\ref{tab:BaselineSH0ES_results}.}
\label{tab:BaselineSH0ES_chisquare}
\end{table*}

%%%%%%%%%%%%%%%%%%%%%%%%%%%%%%%%%%%%%%%%%%%%%%%%%%%%%%%%%%%%%%%%%
\newpage
%%%%%%%%%%%%%%%%%%%%%%%%%%%%%%%%%%%%%%%%%%%%%%%%%%%%%%%%%%%%%%%%%
\begin{table*}[!ht]
\renewcommand{\arraystretch}{0.85}
\begin{center}
\resizebox{1\textwidth}{!}{
\begin{tabular}{|c  |c | c |  c | c | c  |c  |}
 \multicolumn{1}{c}{} & \multicolumn{4}{c}{Baseline (No pol.)}
\\\hline
{\scriptsize Experiment} & {\scriptsize $\Lambda$CDM}  & {\scriptsize type-I RRVM} & {\scriptsize type-I RRVM$_{\rm thr.}$}  &  {\scriptsize type-II RRVM} &  {\scriptsize XCDM}
\\\hline \hline
{\scriptsize CMB}  & {\scriptsize $1184.03$} & {\scriptsize  $1185.16$} & {\scriptsize  $1183.39$}  & {\scriptsize  $1184.93$} & {\scriptsize  $1186.79$}
\\\hline
{\scriptsize SNIa} & {\scriptsize $1405.84$}  & {\scriptsize  $1405.51$} & {\scriptsize  $1403.41$}  &  {\scriptsize  $1405.60$} & {\scriptsize  $1402.65$}
\\\hline
{\scriptsize $f\sigma_8$} & {\scriptsize $16.03$}  & {\scriptsize $14.99$} & {\scriptsize $8.27$}  &  {\scriptsize  $15.35$} &  {\scriptsize  $13.38$}
\\\hline
{\scriptsize BAO-$f\sigma_8$ (correl.)} & {\scriptsize $19.44$}  & {\scriptsize  $19.12$} & {\scriptsize $14.12$ }  &  {\scriptsize $19.20$} &  {\scriptsize  $16.99$}
\\\hline
{\scriptsize $H(z)$ } & {\scriptsize $13.20$}  & {\scriptsize  $13.29$} & {\scriptsize $13.35$ }  &  {\scriptsize  $12.85$} & {\scriptsize  $13.38$}
\\\hline
{\scriptsize BAO} & {\scriptsize $10.91$}  & {\scriptsize $10.97$} & {\scriptsize  $10.65$}  &  {\scriptsize  $10.96$} & {\scriptsize  $10.47$}
\\\hline
{\scriptsize $\chi^2_{\rm total}$} & {\scriptsize $2649.45$}  & {\scriptsize $2649.04$} & {\scriptsize $2633.19$}  &  {\scriptsize  $2648.89$} & {\scriptsize  $2643.66$}
\\\hline
\end{tabular}}
\end{center}
\caption{\scriptsize Same as Table\,\ref{tab:Baseline_chisquare} but for the Baseline (No pol.) dataset and making use of the parameters provided in Table\,\ref{tab:BaselineNoPol_results}.}
\label{tab:BaselineNoPol_chisquare}
\end{table*}

%%%%%%%%%%%%%%%%%%%%%%%%%%%%%%%%%%%%%%%%%%%%%%%%%%%%%%%%%%%%%%%%%
%%%%%%%%%%%%%%%%%%%%%%%%%%%%%%%%%%%%%%%%%%%%%%%%%%%%%%%%%%%%%%%%%
\begin{table*}[!ht]
\renewcommand{\arraystretch}{0.85}
\begin{center}
\resizebox{1\textwidth}{!}{
\begin{tabular}{|c  |c | c |  c | c | c  |c  |}
 \multicolumn{1}{c}{} & \multicolumn{4}{c}{Baseline (No pol.) + SH0ES }
\\\hline
{\scriptsize Experiment} & {\scriptsize $\Lambda$CDM}  & {\scriptsize type-I RRVM} & {\scriptsize type-I RRVM$_{\rm thr.}$}  &  {\scriptsize type-II RRVM} &  {\scriptsize XCDM}
\\\hline \hline
{\scriptsize CMB}  & {\scriptsize $1186.28$} & {\scriptsize  $1185.07$} & {\scriptsize  $1186.32$}  & {\scriptsize  $1189.76$} & {\scriptsize  $1187.75$}
\\\hline
{\scriptsize SNIa} & {\scriptsize $1490.20$}  & {\scriptsize  $1489.18$} & {\scriptsize  $1490.80$}  &  {\scriptsize  $1469.30$} & {\scriptsize  $1490.24$}
\\\hline
{\scriptsize $f\sigma_8$} & {\scriptsize $14.15$}  & {\scriptsize $15.17$} & {\scriptsize $8.16$}  &  {\scriptsize  $15.07$} &  {\scriptsize $13.10$}
\\\hline
{\scriptsize BAO-$f\sigma_8$ (correl.)} & {\scriptsize $20.49$}  & {\scriptsize  $21.50$} & {\scriptsize $14.07$ }  &  {\scriptsize $19.25$} &  {\scriptsize $19.26$ }
\\\hline
{\scriptsize $H(z)$ } & {\scriptsize $12.97$}  & {\scriptsize  $12.91$} & {\scriptsize $12.91$ }  &  {\scriptsize  $12.79$} & {\scriptsize $13.01$ }
\\\hline
{\scriptsize BAO } & {\scriptsize $10.86$}  & {\scriptsize $10.86$} & {\scriptsize  $10.48$}  &  {\scriptsize  $10.86$} & {\scriptsize  $10.64$}
\\\hline
{\scriptsize $\chi^2_{\rm total}$} & {\scriptsize $2734.95$}  & {\scriptsize $2734.69$} & {\scriptsize $2722.74$}  &  {\scriptsize  $2717.20$} & {\scriptsize $2734.00$}
\\\hline
\end{tabular}}
\end{center}
\caption{\scriptsize Same as Table\,\ref{tab:Baseline_chisquare} but for the Baseline (No pol.)+SH0ES dataset. In this case we have used the parameters listed in Table\,\ref{tab:BaselineNoPolSH0ES_results}.}
\label{tab:BaselineNoPolSH0ES_chisquare}
\end{table*}
\FloatBarrier
\end{appendices}

%\newpage

%\bibliographystyle{ieeetr}
%\bibliography{references}

\begin{thebibliography}{100}

\bibitem{peebles:1993}
P.~J.~E. Peebles, {\em {Principles of physical cosmology}}.
\newblock Princeton University Press, 1993.

\bibitem{Peebles:1984ge}
P.~J.~E. Peebles, ``{Tests of Cosmological Models Constrained by Inflation},''
  {\em Astrophys. J.}, vol.~284, pp.~439--444, 1984.

\bibitem{Turner:2022gvw}
M.~S. Turner, ``{The Road to Precision Cosmology},'' 2201.04741 [astro-ph.CO]
  (to appear in Annu. Rev. Nucl. Part. Sci. 2022).
\newblock arXiv:2201.04741.

\bibitem{Amendola:2015ksp}
L.~Amendola and S.~Tsujikawa, {\em {Dark Energy}: {Theory and Observations}}.
\newblock Cambridge University Press, 2015.

\bibitem{Peebles:2002gy}
P.~J.~E. Peebles and B.~Ratra, ``{The Cosmological Constant and Dark Energy},''
  {\em Rev. Mod. Phys.}, vol.~75, pp.~559--606, 2003.
\newblock astro-ph/0207347.

\bibitem{Padmanabhan:2002ji}
T.~Padmanabhan, ``{Cosmological constant: The Weight of the vacuum},'' {\em
  Phys. Rept.}, vol.~380, pp.~235--320, 2003.
\newblock hep-th/0212290.

\bibitem{Riess:1998cb}
A.~G. Riess {\em et~al.}, ``Observational evidence from supernovae for an
  accelerating universe and a cosmological constant,'' {\em Astron. J.},
  vol.~116, pp.~1009--1038, 1998.
\newblock astro-ph/9805201.

\bibitem{Perlmutter:1998np}
S.~Perlmutter {\em et~al.}, ``{Measurements of $\Omega$ and $\Lambda$ from 42
  high redshift supernovae},'' {\em Astrophys. J.}, vol.~517, pp.~565--586,
  1999.
\newblock astro-ph/9812133.

\bibitem{Boomerang:2000efg}
P.~de~Bernardis {\em et~al.}, ``{A Flat universe from high resolution maps of
  the cosmic microwave background radiation},'' {\em Nature}, vol.~404,
  pp.~955--959, 2000.
\newblock astro-ph/0004404.

\bibitem{Eisenstein:2005su}
D.~J. Eisenstein {\em et~al.}, ``{Detection of the Baryon Acoustic Peak in the
  Large-Scale Correlation Function of SDSS Luminous Red Galaxies},'' {\em
  Astrophys. J.}, vol.~633, pp.~560--574, 2005.
\newblock astro-ph/0501171.

\bibitem{SDSS:2006lmn}
M.~Tegmark {\em et~al.}, ``{Cosmological Constraints from the SDSS Luminous Red
  Galaxies},'' {\em Phys. Rev. D}, vol.~74, p.~123507, 2006.
\newblock astro-ph/0608632.

\bibitem{WMAP:2008lyn}
E.~Komatsu {\em et~al.}, ``{Five-Year Wilkinson Microwave Anisotropy Probe
  (WMAP) Observations: Cosmological Interpretation},'' {\em Astrophys. J.
  Suppl.}, vol.~180, pp.~330--376, 2009.
\newblock arXiv:0803.0547.

\bibitem{Riess:2011yx}
A.~G. Riess, L.~Macri, S.~Casertano, H.~Lampeitl, H.~C. Ferguson, A.~V.
  Filippenko, S.~W. Jha, W.~Li, and R.~Chornock, ``{A 3\% Solution:
  Determination of the Hubble Constant with the Hubble Space Telescope and Wide
  Field Camera 3},'' {\em Astrophys. J.}, vol.~730, p.~119, 2011.
\newblock arXiv:1103.2976 [Erratum: Astrophys. J.732,129(2011)].

\bibitem{Planck:2015fie}
P.~A.~R. Ade {\em et~al.}, ``{Planck 2015 results. XIII. Cosmological
  parameters},'' {\em Astron. Astrophys.}, vol.~594, p.~A13, 2016.
\newblock arXiv:1502.01589.

\bibitem{Planck:2018vyg}
N.~Aghanim {\em et~al.}, ``{Planck 2018 results. VI. Cosmological
  parameters},'' {\em Astron. Astrophys.}, vol.~641, p.~A6, 2020.
\newblock arXiv:1807.06209.

\bibitem{Weinberg:1988cp}
S.~Weinberg, ``{The Cosmological Constant Problem},'' {\em Rev. Mod. Phys.},
  vol.~61, pp.~1--23, 1989.

\bibitem{Sola:2013gha}
J.~Sol{\`{a}}, ``{Cosmological constant and vacuum energy: old and new
  ideas},'' {\em J. Phys. Conf. Ser.}, vol.~453, p.~012015, 2013.
\newblock arXiv:1306.1527.

\bibitem{Steinhardt:2003st}
P.~J. Steinhardt, ``{A quintessential introduction to dark energy},'' {\em
  Phil.\ Trans.\ Roy.\ Soc.\ Lond.\ A}, vol.~361, pp.~2497--2513, 2003.

\bibitem{Sola:2015rra}
J.~Sol\`a and A.~Gómez-Valent, ``{The $\bar{\Lambda}{\rm CDM}$ cosmology: From
  inflation to dark energy through running $\Lambda$},'' {\em Int. J. Mod.
  Phys. D}, vol.~24, p.~1541003, 2015.
\newblock arXiv:1501.03832.

\bibitem{SolaPeracaula:2022hpd}
J.~Sol\`a~Peracaula, ``{The cosmological constant problem and running vacuum in
  the expanding universe},'' {\em Phil. Trans. Roy. Soc. Lond. A}, vol.~380,
  p.~20210182, 2022.
\newblock arXiv:2203.13757.

\bibitem{lemaitre1934evolution}
G.~Lema{\^\i}tre, ``Evolution of the expanding universe,'' {\em Proceedings of
  the National Academy of Sciences}, vol.~20, no.~1, pp.~12--17, 1934.

\bibitem{Gron:2018mpg}
O.~G. Gr\o{}n, ``{A hundred years with the cosmological constant},'' {\em Eur.
  J. Phys.}, vol.~39, no.~4, p.~043001, 2018.

\bibitem{zeldovich:1967}
Y.~Zel'dovich, ``{Cosmological constant and elementary particles},'' {\em Sov.
  Phys. JETP. Lett.}, vol.~6, p.~3167, 1967.

\bibitem{Zeldovich:1968}
Y.~Zel'dovich, ``{The cosmological constant and the theory of elementary
  particles},'' {\em Sov. Phys. Ups.}, vol.~11, p.~381, 1968.

\bibitem{Copeland:2006wr}
E.~J. Copeland, M.~Sami, and S.~Tsujikawa, ``{Dynamics of dark energy},'' {\em
  Int.\ J.\ Mod.\ Phys.\ D}, vol.~15, pp.~1753--1936, 2006.
\newblock hep-th/0603057.

\bibitem{Tsujikawa:2013fta}
S.~Tsujikawa, ``{Quintessence: A Review},'' {\em Class. Quant. Grav.}, vol.~30,
  p.~214003, 2013.
\newblock arXiv:1304.1961.

\bibitem{Gron:1986fv}
O.~Gr\o{}n, ``{Repulsive Gravitation and Inflationary Universe Models},'' {\em
  Am. J. Phys.}, vol.~54, pp.~46--52, 1986.

\bibitem{Wang:2017oiy}
Q.~Wang, Z.~Zhu, and W.~G. Unruh, ``How the huge energy of quantum vacuum
  gravitates to drive the slow accelerating expansion of the universe,'' {\em
  Physical Review D}, vol.~95, no.~10, p.~103504, 2017.
\newblock arXiv:1703.00543.

\bibitem{Moreno-Pulido:2020anb}
C.~Moreno-Pulido and J.~Sol\`a, ``{Running vacuum in quantum field theory in
  curved spacetime: renormalizing $\rho_{vac}$ without $\sim m^4$ terms},''
  {\em Eur. Phys. J. C}, vol.~80, no.~8, p.~692, 2020.
\newblock arXiv:2005.03164.

\bibitem{Moreno-Pulido:2022phq}
C.~Moreno-Pulido and J.~Sol\`a~Peracaula, ``{Renormalizing the vacuum energy in
  cosmological spacetime: implications for the cosmological constant
  problem},'' {\em Eur. Phys. J. C}, vol.~82, no.~6, p.~551, 2022.
\newblock arXiv:2201.05827.

\bibitem{Moreno-Pulido:2022upl}
C.~Moreno-Pulido and J.~Sol{\`{a}}~Peracaula, ``{Equation of state of the
  running vacuum},'' {\em Eur. Phys. J. C}, vol.~82, no.~12, p.~1137, 2022.
\newblock arXiv:2207.07111.

\bibitem{Moreno-Pulido:2023ryo}
C.~Moreno-Pulido, J.~Sol{\`{a}}~Peracaula, and S.~Cheraghchi, ``{Running vacuum
  in QFT in FLRW spacetime: The dynamics of $\rho_{\rm vac}(H)$ from the
  quantized matter fields},'' 2023.
\newblock arXiv:2301.05205.

\bibitem{Perivolaropoulos:2021jda}
L.~Perivolaropoulos and F.~Skara, ``{Challenges for $\Lambda$CDM: An update},''
  {\em New Astron. Rev.}, vol.~95, p.~101659, 2022.
\newblock arXiv:2105.05208.

\bibitem{DiValentino:2020zio}
E.~Di~Valentino {\em et~al.}, ``{Snowmass2021 - Letter of interest cosmology
  intertwined II: The hubble constant tension},'' {\em Astropart. Phys.},
  vol.~131, p.~102605, 2021.
\newblock arXiv:2008.11284.

\bibitem{Krishnan:2020vaf}
C.~Krishnan, E.~O. Colg\'ain, M.~M. Sheikh-Jabbari, and T.~Yang, ``{Running
  Hubble Tension and a H0 Diagnostic},'' {\em Phys. Rev. D}, vol.~103, no.~10,
  p.~103509, 2021.
\newblock arXiv:2011.02858.

\bibitem{Dainotti:2021pqg}
M.~G. Dainotti, B.~De~Simone, T.~Schiavone, G.~Montani, E.~Rinaldi, and
  G.~Lambiase, ``{On the Hubble constant tension in the SNe Ia Pantheon
  sample},'' {\em Astrophys. J.}, vol.~912, no.~2, p.~150, 2021.
\newblock arXiv:2103.02117.

\bibitem{DiValentino:2020vvd}
E.~Di~Valentino {\em et~al.}, ``{Cosmology Intertwined III: $f \sigma_8$ and
  $S_8$},'' {\em Astropart. Phys.}, vol.~131, p.~102604, 2021.
\newblock arXiv:2008.11285.

\bibitem{Macaulay:2013swa}
E.~Macaulay, I.~K. Wehus, and H.~K. Eriksen, ``{Lower Growth Rate from Recent
  Redshift Space Distortion Measurements than Expected from Planck},'' {\em
  Phys. Rev. Lett.}, vol.~111, no.~16, p.~161301, 2013.
\newblock arXiv:1303.6583.

\bibitem{Nesseris:2017vor}
S.~Nesseris, G.~Pantazis, and L.~Perivolaropoulos, ``{Tension and constraints
  on modified gravity parametrizations of $G_{\textrm{eff}}(z)$ from growth
  rate and Planck data},'' {\em Phys. Rev.}, vol.~D96, no.~2, p.~023542, 2017.
\newblock arXiv:1703.10538.

\bibitem{Lin:2017bhs}
W.~Lin and M.~Ishak, ``{Cosmological discordances II: Hubble constant, Planck
  and large-scale-structure data sets},'' {\em Phys. Rev. D}, vol.~96, no.~8,
  p.~083532, 2017.
\newblock arXiv:1708.09813.

\bibitem{Gomez-Valent:2018nib}
A.~G{\'{o}}mez-Valent and J.~Sol{\`{a}}~Peracaula, ``{Density perturbations for
  running vacuum: a successful approach to structure formation and to the
  $\sigma_8$-tension},'' {\em Mon. Not. Roy. Astron. Soc.}, vol.~478, no.~1,
  pp.~126--145, 2018.
\newblock arXiv:1801.08501.

\bibitem{Garcia-Quintero:2019cgt}
C.~Garcia-Quintero, M.~Ishak, L.~Fox, and W.~Lin, ``{Cosmological discordances.
  III. More on measure properties, large-scale-structure constraints, the
  Hubble constant and Planck data},'' {\em Phys. Rev. D}, vol.~100, no.~12,
  p.~123538, 2019.
\newblock arXiv:1910.01608.

\bibitem{KiDS:2020suj}
M.~Asgari {\em et~al.}, ``{KiDS-1000 Cosmology: Cosmic shear constraints and
  comparison between two point statistics},'' {\em Astron. Astrophys.},
  vol.~645, p.~A104, 2021.
\newblock arXiv:2007.15633.

\bibitem{DES:2021wwk}
T.~M.~C. Abbott {\em et~al.}, ``{Dark Energy Survey Year 3 results:
  Cosmological constraints from galaxy clustering and weak lensing},'' {\em
  Phys. Rev. D}, vol.~105, no.~2, p.~023520, 2022.
\newblock arXiv:2105.13549.

\bibitem{Nguyen:2023fip}
N.-M. Nguyen, D.~Huterer, and Y.~Wen, ``{Evidence for suppression of structure
  growth in the concordance cosmological model},'' 2 2023.
\newblock arXiv:2302.01331.

\bibitem{Adil:2023jtu}
S.~A. Adil, O.~Akarsu, M.~Malekjani, E.~O. Colg\'ain, S.~Pourojaghi, A.~A. Sen,
  and M.~M. Sheikh-Jabbari, ``{$S_8$ increases with effective redshift in
  $\Lambda$CDM cosmology},'' 2023.
\newblock arXiv:2303.06928.

\bibitem{Heisenberg:2022gqk}
L.~Heisenberg, H.~Villarrubia-Rojo, and J.~Zosso, ``{Can late-time extensions
  solve the H0 and \ensuremath{\sigma}8 tensions?},'' {\em Phys. Rev. D},
  vol.~106, p.~043503, 2022.
\newblock arXiv:2202.01202.

\bibitem{Marra:2021fvf}
V.~Marra and L.~Perivolaropoulos, ``{Rapid transition of Geff at
  zt\ensuremath{\simeq}0.01 as a possible solution of the Hubble and growth
  tensions},'' {\em Phys. Rev. D}, vol.~104, no.~2, p.~L021303, 2021.
\newblock arXiv:2102.06012.

\bibitem{Alestas:2020zol}
G.~Alestas, L.~Kazantzidis, and L.~Perivolaropoulos, ``{$w-M$ phantom
  transition at $z_t$ \ensuremath{<}0.1 as a resolution of the Hubble
  tension},'' {\em Phys. Rev. D}, vol.~103, no.~8, p.~083517, 2021.
\newblock 2012.13932.

\bibitem{Perivolaropoulos:2021bds}
L.~Perivolaropoulos and F.~Skara, ``{Hubble tension or a transition of the
  Cepheid SnIa calibrator parameters?},'' {\em Phys. Rev. D}, vol.~104, no.~12,
  p.~123511, 2021.
\newblock arXiv:2109.04406.

\bibitem{Alestas:2021luu}
G.~Alestas, D.~Camarena, E.~Di~Valentino, L.~Kazantzidis, V.~Marra,
  S.~Nesseris, and L.~Perivolaropoulos, ``{Late-transition versus smooth
  H(z)-deformation models for the resolution of the Hubble crisis},'' {\em
  Phys. Rev. D}, vol.~105, no.~6, p.~063538, 2022.
\newblock arXiv:2110.04336.

\bibitem{Perivolaropoulos:2022khd}
L.~Perivolaropoulos and F.~Skara, ``{A Reanalysis of the Latest SH0ES Data for
  H$_{0}$: Effects of New Degrees of Freedom on the Hubble Tension},'' {\em
  Universe}, vol.~8, no.~10, p.~502, 2022.
\newblock arXiv:2208.11169.

\bibitem{Grande:2006nn}
J.~Grande, J.~Sol\`a, and H.~Stefancic, ``{LXCDM: A Cosmon model solution to
  the cosmological coincidence problem?},'' {\em JCAP}, vol.~08, p.~011, 2006.
\newblock gr-qc/0604057.

\bibitem{Grande:2008re}
J.~Grande, A.~Pelinson, and J.~Sol\`a, ``{Dark energy perturbations and cosmic
  coincidence},'' {\em Phys. Rev. D}, vol.~79, p.~043006, 2009.
\newblock arXiv:0809.3462.

\bibitem{Abdalla:2022yfr}
E.~Abdalla {\em et~al.}, ``{Cosmology intertwined: A review of the particle
  physics, astrophysics, and cosmology associated with the cosmological
  tensions and anomalies},'' {\em JHEAp}, vol.~34, pp.~49--211, 2022.
\newblock arXiv:2203.06142.

\bibitem{Dainotti:2023yrk}
M.~Dainotti, B.~De~Simone, G.~Montani, T.~Schiavone, and G.~Lambiase, ``{The
  Hubble constant tension: current status and future perspectives through new
  cosmological probes},'' 2023.
\newblock arXiv:2301.10572.

\bibitem{Shapiro:2000dz}
I.~L. Shapiro and J.~Sol\`a, ``{Scaling behavior of the cosmological constant:
  Interface between quantum field theory and cosmology},'' {\em JHEP}, vol.~02,
  p.~006, 2002.
\newblock hep-th/0012227.

\bibitem{Babic:2004ev}
A.~Babic, B.~Guberina, R.~Horvat, and H.~Stefancic, ``{Renormalization-group
  running cosmologies. A Scale-setting procedure},'' {\em Phys. Rev.},
  vol.~D71, p.~124041, 2005.
\newblock astro-ph/0407572.

\bibitem{Alvarez:2020xmk}
P.~D. Alvarez, B.~Koch, C.~Laporte, and A.~Rinc\'on, ``{Can scale-dependent
  cosmology alleviate the $H_0$ tension?},'' {\em JCAP}, vol.~06, p.~019, 2021.
\newblock arXiv:2009.02311.

\bibitem{Ozer:1985ws}
M.~Ozer and M.~Taha, ``{A Solution to the Main Cosmological Problems},'' {\em
  Phys.\ Lett.\ B}, vol.~171, pp.~363--365, 1986.

\bibitem{Bertolami:1986bg}
O.~Bertolami, ``{Time dependent cosmological term},'' {\em Nuovo Cim.\ B},
  vol.~93, pp.~36--42, 1986.

\bibitem{Freese:1986dd}
K.~Freese, F.~C. Adams, J.~A. Frieman, and E.~Mottola, ``{Cosmology with
  Decaying Vacuum Energy},'' {\em Nucl.\ Phys.\ B}, vol.~287, pp.~797--814,
  1987.

\bibitem{Peebles:1987ek}
P.~J.~E. Peebles and B.~Ratra, ``{Cosmology with a Time Variable Cosmological
  Constant},'' {\em Astrophys. J.}, vol.~325, p.~L17, 1988.

\bibitem{Chen:1990jw}
W.~Chen and Y.~S. Wu, ``{Implications of a cosmological constant varying as
  $R^{-2}$},'' {\em Phys. Rev. D}, vol.~41, pp.~695--698, 1990.
\newblock [Erratum: Phys.Rev.D 45, 4728 (1992)].

\bibitem{Abdel-Rahman:1992msa}
A.~M.~M. Abdel-Rahman, ``{Singularity - free decaying vacuum cosmologies},''
  {\em Phys. Rev. D}, vol.~45, pp.~3497--3511, 1992.

\bibitem{Carvalho:1991ut}
J.~Carvalho, J.~Lima, and I.~Waga, ``{On the cosmological consequences of a
  time dependent lambda term},'' {\em Phys.\ Rev.\ D}, vol.~46, pp.~2404--2407,
  1992.

\bibitem{Arcuri:1993pb}
R.~C. Arcuri and I.~Waga, ``{Growth of density inhomogeneities in Newtonian
  cosmological models with variable Lambda},'' {\em Phys. Rev. D}, vol.~50,
  pp.~2928--2931, 1994.

\bibitem{Waga:1992hj}
I.~Waga, ``{Decaying vacuum flat cosmological models: Expressions for some
  observable quantities and their properties},'' {\em Astrophys. J.}, vol.~414,
  pp.~436--448, 1993.

\bibitem{Lima:1994gi}
J.~A.~S. Lima and J.~M.~F. Maia, ``{Deflationary cosmology with decaying vacuum
  energy density},'' {\em Phys. Rev. D}, vol.~49, pp.~5597--5600, 1994.

\bibitem{Lima:1995ea}
J.~A.~S. Lima and M.~Trodden, ``{Decaying vacuum energy and deflationary
  cosmology in open and closed universes},'' {\em Phys. Rev. D}, vol.~53,
  pp.~4280--4286, 1996.
\newblock astro-ph/9508049.

\bibitem{Arbab:1997ph}
A.~I. Arbab, ``{Cosmological models with variable cosmological and
  gravitational constants and bulk viscous models},'' {\em Gen. Rel. Grav.},
  vol.~29, pp.~61--74, 1997.

\bibitem{Espana-Bonet:2003qjh}
C.~Espana-Bonet~et al, ``{Testing the running of the cosmological constant with
  type Ia supernovae at high z},'' {\em JCAP}, vol.~02, p.~006, 2004.
\newblock hep-ph/0311171.

\bibitem{Wang:2004cp}
P.~Wang and X.-H. Meng, ``{Can vacuum decay in our universe?},'' {\em Class.
  Quant. Grav.}, vol.~22, pp.~283--294, 2005, astro-ph/0408495.

\bibitem{Borges:2005qs}
H.~A. Borges and S.~Carneiro, ``{Friedmann cosmology with decaying vacuum
  density},'' {\em Gen. Rel. Grav.}, vol.~37, pp.~1385--1394, 2005.
\newblock gr-qc/0503037.

\bibitem{Alcaniz:2005dg}
J.~S. Alcaniz and J.~A.~S. Lima, ``{Interpreting cosmological vacuum decay},''
  {\em Phys. Rev. D}, vol.~72, p.~063516, 2005.
\newblock astro-ph/0507372.

\bibitem{Barrow:2006hia}
J.~D. Barrow and T.~Clifton, ``{Cosmologies with energy exchange},'' {\em Phys.
  Rev. D}, vol.~73, p.~103520, 2006.
\newblock astro-ph/0604063.

\bibitem{Costa:2009wv}
F.~E.~M. Costa and J.~S. Alcaniz, ``{Cosmological consequences of a possible
  $\Lambda$-dark matter interaction},'' {\em Phys. Rev. D}, vol.~81, p.~043506,
  2010.
\newblock arXiv:0908.4251.

\bibitem{Bessada:2013maa}
D.~Bessada and O.~D. Miranda, ``{Probing a cosmological model with a $\Lambda =
  \Lambda_0 + 3\beta H^2$ decaying vacuum},'' {\em Phys. Rev. D}, vol.~88,
  no.~8, p.~083530, 2013.
\newblock arXiv:1310.8571.

\bibitem{Gomez-Valent:2014fda}
A.~G\'omez-Valent and J.~Sol\`a, ``{Vacuum models with a linear and a quadratic
  term in H: structure formation and number counts analysis},'' {\em Mon. Not.
  Roy. Astron. Soc.}, vol.~448, pp.~2810--2821, 2015.
\newblock arXiv:1412.3785.

\bibitem{Overduin:1998zv}
J.~Overduin and F.~Cooperstock, ``{Evolution of the scale factor with a
  variable cosmological term},'' {\em Phys.\ Rev.\ D}, vol.~58, p.~043506,
  1998.
\newblock astro-ph/9805260.

\bibitem{Basilakos:2019acj}
S.~Basilakos, N.~E. Mavromatos, and J.~Sol\`a~Peracaula, ``{Gravitational and
  Chiral Anomalies in the Running Vacuum Universe and Matter-Antimatter
  Asymmetry},'' {\em Phys. Rev.}, vol.~D101, no.~4, p.~045001, 2020.
\newblock arXiv:1907.04890.

\bibitem{Basilakos:2020qmu}
S.~Basilakos, N.~E. Mavromatos, and J.~Sol\`a~Peracaula, ``{Quantum Anomalies
  in String-Inspired Running Vacuum Universe: Inflation and Axion Dark
  Matter},'' {\em Phys. Lett.}, vol.~B803, p.~135342, 2020.
\newblock arXiv:2001.03465.

\bibitem{Mavromatos:2020kzj}
N.~E. Mavromatos and J.~Sol\`a~Peracaula, ``{Stringy-running-vacuum-model
  inflation: from primordial gravitational waves and stiff axion matter to
  dynamical dark energy},'' {\em Eur. Phys. J. ST}, vol.~230, no.~9,
  pp.~2077--2110, 2021.
\newblock arXiv:2012.07971.

\bibitem{Mavromatos:2021urx}
N.~E. Mavromatos and J.~Sol\`a~Peracaula, ``{Inflationary physics and
  trans-Planckian conjecture in the stringy running vacuum model: from the
  phantom vacuum to the true vacuum},'' {\em Eur. Phys. J. Plus}, vol.~136,
  no.~11, p.~1152, 2021.
\newblock arXiv:2105.02659.

\bibitem{SolaPeracaula:2021gxi}
J.~Sol\`a~Peracaula, A.~G\'omez-Valent, J.~de~Cruz~P\'erez, and
  C.~Moreno-Pulido, ``{Running vacuum against the $H_0$ and $\sigma_8$
  tensions},'' {\em EPL}, vol.~134, no.~1, p.~19001, 2021.
\newblock arXiv:2102.12758.

\bibitem{Sola:2017znb}
J.~Sol{\`{a}}~Peracaula, A.~G{\'{o}}mez-Valent, and J.~de~Cruz~P{\'{e}}rez,
  ``{The $H_0$ tension in light of vacuum dynamics in the Universe},'' {\em
  Phys. Lett.}, vol.~B774, pp.~317--324, 2017.
\newblock arXiv:1705.06723.

\bibitem{Gomez-Valent:2017idt}
A.~G\'omez-Valent and J.~Sol\`a, ``{Relaxing the $\sigma_8$-tension through
  running vacuum in the Universe},'' {\em EPL}, vol.~120, no.~3, p.~39001,
  2017.
\newblock arXiv:1711.00692.

\bibitem{Sola:2016ecz}
J.~Sol{\`{a}}~Peracaula, J.~de~Cruz~P{\'{e}}rez, and A.~G{\'{o}}mez-Valent,
  ``{Dynamical dark energy vs. $\Lambda$ = const in light of observations},''
  {\em EPL}, vol.~121, no.~3, p.~39001, 2018.
\newblock arXiv:1606.00450.

\bibitem{Sola:2016zeg}
J.~Sol\`a, ``{Cosmological constant vis-a-vis dynamical vacuum: bold
  challenging the $\Lambda$CDM},'' {\em Int. J. Mod. Phys. A}, vol.~31, no.~23,
  p.~1630035, 2016.
\newblock arXiv:1612.02449.

\bibitem{sola2017first}
J.~Sol\`a, A.~Gómez-Valent, and J.~de~Cruz~Pérez, ``{First evidence of
  running cosmic vacuum: challenging the concordance model},'' {\em Astrophys.
  J.}, vol.~836, no.~1, p.~43, 2017.
\newblock arXiv:1602.02103.

\bibitem{Sola:2017jbl}
J.~Sol{\`{a}}~Peracaula, J.~de~Cruz~P{\'{e}}rez, and A.~G{\'{o}}mez-Valent,
  ``{Possible signals of vacuum dynamics in the Universe},'' {\em Mon. Not.
  Roy. Astron. Soc.}, vol.~478, no.~4, pp.~4357--4373, 2018.
\newblock arXiv:1703.08218.

\bibitem{Sola:2015wwa}
J.~Sol\`a, A.~G\'omez-Valent, and J.~de~Cruz~P\'erez, ``{Hints of dynamical
  vacuum energy in the expanding Universe},'' {\em Astrophys. J.}, vol.~811,
  p.~L14, 2015.
\newblock arXiv:1506.05793.

\bibitem{Gomez-Valent:2014rxa}
A.~Gómez-Valent, J.~Sol\`a, and S.~Basilakos, ``{Dynamical vacuum energy in
  the expanding Universe confronted with observations: a dedicated study},''
  {\em JCAP}, vol.~01, p.~004, 2015.
\newblock arXiv:1409.7048.

\bibitem{SolaPeracaula:2022mlg}
J.~Sol\`a~Peracaula, ``{Running Vacuum and the $\Lambda$CDM tensions},'' {\em
  PoS}, vol.~CORFU2021, p.~106, 2022, doi:10.22323/1.406.0106.

\bibitem{SolaPeracaula:2018xsi}
J.~Sol{\`{a}}~Peracaula, ``{Tensions in the $\Lambda$CDM and vacuum
  dynamics},'' {\em Int. J. Mod. Phys.}, vol.~A33, no.~31, p.~1844009, 2018.

\bibitem{Sola:2007sv}
J.~Sol\`a, ``{Dark energy: A Quantum fossil from the inflationary Universe?},''
  {\em J. Phys.}, vol.~A41, p.~164066, 2008.
\newblock arXiv:0710.4151.

\bibitem{Basilakos:2012ra}
S.~Basilakos, D.~Polarski, and J.~Sol\`a, ``{Generalizing the running vacuum
  energy model and comparing with the entropic-force models},'' {\em Phys. Rev.
  D}, vol.~86, p.~043010, 2012.
\newblock arXiv:1204.4806.

\bibitem{Basilakos:2014tha}
S.~Basilakos and J.~Sol\`a, ``{Entropic-force dark energy reconsidered},'' {\em
  Phys. Rev. D}, vol.~90, no.~2, p.~023008, 2014.
\newblock arXiv:1402.6594.

\bibitem{Gomez-Valent:2015pia}
A.~G\'omez-Valent, E.~Karimkhani, and J.~Sol\`a, ``{Background history and
  cosmic perturbations for a general system of self-conserved dynamical dark
  energy and matter},'' {\em JCAP}, vol.~12, p.~048, 2015.
\newblock arXiv:1509.03298.

\bibitem{Rezaei:2019xwo}
M.~Rezaei, M.~Malekjani, and J.~Sol\`a, ``{Can dark energy be expressed as a
  power series of the Hubble parameter?},'' {\em Phys. Rev. D}, vol.~100,
  no.~2, p.~023539, 2019.
\newblock arXiv:1905.00100.

\bibitem{Rezaei:2022bkb}
M.~Rezaei and J.~Sol\`a~Peracaula, ``{Running vacuum versus holographic dark
  energy: a cosmographic comparison},'' {\em Eur. Phys. J. C}, vol.~82, p.~765,
  2022.
\newblock arXiv:2207.14250.

%\bibitem{Sola:2021txs}
%J.~Sol\`a, A.~G\'omez-Valent, J.~de~Cruz~P\'erez, and C.~Moreno-Pulido,
%  ``{Running vacuum against the $H_0$ and $\sigma_8$ tensions},'' {\em EPL},
%  vol.~134, no.~1, p.~19001, 2021.
%\newblock arXiv:2102.12758.

\bibitem{Lesgourgues:2011re}
J.~Lesgourgues, ``{The Cosmic Linear Anisotropy Solving System (CLASS) I:
  Overview},'' 4 2011.
\newblock arXiv:1104.2932.

\bibitem{Blas:2011rf}
D.~Blas, J.~Lesgourgues, and T.~Tram, ``{The Cosmic Linear Anisotropy Solving
  System (CLASS) II: Approximation schemes},'' {\em JCAP}, vol.~1107, p.~034,
  2011.
\newblock arXiv:1104.2933.

\bibitem{Ma:1995ey}
C.-P. Ma and E.~Bertschinger, ``{Cosmological perturbation theory in the
  synchronous and conformal Newtonian gauges},'' {\em Astrophys. J.}, vol.~455,
  pp.~7--25, 1995.
\newblock astro-ph/9506072.

\bibitem{Wang:2013qy}
Y.~Wang, D.~Wands, L.~Xu, J.~De-Santiago, and A.~Hojjati, ``{Cosmological
  constraints on a decomposed Chaplygin gas},'' {\em Phys. Rev. D}, vol.~87,
  no.~8, p.~083503, 2013.
\newblock arXiv:1301.5315.

\bibitem{Wang:2014xca}
Y.~Wang, D.~Wands, G.-B. Zhao, and L.~Xu, ``{Post-$Planck$ constraints on
  interacting vacuum energy},'' {\em Phys. Rev. D}, vol.~90, no.~2, p.~023502,
  2014.
\newblock arXiv:1404.5706.

\bibitem{Salvatelli:2014zta}
V.~Salvatelli, N.~Said, M.~Bruni, A.~Melchiorri, and D.~Wands, ``{Indications
  of a late-time interaction in the dark sector},'' {\em Phys. Rev. Lett.},
  vol.~113, no.~18, p.~181301, 2014.
\newblock arXiv:1406.7297.

\bibitem{Martinelli:2019dau}
M.~Martinelli, N.~B. Hogg, S.~Peirone, M.~Bruni, and D.~Wands, ``{Constraints
  on the interacting vacuum–geodesic CDM scenario},'' {\em Mon. Not. Roy.
  Astron. Soc.}, vol.~488, no.~3, pp.~3423--3438, 2019.
\newblock arXiv:1902.10694.

\bibitem{Hogg:2020rdp}
N.~B. Hogg, M.~Bruni, R.~Crittenden, M.~Martinelli, and S.~Peirone, ``{Latest
  evidence for a late time vacuum\textendash{}geodesic CDM interaction},'' {\em
  Phys. Dark Univ.}, vol.~29, p.~100583, 2020.
\newblock arXiv:2002.10449.

\bibitem{Goh:2022gxo}
L.~W.~K. Goh, A.~G\'omez-Valent, V.~Pettorino, and M.~Kilbinger,
  ``{Constraining constant and tomographic coupled dark energy with
  low-redshift and high-redshift probes},'' {\em Phys. Rev. D}, vol.~107,
  no.~8, p.~083503, 2023.
\newblock arXiv:2211.13588.

\bibitem{Fritzsch:2012qc}
H.~Fritzsch and J.~Sol\`a, ``{Matter Non-conservation in the Universe and
  Dynamical Dark Energy},'' {\em Class. Quant. Grav.}, vol.~29, p.~215002,
  2012.
\newblock arXiv:1202.5097.

\bibitem{Fritzsch:2015lua}
H.~Fritzsch and J.~Sol\`a, ``{Fundamental constants and cosmic vacuum: the
  micro and macro connection},'' {\em Mod. Phys. Lett. A}, vol.~30, no.~22,
  p.~1540034, 2015.
\newblock arXiv:1502.01411.

\bibitem{BransDicke1961}
C.~Brans and R.~Dicke, ``Mach's principle and a relativistic theory of
  gravitation,'' {\em Phys. Rev}, vol.~124, p.~925, 1961.

\bibitem{Uzan:2010pm}
J.-P. Uzan, ``{Varying Constants, Gravitation and Cosmology},'' {\em Living
  Rev. Rel.}, vol.~14, p.~2, 2011.
\newblock arXiv:1009.5514.

\bibitem{Kramer:2021jcw}
M.~Kramer {\em et~al.}, ``{Strong-Field Gravity Tests with the Double
  Pulsar},'' {\em Phys. Rev. X}, vol.~11, no.~4, p.~041050, 2021.
\newblock arXiv:2112.06795.

\bibitem{Zhu:2018etc}
W.~W. Zhu {\em et~al.}, ``{Tests of Gravitational Symmetries with Pulsar Binary
  J1713+0747},'' {\em Mon. Not. Roy. Astron. Soc.}, vol.~482, no.~3,
  pp.~3249--3260, 2018.
\newblock arXiv:1802.09206.

\bibitem{Genova2018}
A.~Genova {\em et~al.}, ``{Solar system expansion and strong equivalence
  principle as seen by the NASA MESSENGER mission},''
\newblock Nat Commun 9, 289 (2018).

\bibitem{SolaPeracaula:2019zsl}
J.~Sol\`a~Peracaula, A.~G\'omez-Valent, J.~de~Cruz~P\'erez, and
  C.~Moreno-Pulido, ``{Brans\textendash{}Dicke Gravity with a Cosmological
  Constant Smoothes Out $\Lambda$CDM Tensions},'' {\em Astrophys. J. Lett.},
  vol.~886, no.~1, p.~L6, 2019.
\newblock arXiv:1909.02554.

\bibitem{SolaPeracaula:2020vpg}
J.~Sol\`a~Peracaula, A.~G\'omez-Valent, J.~de~Cruz~P\'erez, and
  C.~Moreno-Pulido, ``{Brans\textendash{}Dicke cosmology with a $\Lambda$-term:
  a possible solution to $\Lambda$CDM tensions},'' {\em Class. Quant. Grav.},
  vol.~37, no.~24, p.~245003, 2020.
\newblock arXiv:2006.04273.

\bibitem{deCruzPerez:2023wzd}
J.~de~Cruz~P\'erez, J.~Sol\`a~Peracaula, and C.~P. Singh, ``{Running vacuum in
  Brans-Dicke theory: a possible cure for the $\sigma_8$ and $H_0$ tensions},''
  2 2023.
\newblock arXiv:2302.04807.

\bibitem{Clifton:2011jh}
T.~Clifton, P.~G. Ferreira, A.~Padilla, and C.~Skordis, ``{Modified Gravity and
  Cosmology},'' {\em Phys. Rept.}, vol.~513, pp.~1--189, 2012.
\newblock arXiv:1106.2476.

\bibitem{Avilez:2013dxa}
A.~Avilez and C.~Skordis, ``{Cosmological constraints on Brans-Dicke theory},''
  {\em Phys. Rev. Lett.}, vol.~113, no.~1, p.~011101, 2014.
\newblock arXiv:1303.4330.

\bibitem{Gomez-Valent:2021joz}
A.~G\'omez-Valent and P.~Hassan~Puttasiddappa, ``{Difficulties in reconciling
  non-negligible differences between the local and cosmological values of the
  gravitational coupling in extended Brans-Dicke theories},'' {\em JCAP},
  vol.~09, p.~040, 2021.
\newblock arXiv:2105.14819.

\bibitem{Shapiro:2004ch}
I.~L. Shapiro, J.~Sola, and H.~Stefancic, ``{Running G and Lambda at low
  energies from physics at M(X): Possible cosmological and astrophysical
  implications},'' {\em JCAP}, vol.~01, p.~012, 200.
\newblock hep-ph/0410095.

\bibitem{Carter:2018vce}
P.~Carter, F.~Beutler, W.~J. Percival, C.~Blake, J.~Koda, and A.~J. Ross,
  ``{Low Redshift Baryon Acoustic Oscillation Measurement from the
  Reconstructed 6-degree Field Galaxy Survey},'' {\em Mon. Not. Roy. Astron.
  Soc.}, vol.~481, no.~2, pp.~2371--2383, 2018.
\newblock arXiv:1803.01746.

\bibitem{Gil-Marin:2016wya}
H.~Gil-Mar{\'{i}}n, W.~J. Percival, L.~Verde, J.~R. Brownstein, C.-H. Chuang,
  F.-S. Kitaura, S.~A. Rodr{\'{i}}guez-Torres, and M.~D. Olmstead, ``{The
  clustering of galaxies in the SDSS-III Baryon Oscillation Spectroscopic
  Survey: RSD measurement from the power spectrum and bispectrum of the DR12
  BOSS galaxies},'' {\em Mon. Not. Roy. Astron. Soc.}, vol.~465, no.~2,
  pp.~1757--1788, 2017.
\newblock arXiv:1606.00439.

\bibitem{Kazin:2014qga}
E.~A. Kazin {\em et~al.}, ``{The WiggleZ Dark Energy Survey: improved distance
  measurements to z = 1 with reconstruction of the baryonic acoustic
  feature},'' {\em Mon. Not. Roy. Astron. Soc.}, vol.~441, no.~4,
  pp.~3524--3542, 2014.
\newblock arXiv:1401.0358.

\bibitem{DES:2021esc}
T.~M.~C. Abbott {\em et~al.}, ``{Dark Energy Survey Year 3 results: A 2.7\%
  measurement of baryon acoustic oscillation distance scale at redshift
  0.835},'' {\em Phys. Rev. D}, vol.~105, no.~4, p.~043512, 2022.
\newblock arXiv:2107.04646.

\bibitem{Hou:2020rse}
J.~Hou {\em et~al.}, ``{The Completed SDSS-IV extended Baryon Oscillation
  Spectroscopic Survey: BAO and RSD measurements from anisotropic clustering
  analysis of the Quasar Sample in configuration space between redshift 0.8 and
  2.2},'' {\em Mon. Not. Roy. Astron. Soc.}, vol.~500, no.~1, pp.~1201--1221,
  2020.
\newblock arXiv:2007.08998.

\bibitem{duMasdesBourboux:2020pck}
H.~du~Mas~des Bourboux {\em et~al.}, ``{The Completed SDSS-IV Extended Baryon
  Oscillation Spectroscopic Survey: Baryon Acoustic Oscillations with
  Ly\ensuremath{\alpha} Forests},'' {\em Astrophys. J.}, vol.~901, no.~2,
  p.~153, 2020.
\newblock arXiv:2007.08995.

\bibitem{Zhang:2012mp}
C.~Zhang, H.~Zhang, S.~Yuan, T.-J. Zhang, and Y.-C. Sun, ``{Four new
  observational $H(z)$ data from luminous red galaxies in the Sloan Digital Sky
  Survey data release seven},'' {\em Res. Astron. Astrophys.}, vol.~14, no.~10,
  pp.~1221--1233, 2014.
\newblock arXiv:1207.4541.

\bibitem{Jimenez:2003iv}
R.~Jim{\'{e}}nez, L.~Verde, T.~Treu, and D.~Stern, ``{Constraints on the
  equation of state of dark energy and the Hubble constant from stellar ages
  and the CMB},'' {\em Astrophys. J.}, vol.~593, pp.~622--629, 2003.
\newblock astro-ph/0302560.

\bibitem{Simon:2004tf}
J.~Simon, L.~Verde, and R.~Jim\'enez, ``{Constraints on the redshift dependence
  of the dark energy potential},'' {\em Phys. Rev.}, vol.~D71, p.~123001, 2005.
\newblock astro-ph/0412269.

\bibitem{Moresco:2012jh}
M.~Moresco {\em et~al.}, ``{Improved constraints on the expansion rate of the
  Universe up to z~1.1 from the spectroscopic evolution of cosmic
  chronometers},'' {\em JCAP}, vol.~1208, p.~006, 2012.
\newblock arXiv:1201.3609.

\bibitem{Moresco:2016mzx}
M.~Moresco, L.~Pozzetti, A.~Cimatti, R.~Jim{\'{e}}nez, C.~Maraston, L.~Verde,
  D.~Thomas, A.~Citro, R.~Tojeiro, and D.~Wilkinson, ``{A 6\% measurement of
  the Hubble parameter at $z\sim0.45$: direct evidence of the epoch of cosmic
  re-acceleration},'' {\em JCAP}, vol.~1605, no.~05, p.~014, 2016.
\newblock arXiv:1601.01701.

\bibitem{Ratsimbazafy:2017vga}
A.~Ratsimbazafy, S.~Loubser, S.~Crawford, C.~Cress, B.~Bassett, R.~Nichol, and
  P.~V\"ais\"anen, ``{Age-dating Luminous Red Galaxies observed with the
  Southern African Large Telescope},'' {\em Mon. Not. Roy. Astron. Soc.},
  vol.~467, no.~3, pp.~3239--3254, 2017.
\newblock arXiv:1702.00418.

\bibitem{Stern:2009ep}
D.~Stern, R.~Jim{\'{e}}nez, L.~Verde, M.~Kamionkowski, and S.~A. Stanford,
  ``{Cosmic Chronometers: Constraining the Equation of State of Dark Energy. I:
  H(z) Measurements},'' {\em JCAP}, vol.~1002, p.~008, 2010.
\newblock arXiv:0907.3149.

\bibitem{Borghi:2021rft}
N.~Borghi, M.~Moresco, and A.~Cimatti, ``{Toward a Better Understanding of
  Cosmic Chronometers: A New Measurement of H(z) at z \ensuremath{\sim} 0.7},''
  {\em Astrophys. J. Lett.}, vol.~928, no.~1, p.~L4, 2022.
\newblock arXiv:2110.04304.

\bibitem{Moresco:2015cya}
M.~Moresco, ``{Raising the bar: new constraints on the Hubble parameter with
  cosmic chronometers at $z \sim 2$},'' {\em Mon. Not. Roy. Astron. Soc.},
  vol.~450, no.~1, pp.~L16--L20, 2015.
\newblock arXiv:1503.01116.

\bibitem{Moresco:2020fbm}
M.~Moresco, R.~Jimenez, L.~Verde, A.~Cimatti, and L.~Pozzetti, ``{Setting the
  Stage for Cosmic Chronometers. II. Impact of Stellar Population Synthesis
  Models Systematics and Full Covariance Matrix},'' {\em Astrophys. J.},
  vol.~898, no.~1, p.~82, 2020.
\newblock arXiv:2003.07362.

\bibitem{Avila:2021dqv}
F.~Avila, A.~Bernui, E.~de~Carvalho, and C.~P. Novaes, ``{The growth rate of
  cosmic structures in the local Universe with the ALFALFA survey},'' {\em Mon.
  Not. Roy. Astron. Soc.}, vol.~505, no.~3, pp.~3404--3413, 2021.
\newblock arXiv:2105.10583.

\bibitem{Said:2020epb}
K.~Said, M.~Colless, C.~Magoulas, J.~R. Lucey, and M.~J. Hudson, ``{Joint
  analysis of 6dFGS and SDSS peculiar velocities for the growth rate of cosmic
  structure and tests of gravity},'' {\em Mon. Not. Roy. Astron. Soc.},
  vol.~497, no.~1, pp.~1275--1293, 2020.
\newblock arXiv:2007.04993.

\bibitem{Simpson:2015yfa}
F.~Simpson, C.~Blake, J.~A. Peacock, I.~Baldry, J.~Bland-Hawthorn, A.~Heavens,
  C.~Heymans, J.~Loveday, and P.~Norberg, ``{Galaxy and mass assembly: Redshift
  space distortions from the clipped galaxy field},'' {\em Phys. Rev.},
  vol.~D93, no.~2, p.~023525, 2016.
\newblock arXiv:1505.03865.

\bibitem{Blake:2013nif}
C.~Blake {\em et~al.}, ``{Galaxy And Mass Assembly (GAMA): improved cosmic
  growth measurements using multiple tracers of large-scale structure},'' {\em
  Mon. Not. Roy. Astron. Soc.}, vol.~436, p.~3089, 2013.
\newblock arXiv:1309.5556.

\bibitem{Blake:2011rj}
C.~Blake {\em et~al.}, ``{The WiggleZ Dark Energy Survey: the growth rate of
  cosmic structure since redshift z=0.9},'' {\em Mon. Not. Roy. Astron. Soc.},
  vol.~415, p.~2876, 2011.
\newblock arXiv:1104.2948.

\bibitem{Mohammad:2018mdy}
F.~G. Mohammad {\em et~al.}, ``{The VIMOS Public Extragalactic Redshift Survey
  (VIPERS): Unbiased clustering estimate with VIPERS slit assignment},'' {\em
  Astron. Astrophys.}, vol.~619, p.~A17, 2018.
\newblock arXiv:1807.05999.

\bibitem{Guzzo:2008ac}
L.~Guzzo {\em et~al.}, ``{A test of the nature of cosmic acceleration using
  galaxy redshift distortions},'' {\em Nature}, vol.~451, pp.~541--545, 2008.
\newblock arXiv:0802.1944.

\bibitem{Song:2008qt}
Y.-S. Song and W.~J. Percival, ``{Reconstructing the history of structure
  formation using Redshift Distortions},'' {\em JCAP}, vol.~0910, p.~004, 2009.
\newblock arXiv:0807.0810.

\bibitem{Okumura:2015lvp}
T.~Okumura {\em et~al.}, ``{The Subaru FMOS galaxy redshift survey (FastSound).
  IV. New constraint on gravity theory from redshift space distortions at
  $z\sim 1.4$},'' {\em Publ. Astron. Soc. Jap.}, vol.~68, no.~3, p.~38, 2016.
\newblock arXiv:1511.08083.

\bibitem{Turner:1998ex}
M.~S. Turner and M.~J. White, ``{CDM models with a smooth component},'' {\em
  Phys. Rev.}, vol.~D56, no.~8, p.~R4439, 1997.
\newblock astro-ph/9701138.

\bibitem{Scolnic:2021amr}
D.~Scolnic {\em et~al.}, ``{The Pantheon+ Analysis: The Full Data Set and
  Light-curve Release},'' {\em Astrophys. J.}, vol.~938, no.~2, p.~113, 2022.
\newblock arXiv:2112.03863.

\bibitem{Favale:2023lnp}
A.~Favale, A.~G\'omez-Valent, and M.~Migliaccio, ``{Cosmic chronometers to
  calibrate the ladders and measure the curvature of the Universe. A
  model-independent study},'' 01 2023.
\newblock arXiv:2301.09591.

\bibitem{Riess:2021jrx}
A.~G. Riess {\em et~al.}, ``{A Comprehensive Measurement of the Local Value of
  the Hubble Constant with 1 km/s/Mpc Uncertainty from the Hubble Space
  Telescope and the SH0ES Team},'' {\em Astrophys. J. Lett.}, vol.~934, no.~1,
  p.~L7, 2022.
\newblock arXiv:2112.04510.

\bibitem{Brout:2022vxf}
D.~Brout {\em et~al.}, ``{The Pantheon+ Analysis: Cosmological Constraints},''
  {\em Astrophys. J.}, vol.~938, no.~2, p.~110, 2022.
\newblock arXiv:2202.04077.

\bibitem{Aghanim:2018oex}
N.~Aghanim {\em et~al.}, ``{Planck 2018 results. VIII. Gravitational
  lensing},'' {\em Astron. Astrophys.}, vol.~641, p.~A8, 2020.
\newblock arXiv:1807.06210.

\bibitem{Jimenez:2001gg}
R.~Jim{\'{e}}nez and A.~Loeb, ``{Constraining cosmological parameters based on
  relative galaxy ages},'' {\em Astrophys. J.}, vol.~573, pp.~37--42, 2002.
\newblock astro-ph/0106145.

\bibitem{Alcock:1979mp}
C.~Alcock and B.~Paczynski, ``{An evolution free test for non-zero cosmological
  constant},'' {\em Nature}, vol.~281, pp.~358--359, 1979.

\bibitem{Kazantzidis:2018rnb}
L.~Kazantzidis and L.~Perivolaropoulos, ``{Evolution of the $f\sigma_8$ tension
  with the Planck15/$\Lambda$CDM determination and implications for modified
  gravity theories},'' {\em Phys. Rev.}, vol.~D97, no.~10, p.~103503, 2018.
\newblock arXiv:1803.01337.

\bibitem{Freedman:2019jwv}
W.~L. Freedman {\em et~al.}, ``{The Carnegie-Chicago Hubble Program. VIII. An
  Independent Determination of the Hubble Constant Based on the Tip of the Red
  Giant Branch},'' {\em Astrophys. J.}, vol.~882, p.~34, 2019.
\newblock arXiv:1907.05922.

\bibitem{Freedman:2021ahq}
W.~L. Freedman, ``{Measurements of the Hubble Constant: Tensions in
  Perspective},'' {\em Astrophys. J.}, vol.~919, no.~1, p.~16, 2021.
\newblock arXiv:2106.15656.

\bibitem{Cao:2023eja}
S.~Cao and B.~Ratra, ``{$H_0=69.8\pm1.3$$\rm{km \ s^{-1} \ Mpc^{-1}}$,
  $\Omega_{m0}=0.288\pm0.017$, and other constraints from lower-redshift,
  non-CMB and non-distance-ladder, expansion-rate data},'' 2023.
\newblock arXiv:2302.14203.

\bibitem{Yuan:2019npk}
W.~Yuan, A.~G. Riess, L.~M. Macri, S.~Casertano, and D.~Scolnic, ``{Consistent
  Calibration of the Tip of the Red Giant Branch in the Large Magellanic Cloud
  on the Hubble Space Telescope Photometric System and a Re-determination of
  the Hubble Constant},'' {\em Astrophys. J.}, vol.~886, p.~61, 2019.
\newblock arXiv:1908.00993.

\bibitem{Soltis:2020gpl}
J.~Soltis, S.~Casertano, and A.~G. Riess, ``{The Parallax of $\omega$ Centauri
  Measured from Gaia EDR3 and a Direct, Geometric Calibration of the Tip of the
  Red Giant Branch and the Hubble Constant},'' {\em Astrophys. J. Lett.},
  vol.~908, no.~1, p.~L5, 2021.
\newblock arXiv:2012.09196.

\bibitem{Anand:2021sum}
G.~S. Anand, R.~B. Tully, L.~Rizzi, A.~G. Riess, and W.~Yuan, ``{Comparing Tip
  of the Red Giant Branch Distance Scales: An Independent Reduction of the
  Carnegie-Chicago Hubble Program and the Value of the Hubble Constant},'' {\em
  Astrophys. J.}, vol.~932, no.~1, p.~15, 2022.
\newblock arXiv:2108.00007.

\bibitem{Scolnic:2023mrv}
D.~Scolnic, A.~G. Riess, J.~Wu, S.~Li, G.~S. Anand, R.~Beaton, S.~Casertano,
  R.~Anderson, S.~Dhawan, and X.~Ke, ``{CATS: The Hubble Constant from
  Standardized TRGB and Type Ia Supernova Measurements},''
\newblock arXiv:2304.06693.

\bibitem{1953JChPh..21.1087M}
N.~Metropolis, A.~W. Rosenbluth, M.~N. Rosenbluth, A.~H. Teller, and E.~Teller,
  ``{Equation of state calculations by fast computing machines},'' {\em J.
  Chem. Phys.}, vol.~21, pp.~1087--1092, 1953.

\bibitem{Hastings:1970aa}
W.~Hastings, ``{Monte Carlo Sampling Methods Using Markov Chains and Their
  Applications},'' {\em Biometrika}, vol.~57, pp.~97--109, 1970.

\bibitem{Audren:2012wb}
B.~Audren, J.~Lesgourgues, K.~Benabed, and S.~Prunet, ``{Conservative
  Constraints on Early Cosmology: an illustration of the Monte Python
  cosmological parameter inference code},'' {\em JCAP}, vol.~02, p.~001, 2013.
\newblock arXiv:1210.7183.

\bibitem{Brinckmann:2018cvx}
T.~Brinckmann and J.~Lesgourgues, ``{MontePython 3: boosted MCMC sampler and
  other features},'' {\em Phys. Dark Univ.}, vol.~24, p.~100260, 2019.
\newblock arXiv:1804.07261, "\url{https://baudren.github.io/montepython.html}".

\bibitem{R1:1997}
S.~Brooks and A.~Gelman, ``{General Methods for Monitoring Convergence of
  Iterative Simulations},'' {\em Journal of Computational and Graphical
  Statistics}, vol.~7, p.~434, 1997.

\bibitem{R2:1992}
A.~Gelman and D.~Rubin, ``{Inference from Iterative Simulation Using Multiple
  Sequences},'' {\em Statistical Science}, vol.~7, p.~457, 1992.

\bibitem{Lewis:2019xzd}
A.~Lewis, ``{GetDist: a Python package for analysing Monte Carlo samples},''
  2019.
\newblock arXiv:1910.13970,\url{https://getdist.readthedocs.io}.

\bibitem{Fixsen:2009ug}
D.~J. Fixsen, ``{The Temperature of the Cosmic Microwave Background},'' {\em
  Astrophys. J.}, vol.~707, pp.~916--920, 2009.
\newblock arXiv:0911.1955.

\bibitem{DIC}
D.~J. Spiegelhalter, N.~G. Best, B.~P. Carlin, and A.~van~der Linde, ``Bayesian
  measures of model complexity and fit,'' {\em J. Roy. Stat. Soc.}, vol.~64,
  p.~583, 2002.

\bibitem{Liddle:2007fy}
A.~R. Liddle, ``{Information criteria for astrophysical model selection},''
  {\em Mon. Not. Roy. Astron. Soc.}, vol.~377, pp.~L74--L78, 2007,
  astro-ph/0701113.

\bibitem{Akaike}
H.~Akaike, ``{A new look at the statistical model identification},'' {\em IEEE
  Trans. Autom. Control}, vol.~19, pp.~716--723, 1974.

\bibitem{Gomez-Valent:2021cbe}
A.~G\'omez-Valent, Z.~Zheng, L.~Amendola, V.~Pettorino, and C.~Wetterich,
  ``{Early dark energy in the pre- and postrecombination epochs},'' {\em Phys.
  Rev. D}, vol.~104, no.~8, p.~083536, 2021.
\newblock arXiv:2107.11065.

\bibitem{Poulin:2022sgp}
V.~Poulin, J.~L. Bernal, E.~Kovetz, and M.~Kamionkowski, ``{The Sigma-8 Tension
  is a Drag},'' 9 2022.
\newblock arXiv:2209.06217.

\bibitem{Gomez-Valent:2021hda}
A.~G\'omez-Valent, ``{Measuring the sound horizon and absolute magnitude of
  SNIa by maximizing the consistency between low-redshift data sets},'' {\em
  Phys. Rev. D}, vol.~105, no.~4, p.~043528, 2022.
\newblock arXiv:2111.15450.

\bibitem{Benevento:2022cql}
G.~Benevento, J.~A. Kable, G.~E. Addison, and C.~L. Bennett, ``{An Exploration
  of an Early Gravity Transition in Light of Cosmological Tensions},'' {\em
  Astrophys. J.}, vol.~935, no.~2, p.~156, 2022.
\newblock arXiv:2202.09356.

\bibitem{Ballesteros:2020sik}
G.~Ballesteros, A.~Notari, and F.~Rompineve, ``{The $H_0$ tension: $\Delta G_N$
  vs. $\Delta N_{\rm eff}$},'' {\em JCAP}, vol.~11, p.~024, 2020.
\newblock arXiv:2004.05049.

\bibitem{Braglia:2020iik}
M.~Braglia, M.~Ballardini, W.~T. Emond, F.~Finelli, A.~E. Gumrukcuoglu,
  K.~Koyama, and D.~Paoletti, ``{Larger value for $H_0$ by an evolving
  gravitational constant},'' {\em Phys. Rev. D}, vol.~102, no.~2, p.~023529,
  2020.
\newblock arXiv:2004.11161.

\bibitem{Sola:2016hnq}
J.~Sol{\`{a}}~Peracaula, A.~G{\'{o}}mez-Valent, and J.~de~Cruz~P{\'{e}}rez,
  ``{Dynamical dark energy: scalar fields and running vacuum},'' {\em Mod.
  Phys. Lett.}, vol.~A32, no.~9, p.~1750054, 2017.
\newblock arXiv:1610.08965.

\bibitem{Sola:2018sjf}
J.~Sol{\`{a}}~Peracaula, A.~G{\'{o}}mez-Valent, and J.~de~Cruz~P{\'{e}}rez,
  ``{Signs of Dynamical Dark Energy in Current Observations},'' {\em Phys. Dark
  Univ.}, vol.~25, p.~100311, 2019.
\newblock arXiv:1811.03505.

\bibitem{Sola:2011qr}
J.~Sol\`a, ``{Cosmologies with a time dependent vacuum},'' {\em J. Phys. Conf.
  Ser.}, vol.~283, p.~012033, 2011.
\newblock arXiv:1102.1815.

\bibitem{Sola:2014tta}
J.~Sol\`a, ``{Vacuum energy and cosmological evolution},'' {\em AIP Conf.
  Proc.}, vol.~1606, no.~1, pp.~19--37, 2015.
\newblock arXiv:1402.7049.

\end{thebibliography}

\end{document}